\documentclass[%
 reprint,
superscriptaddress,
showpacs,preprintnumbers,
 amsmath,amssymb,
aps,
pre,
floatfix,
]{revtex4-1}

\usepackage[utf8]{inputenc}

\usepackage{graphicx}
\usepackage{bm}% bold math
\usepackage{hyperref}% add hypertext capabilities
\usepackage{empheq}
\usepackage{xspace}
\usepackage{color}
\usepackage{amssymb}
\usepackage{booktabs}
\usepackage{multirow}
\usepackage{textcomp}

\hypersetup{
bookmarksnumbered = true,%     % Signets numérotés
unicode=false,          % non-Latin characters in Acrobat’s bookmarks
pdftoolbar=true,        % show Acrobat’s toolbar?
pdfmenubar=true,        % show Acrobat’s menu?
pdffitwindow=false,     % window fit to page when opened
pdfstartview={Fit},    % Comment le PDF s'ouvre autres option FitV FitH Fit FitBH FitBV
pdftitle={Passive force recovery in skeletal muscles I. Mechanical model},    % title
pdfauthor={M. Caruel},     % author
pdfsubject={Subject},   % subject of the document
pdfcreator={Creator},   % creator of the document
pdfproducer={Producer}, % producer of the document
pdfkeywords={keyword1} {key2} {key3}, % list of keywords
pdfnewwindow=true,      % links in new window
colorlinks=true,       % false: boxed links; true: colored links
linkcolor=black,          % color of internal links
citecolor=black,        % color of links to bibliography
filecolor=black,      % color of file links
urlcolor=black,           % color of external links
pdfdisplaydoctitle = true
}
\newcommand{\sech}{\mathrm{sech}}

\newcommand{\mean}[1]{\left\langle #1 \right\rangle}

\usepackage[normalem]{ulem}
\usepackage{color}
\definecolor{myorange}{rgb}{0.9568,0.4941,0.1961}
\definecolor{myred}{rgb}{0.9098,0.1294,0.2078}
\definecolor{myblue}{rgb}{0.0352,0.4981,0.6509}
\definecolor{mygreen}{rgb}{0.2235,0.6353,0.2588}

\usepackage{marvosym}

\begin{document}

\title{Statistical mechanics  of the Huxley-Simmons model}

\author{M. Caruel}
\email{matthieu.caruel@u-pec.fr}
\affiliation{MSME, CNRS-UMR 8208, 61 Avenue du G\'en\'eral de Gaulle, 94010 Cr\'eteil, France
}
\author{ L. Truskinovsky}
\affiliation{
LMS,  CNRS-UMR  7649,
Ecole Polytechnique, 91128 Palaiseau Cedex,  France
}

\date{\today}

\begin{abstract}
 
The chemomechanical model of Huxley and Simmons (HS) [A. F. Huxley and R. M. Simmons, \href{http://www.nature.com/nature/journal/v233/n5321/abs/233533a0.html}{Nature \textbf{233}, 533 (1971)}] provides a paradigmatic description of  mechanically  induced  collective conformational changes  relevant  in a variety of  biological contexts, from muscles power-stroke and hair cell gating to integrin binding and hairpin unzipping. We develop a statistical mechanical perspective on the HS model  by exploiting a formal analogy with a paramagnetic Ising model.  We  first study the  equilibrium HS model with a finite number of elements and compute  explicitly its mechanical and  thermal properties.  To model kinetics, we derive  a master equation and solve it for several loading protocols. The developed formalism is  applicable to a broad range of  allosteric systems with mean-field interactions.

\end{abstract}

\maketitle

\section{Introduction}

 Passive, mechanically induced  conformational change  in a parallel bundle of bistable elements subjected to finite temperature was first studied  theoretically in the pioneering paper of A.F.~Huxley and  R.M.~Simmons \cite{Huxley_1971}. They modeled in this way the mechanism of fast force recovery in skeletal muscles subjected to shortening under a length clamp (isometric) protocol. Such loading defines a hard device ensemble,  which has to be distinguished from a soft device (isotonic) ensemble exhibiting some rather different properties \cite{Caruel:2013jw}.  
 
The HS model  interpreted the conformational change, appearing  in the muscle context under the name of a power stroke,  in a highly simplified way,  as a  ``digital'' switch  between an extended and a contracted states. HS assumed that the contracted state is   biased by the imposed shortening  and  treated the ensuing collective folding as a deterministic chemical reaction. The information about the energetic preference of the  contracted state and about the corresponding energy barriers was encoded into the reaction rates which became functions of the ``mechanical configuration'' of the system.   
  
In the muscle literature  the chemomechanical description of HS was later refined through the inclusion of numerous additional chemical reactions   between various intermediate configurations  and their kinetics was modelled   phenomenologically \cite{Hill:1976gf, Huxley_1996,Piazzesi_1995,Smith_2008,Linari_2010a}. Almost identical descriptions of  mechanically driven conformational changes were proposed independently  in the  studies of cell adhesion \cite{Bell_1978,Erdmann_2007} and  in the context of hair cell gating \cite{Howard:1988uu,Martin_2000}. Other closely related systems include mechanical denaturation  of RNA (ribonucleic acid) and DNA (deoxyribonucleic acid) hairpins \cite{Liphardt:2001fp,Bosaeus:2012kp,Woodside:2008uz}, unzipping of biological macromolecules \citep{Gupta:2011cp,Prados:2012ks,Munoz:1998vf,Bornschlogl:2006we,Srivastava:2013tm,Gerland:2003ie,Thomas:2012ur}, collective action of SNARE (soluble N-ethylmaleimide sensitive receptor) proteins during opening of synaptic  pores \cite{Sudhof:2013fw} and even formation of  ripples in graphene sheets \citep{Bonilla:2012ib}. For all these systems the HS model can be viewed as a fundamental mean-field prototype. 
 
 The goal of the present  paper is to reassess  the  chemical reaction based approach of HS from the perspective of statistical mechanics for a system with a finite number of elements while emphasizing the role of fluctuations. In such reformulation of the HS model we  follow the pioneering work of T.L. Hill \cite{Hill:1974ks,Hill:1976gf}  and more recent developments  in Refs.~\cite{Marcucci_2010,Caruel:2013jw}.   The  zero temperature  HS  model  was studied   from this viewpoint in Ref.~\cite{Caruel:2015im} where it was  presented as  a version of a  fiber bundle model \cite{Pradhan:2010hd}.  Here we extend the analysis of Ref.~\cite{Caruel:2015im} to finite temperatures  focusing on  thermomechanical coupling   that has not been previously addressed  in the chemomechanical  framework.  

Viewed from an abstract statistical mechanics perspective,  the  HS model  is  quite similar to a paramagnetic Ising model whose thermodynamic and kinetic properties are well known \cite{Balian:2006wd}. The equivalence, however,  is  not complete  due to the presence in the HS model of an elastic  spring,  buttressing each spin element.   Another complication is the length clamp control which is  unconventional  for  magnetic analogs of the HS system.  Among the  new effects revealed by the HS model,  which  would be unusual for paramagnets in an external field,  it is enough to  mention  negative susceptibility and pseudo-critical behavior without genuine cooperativity.  

Given that the explicit formulas for the equilibrium free energy of a spin system with mean-field interactions are rather straightforward, we can easily   access  both mechanical and  thermal properties  of the HS model including the heat release associated with mechanical loading.  We can also  specify the  entropic contributions to mechanical and thermal susceptibilities  and   distinguish adiabatic from isothermal responses.  
 
To complement the equilibrium picture, we  study  in this paper the stochastic dynamics  of a HS system with a finite number of elements.  The starting point here is a  thermally induced  random walk in  the  energy landscape biased by the mechanical loading \cite{Vilfan_2003, Erdmann:2013wr}. We show that due to the mean-field nature of the interactions,  the kinetic properties  of the HS system are fully determined by the behavior of a single element.
This justifies the approach of  HS  who could  model the evolution of the first moment of the underlying probability distribution by  a single  reaction equation.

  While we did not attempt  in this paper to conduct a systematic quantitative  comparison of our statistical HS model with experiment, we included  at the end of the paper a  brief discussion of  the   relevance  of our results  for  skeletal muscles and for several other allosteric system with mean-field coupling. 
 
  The paper is organized as follows. 
  In Sec.~\ref{sec:equilibrium} we study the equilibrium properties of \( N \) bi stable elements connected in parallel and loaded in a hard device.  Sec.~\ref{sec:kinetics} contains the analysis of the  mechanical transients in this  system.  The applicability of the original HS model  for the description of skeletal muscles is discussed in Sec.~\ref{sec:skeletal_muscles}.  Various non-muscle applications are briefly reviewed in Section~\ref{sec:non_muscle_applications}.
 In Sec.~\ref{sec:conclusions} we summarize  our results and identify some open problems.

%===============================================
\begin{figure}[t]
	\centering
		\includegraphics[]{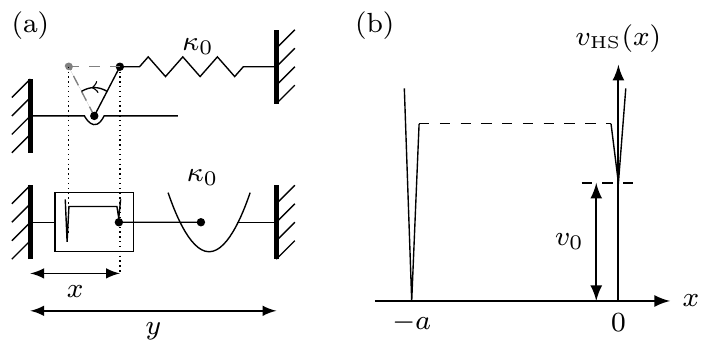}
 	\caption{ Huxley-Simmons (HS) model of a single cross-bridge:  (a) mechanical representation a myosin head and (b) energy landscape representing two chemical states.  The conformation is characterized by the spin variable \( x \) which represents two conformations of the head. The bistable element is linked in series with a linear spring of stiffness \( \kappa_0 \).}
	\label{fig:Single_cross_bridge}
\end{figure}
%============================	===================
\section{Equilibrium} % (fold)
\label{sec:equilibrium}

In this section we study the finite temperature equilibrium mechanical response of  a folding-unfolding  system containing  a finite number of  elements.

\subsection{Single HS element} % (fold)
\label{sub:a_single_hs_element}

The Huxley-Simmons paper \cite{Huxley_1971} deals  essentially with  a  single  folding element (representing a myosin cross-bridge).
The HS element  can be modeled as  an elastic spring  with stiffness \( \kappa_0 \)  (denoted by \( K \) in Ref.~\cite{Huxley_1971}) which is  connected in series  with  a bistable unit, see Fig.~\ref{fig:Single_cross_bridge}.
The two  states represent the two conformations of the myosin head and  the  variable $x$  (denoted by \( -\theta \) in Ref.~\cite{Huxley_1971}) takes the  values $0$ (pre-power-stroke or unfolded conformation) and $-a$  (post-power-stroke or  folded conformation). The discrete ``digital'' nature of the conformational state in the HS model  allows us to interpret \( x \)  as a spin variable. The soft spins (snap-springs)  version of the HS model,  corresponding to the case when each of the two energy wells is represented by a quadratic potential,  was developed in \cite{Marcucci_2010, Caruel:2013jw}, however,  the comparison of the two models shows that the additional effects due to elasticity of the conformational states are of mostly quantitative nature.

We choose \( a \), denoted by \( h \) in Ref.~\cite{Huxley_1971}, as the  ``reference'' size of the conformational change equal to the  distance between two infinitely localized energy wells, and we denote by \( v_{0} \) the intrinsic energy bias distinguishing the two states; see Fig.~\ref{fig:Single_cross_bridge}. The energy of the spin element can be now written as
\begin{equation*}
	v_{\mbox{\tiny  HS}}(x)=
	\begin{cases}
		v_{0}&\text{if } x=0, \\
		0&\text{if } x=-a,\\
		\infty&\text{otherwise}.
	\end{cases}
\end{equation*}
The energy  \( v_{0} \) is an implicit representation of the ATP-fueled activity in this otherwise passive system. The presence of such bias ensures that in the reference state the series spring is stretched and generates (active or tetanized) tension.

It will be  convenient to use dimensionless variables and   we choose  \( a \) as our characteristic distance, assuming  that the non-dimensional spin variable $x$   takes values \( 0 \) or \( -1 \). We   normalize the total energy  of the system by   \( \kappa_0 a^{2} \)  and  obtain 
\begin{equation}\label{eq:u}
	v(x;y) = (1+x)v_{0}+\frac{1}{2}(y-x)^{2},
\end{equation}
where $ x=\{0,-1\} $ and  \( y \) is the length of the combined element that includes a bistable unit and a linear spring. If we define  \( y_0=v_{0}-1/2 \) and use the muscle mechanics jargon, we can say that for \( y>y_0 \) (respectively \( y<y_0 \)) the global minimum of the energy (\ref{eq:u})  corresponds to the pre-power-stroke state (respectively post-power-stroke state); in \cite{Huxley_1971}, the shifted elongation \( y-y_0 \) was denoted by \( y  \).  

Note that our variable \( y \) plays a role of the external (magnetic) field for the spin variable $x$ and therefore our model resembles the zero dimensional Ising model of paramagnetism \cite{Balian:2006wd}.  However, due to the presence of a linear spring this Ising model is unusual:   The external field has its own ``energy'' represented by the  quadratic term in \( y \).  In the original HS experiments  a muscle  was loaded in a hard device which apparently makes this ``energy''  irrelevant.  However,  as we show below, the quadratic term in $y$  brings additional stiffness into the overall mechanical response of the system and is therefore responsible for some interesting effects.

\paragraph{Thermal equilibrium. } Denoting by \( T \) the absolute temperature we can write  the  equilibrium probability density for the configuration of a single element $x$ at fixed \( y \)   in the form
\begin{equation}\label{eq:rho_1}
	\rho_{1}(x;y,\beta) = Z_{1}(y,\beta)^{-1} \exp\left[-\beta v(x;y)\right],
\end{equation}
where
$$ \beta=\frac{\kappa_0 a^{2}}{k_{b}T}$$
 is the nondimensionalized inverse  temperature and \(k_{b}\) is the Boltzmann constant. The   partition function for a single element is  then
\begin{equation}
\label{eq:Z_1}
	Z_{1}(y,\beta)\!  =  
 \exp\!\left[
 		-\frac{\beta}{2}\!\left(y+1\right)^{2}
 	   \right]
	   \!+ \exp\!\left[
		-\beta\!\left(\frac{y^{2}}{2}+v_0\right)
		\right]\!.
\end{equation}
From  \eqref{eq:rho_1} we can compute the average conformation  \( \mean{x}=\sum_{x=\{0,-1\}} x \rho_{1}(x;y,\beta) \),  obtaining,
\begin{equation}
\label{eq:mean_x}
	\mean{x}(y,\beta) 
	=-\frac{1}{2}\left\{1-\tanh\left[\frac{\beta}{2}\left(y-y_0\right)\right]\right\},
\end{equation}
which is the analog of Eq. (15) in Ref.~\cite{Huxley_1971} (where the corresponding variable  was denoted by \(- n_{2} \)).
In paramagnetic interpretation,  \( \mean{x}(y,\beta) \) is the  ``average magnetization'' conjugate to the ``applied magnetic field'' $y$. 

%=============================================
\begin{figure}[t]
	\centering
	\includegraphics{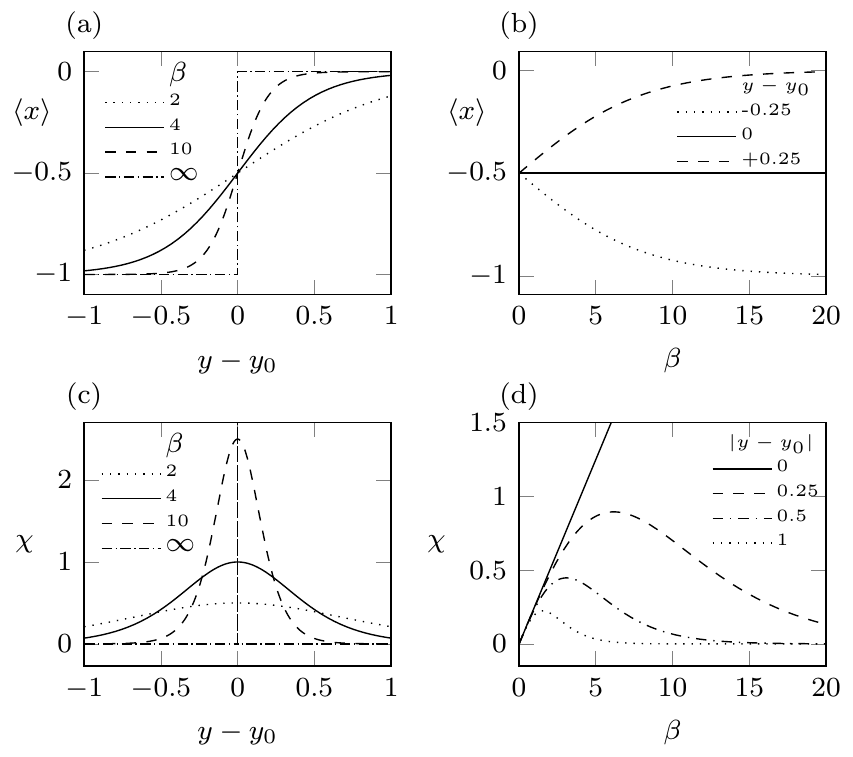}
	\caption{
	Average conformation and susceptibility of a single HS element  in thermal equilibrium.
	{(a) and (b)}, Average configuration as a function of the applied elongation at different temperatures (a) and  as a function of temperature at different  elongations (b);
	{(c) and (d)}, susceptibility as a function of elongation at different temperatures (c), and as function of the temperature for selected values of \( y \) (d).
	}
	\label{fig:average_x_and_susceptibility}
\end{figure}
%=============================================
The dependence of  $\mean{x}$  on the relative elongation \( y-y_0 \) is illustrated in Fig.~\ref{fig:average_x_and_susceptibility}(a). 
In the zero-temperature limit the  system driven through \( y \) follows the  global minimum of the internal energy \eqref{eq:u} and the population of the wells changes discontinuously  at \( y=y_0 \) \cite{Caruel:2015im}.
As the temperature increases, the transition    smoothens and in the limit \( \beta\to 0 \) we have \( \mean{x}=-1/2 \) independently of the elongation as illustrated in Fig.~\ref{fig:average_x_and_susceptibility}(b).

By differentiating  Eq.~\eqref{eq:mean_x} with respect to \( y \) we obtain the explicit representation of the equilibrium susceptibility 
\begin{equation}
	\label{eq:chi}
	\chi(y,\beta) = \frac{\partial}{\partial y}\mean{x}(y,\beta)
	=\beta \mean{\left[x-\mean{x}(y,\beta)\right]^{2}}
	% =\beta \rho_{1}(-1;y,\beta)\left[1-\rho_{1}(-1;y,\beta)\right] \geq 0
\end{equation}
which is always positive, as expected in paramagnetic systems.
Given  that the  elastic element is linear, Eq.~\eqref{eq:chi} does not depend on the particular form of the energy \( v_{\mbox{\tiny HS}} (x)\).
Thus, it also applies to  models   with more than two discrete states \cite{Linari_2010a} and even to models   with continuous energy landscape \cite{Marcucci_2010,Caruel:2013jw}.

Note  that the susceptibility is proportional to the variance of \( x \) which in the HS model takes the form
\begin{equation*}
	\mean{\left[x-\mean{x}(y,\beta)\right]^{2}} = \left(1/4\right)\left\{\sech\left[\beta\left(y-y_0\right)/2\right]\right\}^{2}.
\end{equation*}
Both quantities  will be used in what follows to assess  the intensity of fluctuations.

 In the zero-temperature limit the variance of \( x \) is negligible at large absolute elongations. Instead, at  \( y=y_0 \), the strength of fluctuations is independent of temperature and we obtain that \( \chi=\beta/4 \), which is an analog of the Curie law in paramagnetism \cite{Balian:2006wd}; see Figs.~\ref{fig:average_x_and_susceptibility}(c) and \ref{fig:average_x_and_susceptibility}(d). For   other values of elongation \( y\neq y_0 \), one can define a characteristic temperature \( \beta = \beta^{*}_{\chi}(y) \) 
solving the equation
\(
	\beta_{\chi}^{*}(y-y_0)\tanh\left[\beta_{\chi}^{*}(y-y_0)/2\right] =1.
\)
At this temperature fluctuations are maximized, see Fig.~\ref{fig:average_x_and_susceptibility}(d).  Below the characteristic temperature  the system is essentially ``frozen'' and therefore resistant to fluctuations.
Fluctuations are also irrelevant at large temperatures where the system is maximally disordered.

\paragraph{Mechanical behavior.} The free   energy  of a single HS element  in a hard device can be computed explicitly,
\begin{multline}
	\label{eq:f}
		f(y,\beta) = -\frac{1}{\beta}\log\left[Z_{1}(y,\beta)\right]
		=\frac{1}{2}\,y^{2}+v_0+\frac{y-y_0}{2}\\
		-\frac{1}{\beta}\ln\left\{2\cosh\left[\frac{\beta}{2}\left(y-y_0\right)\right]\right\}.
\end{multline}
%===============================================
\begin{figure*}[]
	\centering
		\includegraphics{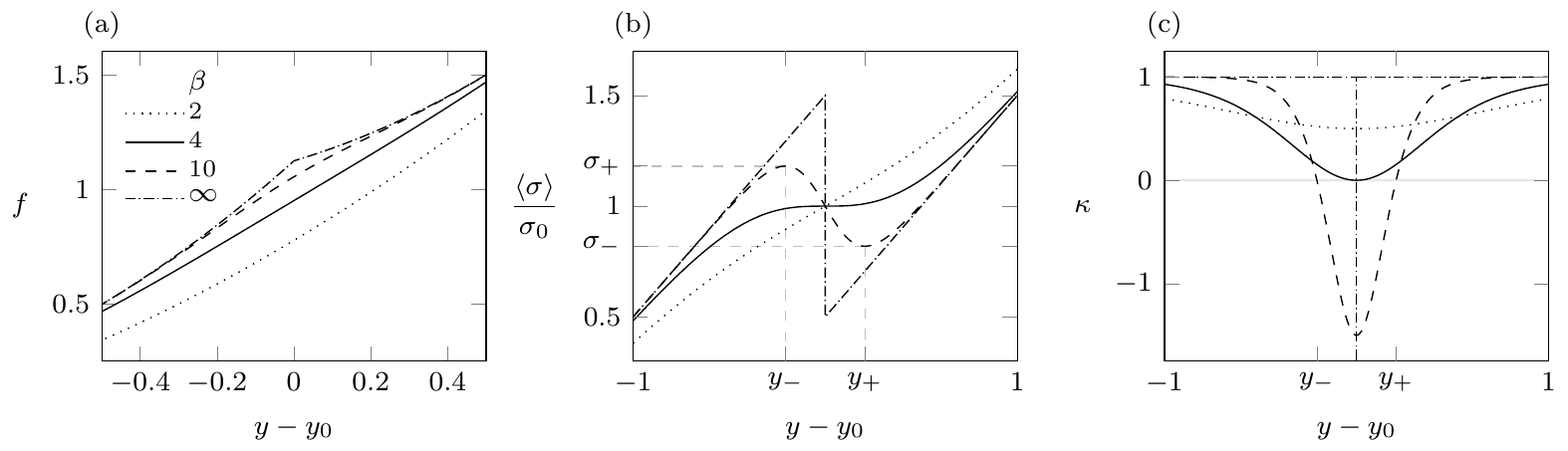}
	\caption{Thermal equilibrium properties of the HS model in a hard device for different values of temperature. (a) Helmholtz free energy; (b) tension-elongation relations; (c) stiffness. Parameters are \( \beta=2 \) (dotted), \( \beta=4 \) (solid), \( \beta=10 \) (dashed) and \( \beta\to\infty \) (dash-dotted). In the limit \( \beta\to\infty \), corresponding to zero temperature, the stiffness \( \kappa \) diverges at \( y=y_0 \), see (c).
	}
	\label{fig:thermal_equilibrium}
\end{figure*}
%===============================================
Its dependence on elongation  is illustrated in Fig.~\ref{fig:thermal_equilibrium}(a). We observe  that for \( \beta\leq 4 \) (large temperatures) the free energy is convex  while for  \( \beta>4 \) (small temperatures) it is  nonconvex. The emergence of a   ``pseudo critical''  temperature  \( \beta=\beta_{c}=4 \) in a   paramagnetic system  is a result of the  presence of the  quadratic energy associated with the ``applied field'' $y$. 

To study the mechanical manifestations of the implied ``criticality'' we introduce  
the tension   \( \sigma=y-x \) experienced by the series  linear  spring.
Due to the presence of the quadratic term \( y^2 \) in the energy,   the conjugate variable to elongation \( y \) is not the average ``magnetization'' \( \mean{x} \) but  the average tension \( \mean{\sigma} \) which is  a linear function of \( \mean{x} \)  independently of the form of the potential \eqref{eq:u}.

The convexity properties of the free energy can be  obtained  through  the study of the averaged tension-elongation relation which corresponds to Eq.~(16) in Ref.~\cite{Huxley_1971},
\begin{equation}
	\label{eq:sigma}
	\mean{\sigma}(y,\beta) =  \frac{\partial f}{\partial y}=  \sigma_0 + y-y_0-\frac{1}{2}\tanh\left[\frac{\beta}{2}(y-y_0)\right],
\end{equation}
where \( \sigma_0=v_0 \). 
The dependence of $\mean{\sigma}$ on the elongation $y-y_0$ is illustrated in Fig.~\ref{fig:thermal_equilibrium}(b) for different values of the temperature. 

We observe that while the relation \( \mean{x}(y) \) at fixed temperature is always monotone,  as it is supposed to be   in a classical paramagnetic spin system,  see Fig.~\ref{fig:average_x_and_susceptibility},  the dependence of the tension  \( \mean{\sigma} \)  on its conjugate variable  \( y \) can be nonmonotone, see  Fig.~\ref{fig:thermal_equilibrium}(b). Behind this nonmonotonicity is the fact that the equilibrium stiffness
\begin{multline}\label{eq:kappa}
		\kappa(y,\beta)= \partial\mean{\sigma}(y,\beta)/\partial y 
		= 1 - \chi(y,\beta) \\
		= 1 - \beta \mean{\left[\sigma-\mean{\sigma}(y,\beta)\right]^{2}} \\
		= 1 -   \left(\beta/4\right)\left\{\sech\left[\beta\left(y-y_0\right)/2\right]\right\}^{2}
\end{multline}
is a sign-indefinite sum of two terms. 

Equation (\ref{eq:kappa}) is a representation of the standard  \cite{Rowlinson:511572,Squire:1969dc} decomposition of an elastic susceptibility into a Cauchy-Born part associated with affine deformation   $\kappa_{\mbox{\tiny CB}}=1$,   and a fluctuation part associated with nonaffine deformation, here    $\kappa_{\mbox{\tiny F}}=\left(\beta/4\right)\left\{\sech\left[\beta\left(y-y_0\right)/2\right]\right\}^{2}$.  Interestingly,  in the HS model the fluctuation-related   term in  (\ref{eq:kappa})  does not disappear in the zero-temperature limit,  producing a singular \( \delta\)-function-type contribution to the affine response at $y=y_0$. At this value of the elongation  the global minimum  of the elastic energy is not unique and  fluctuations are formally present even in  the zero temperature (purely mechanical) model. This can be viewed as  a manifestation of a glassy behavior  \cite{Lutsko:1989bj,Wittmer:2013gz}. 

%=============================================
\begin{figure}[b]
	\centering
	\includegraphics{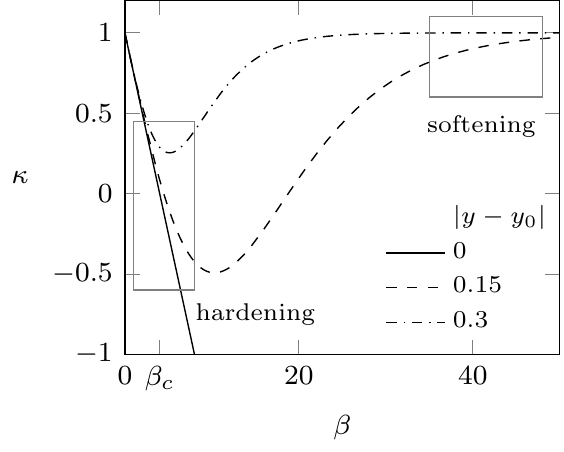}
	\caption{
 Equilibrium stiffness as function of the temperature at different levels of elongation. In the low temperature regimes (large \( \beta \)), an increase of temperature induces softening   while at high temperatures (low \( \beta \)) it induces  hardening. Close to the critical point $\beta_c$, small changes in temperatures have a large impact on the value of the stiffness  which may even change its sign.
	}
	\label{fig:stiffness}
\end{figure}
%=============================================

At finite temperatures the fluctuation-related  contribution to the elastic modulus has a standard temperature dependence  in   pure phases  \( |{y-y_0}|\gg 1 \) (softening). Instead, we observe  a rubber-elasticity-type hardening type behavior  around \( y=y_0 \), see Fig.~\ref{fig:stiffness}.
In this   mixture region  the negative entropic elasticity starts to  dominate the positive enthalpic elasticity  at \( \beta > \beta_{c} \). 

The ``critical'' temperature  \(  \beta_{c} =4\) is defined by the condition that the tension-elongation relation develops  zero stiffness   at \( y=y_0 \). In this state     \( \kappa = 1-\beta/4 \),  which can be again viewed as the analog of the Curie law in magnetism.  Negative stiffness,  resulting from non-additivity of the  system, prevails at subcritical  temperatures;  in this range a  shortening of an element   leads to a tension increase which can be interpreted as a metamaterial behavior \cite{Caruel:2013jw,Wu:2015hi}. At supercritical temperatures the stiffness becomes positive,   reaching asymptotically the value \( \kappa = 1\).  

It is remarkable that while fitting their experimental data  HS  found exactly the critical  value   \( \beta =4 \) (which corresponds to the choice  \( 4/\alpha=8 \) nm in the  units adopted in  Ref.~\cite{Huxley_1971}),  concluding implicitly that  the state of isometric contractions  is only marginally stable.  The advantages of this state are clear from Fig.~\ref{fig:stiffness}: Small variations of temperature   generate large changes in stiffness which can vary from positive to negative values and such  temperature dependence is almost insensitive to the small changes in the stretching around $y_0$.  The ``criticality''  in HS system at $y_0$, however,  is subdued  in the  hard device ensemble, similar to the behavior of a van-der-Waals gas under  controlled volume. The physical picture here is  differs from the case of  a ferromagnetic system  under  applied  magnetic  field  where interactions and cooperativity  play  an important role  and zero susceptibility  signals the presence of a real critical point with diverging fluctuations.

%===============================================
\begin{figure}t
	\centering
	\includegraphics{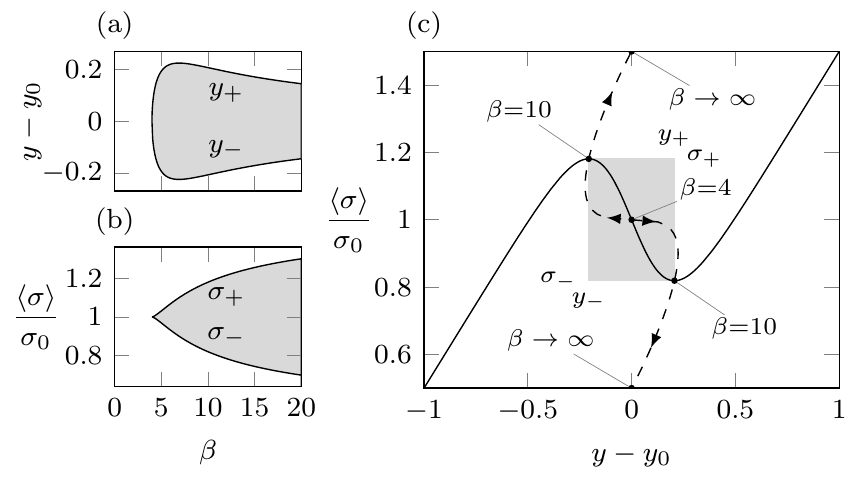}
	\caption{
The temperature dependence of the  parameter domain where the HS system exhibits negative stiffness.
In (a) and (b) we show the horizontal (\( y_{+} \) and \( y_{-} \)) and vertical (\( \sigma_{+} \) and \( \sigma_{-} \)) boundaries of the bistable domain;  in (c) the dashed line is a parametric plot of \( (y_{-}(\beta),\sigma_{+}(\beta)) \) and \( (y_{+}(\beta),\sigma_{-}(\beta)) \)   for ranging from \( \beta=4 \) to \( \beta\to\infty \). Solid line: Tension-elongation relation corresponding to \( \beta=10 \) and the associated bistable domain is represented by the gray area.
	}
	\label{fig:boundaries_of_negative_stiffness}
\end{figure}
%=============================================

To characterize the  metamaterial behavior at temperatures below critical,   we  define an interval \( [y_{-},y_{+}] \)  where the stiffness of the system is negative. The boundaries \( y_{-} \) and \( y_{+} \) correspond to the zeros of the second derivative of the free energy. For \( \beta>4 \) we have
 \begin{align*} 		y_{+}(\beta)-y_0&=\frac{1}{\beta}\log\left[\frac{\sqrt{\beta}+\sqrt{\beta-4}}{\sqrt{\beta}-\sqrt{\beta-4}}\right]\\
		y_{-}(\beta)-y_0&=-\frac{1}{\beta}\log\left[\frac{\sqrt{\beta}+\sqrt{\beta-4}}{\sqrt{\beta}-\sqrt{\beta-4}}\right]
 \end{align*}
In the zero-temperature limit  this interval collapses to a single point \( y=y_0 \). The equilibrium tensions \( \sigma_{-} \) and \( \sigma_{+} \) corresponding to \( y_{+} \) and \( y_{-} \)   are given by
\begin{align*}
		\sigma_{+}(\beta) &= v_{0} + \frac{1}{2}\sqrt{1-4\beta^{-1}}-\frac{1}{\beta}\log\left[\frac{\sqrt{\beta}+\sqrt{\beta-4}}{\sqrt{\beta}-\sqrt{\beta-4}}\right]\\
		\sigma_{-}(\beta) &= v_{0} - \frac{1}{2}\sqrt{1-4\beta^{-1}}+\frac{1}{\beta}\log\left[\frac{\sqrt{\beta}+\sqrt{\beta-4}}{\sqrt{\beta}-\sqrt{\beta-4}}\right].
\end{align*}
They become equal to \( \sigma_{+}=v_0+1/2 \) and to \( \sigma_{-}=v_{0}-1/2 \), when \( \beta\to\infty \),   see Ref.~\cite{Caruel:2015im} for more detail.
The evolution of the domain of metamaterial behavior  with temperature is shown in Fig.~\ref{fig:boundaries_of_negative_stiffness}. 

We finish this subsection with the observation  that since 
 $\mean{\sigma}(y,\beta) = y-\mean{x}(y,\beta),$ we have
$
	\langle\left[\sigma-\mean{\sigma}(y,\beta)\right]^{2}\rangle = \langle\left[x-\mean{x}(y,\beta)\right]^{2}\rangle,
$
which shows that the  fluctuations of  tension originate from the  fluctuations of the conformation. In this sense the ``noisy''   macroscopic force-elongation relations   can be used as an experimental  window into the  microscopic behavior of the system.

 %=============================================
\begin{figure}[b]
	\centering
	\includegraphics{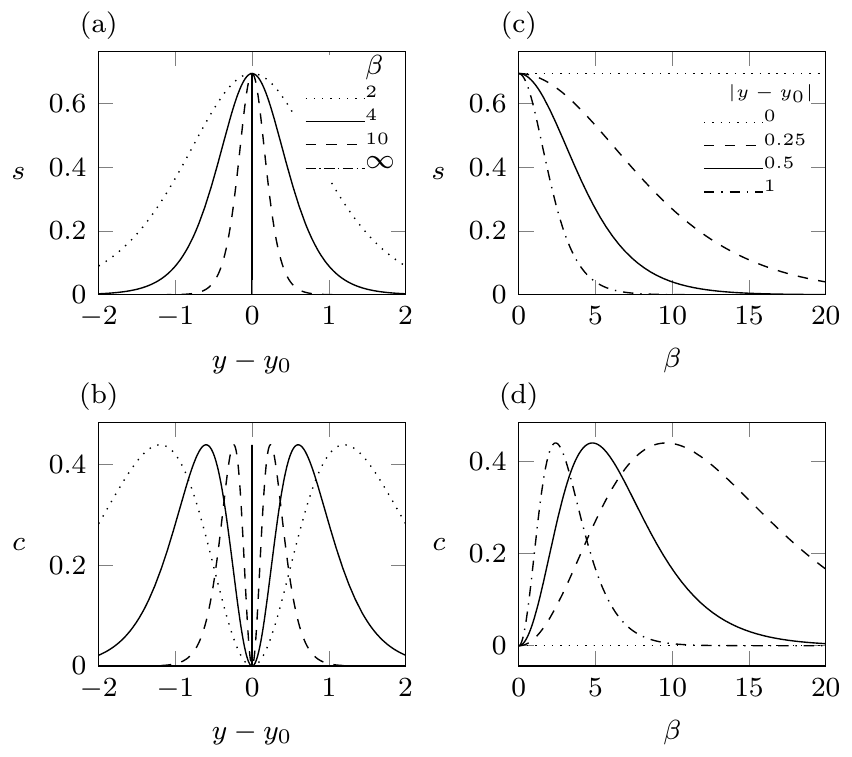}
	\caption{
	Entropy [(a) and (b)] and specific heat [(c) and (d)] in thermal equilibrium represented as function of elongation at different temperatures [(a) and (d)] and as function of temperature for different elongations [(b) and (d)].
	}
	\label{fig:entropy_and_specific_heat}
\end{figure}
%=============================================
\paragraph{Thermal behavior. } While the  experiments on  muscles have been traditionally focused on the mechanical response \cite{Eisenberg_1978,Ford_1977,Ford:1981dl,Kojima_1994,Wakabayashi_1994,Huxley_1994,Piazzesi_2007,Irving_2000}, our study suggests that  measuring the thermal or calorimetric response of such systems may be at least  as informative, see some existing work along these lines on muscles in  Refs.~\cite{Hill_1938,Piazzesi_2003,Decostre_2005,Ranatunga_2007,Coupland_2003,Woledge:2009cv,Linari:1995jr}. The statistical HS model has a considerable predictive power in this respect.  For instance, the entropy of the HS element can be  computed explicitly,
\begin{multline}
	\label{eq:s_def}
	\begin{split}
		s(y,\beta) &= -\beta\frac{\partial}{\partial\beta}\log\left[Z_{1}(y,\beta)\right] + \log\left[Z_{1}(y,\beta)\right] \\
		&= 
		\log\left\{2\cosh\left[\frac{\beta}{2}\left(y-y_0\right)\right]\right\}\\
		&\qquad-\frac{\beta}{2}\left(y-y_0\right)\tanh\left[\frac{\beta}{2}\left(y-y_0\right)\right].
	\end{split}
\end{multline}
and we illustrate the behavior of the function $s(y,\beta) $  in Figs.~\ref{fig:entropy_and_specific_heat}(a) and \ref{fig:entropy_and_specific_heat}(b).  We see that the degree of disorder is maximal in the state of isometric contractions, \( y=y_0 \).  Note also that the entropy  depends   on  a single   normalized coordinate \( \beta\left(y-y_0\right) \)  combining both control parameters,  temperature and  displacement.

A measure of the dependence of the entropy on  temperature   is   the specific heat \cite{Balian:2006wd}
\begin{equation*}
	% \label{eq:c}
	% \begin{split}
		c(y,\beta) 
		= -\beta\frac{\partial}{\partial \beta}s(y,\beta)
		\!=\!\left\{\frac{\beta}{2}\left(y-y_0\right)
	\sech\!\left[\frac{\beta}{2}(y-y_0)\right]\right\}^{2}\!\!\!,
	% \end{split}
\end{equation*}
which is represented as function of \( y-y_0 \) and \( \beta \) in Figs.~\ref{fig:entropy_and_specific_heat}(c) and \ref{fig:entropy_and_specific_heat}(d).
 As  is  typical for paramagnetic systems,  the specific heat  depends only on the combination \( \beta(y-y_0) \). 
Since  at \( y=y_0 \), the entropy is temperature insensitive [\( s(y_0)=\log(2) \)], the specific heat vanishes. Similarly, at large elongations, the systems becomes more and more ordered  and  temperature changes no longer  affect the entropy.  As a result, the specific heat is maximized at a  characteristic  value of the temperature $\beta=\beta^{*}_{c}$ which solves the equation 
\(
	\beta_{c}^{*}(y-y_0)\tanh\left[\beta_{c}^{*}(y-y_0)/2\right] =2.
\)

%\deleted{We now turn to the study of the  variance of the internal energy}  
%\begin{equation}
%	\label{eq:variance_internal_energy}
%	\mean{\left[v-\mean{v}(y,\beta)\right]^{2}} = (y-y_0)^{2}\mean{\left[x-\mean{x}(y,\beta)\right]^{2}}
%\end{equation}
%\deleted{which is  proportional to the specific heat  \cite{Balian:2006wd} 
%}

%=============================================
\begin{figure}[t]
	\centering
	\includegraphics{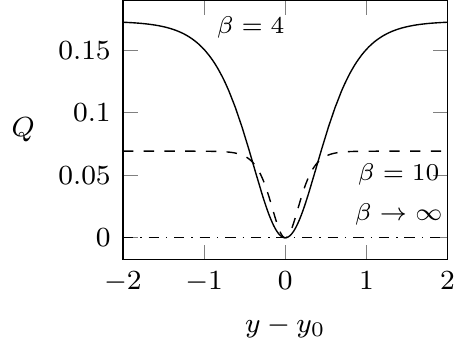}
	\caption{
	Isothermal heat released induced by a displacement from the state \( y=y_0 \) at different temperatures.
	}
	\label{fig:heat_released}
\end{figure}
%=============================================

To study  the heat release associated with the change of length we can use  our knowledge of the entropy variation with $y$. We introduce the heat release \( Q(y,\beta) = -\beta^{-1}\Delta s(y,\beta)  \), where \( \Delta s(y,\beta) = s(y,\beta) - s(y_{\mbox{\tiny in}},\beta) \), is the entropy change from the initial state \( y_{\mbox{\tiny in}} \). The function \( Q \), which  is illustrated  in Fig.~\ref{fig:heat_released}, can be potentially measured by calorimetric techniques if the system is  first driven away from equilibrium adiabatically by a rapid length change and then allowed to relax reaching the original temperature.  
%=============================================
\begin{figure}[b]
	\centering
	\includegraphics{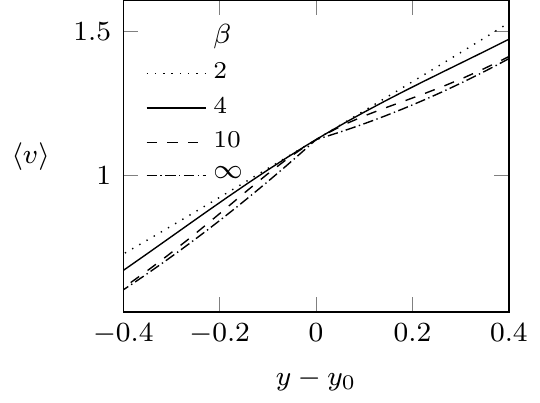}
	\caption{
	Average internal energy for \( \beta=2 \) (dotted), \( \beta=4 \) (solid), \( \beta=10 \) (dashed) and in the athermal limit \( \beta\to\infty \) (dot-dashed).
	}
	\label{fig:average_internal_energy}
\end{figure}
%=============================================

Note that the expression for entropy \eqref{eq:s_def}  can be also rewritten in the form \( s = \beta \mean{v} - \beta f \) where \( \mean{v} \) is the average internal energy
\begin{equation*}
		\mean{v}(y,\beta) 
		= y^{2}/2 + v_0 - (y-y_0)\mean{x}(y,\beta).
\end{equation*}
 In contrast to the equilibrium free energy, which decreases with temperature, see   Fig.~\ref{fig:thermal_equilibrium}(a),  the average internal energy increases with temperature, see Fig.~\ref{fig:average_internal_energy} . In the opposite zero temperature  (athermal) limit (\( \beta\to\infty \)) both the  average internal energy  and the free energy  tend  to the same limiting curve representing the  global minimum of the elastic energy   which is a nonconvex function of elongation  energy \cite{Caruel:2015im}.  Observe, however,   that    the average internal energy approaches the mechanical energy ``from above'' while the free energy approaches the mechanical energy ``from below''.

Another interesting and potentially measurable quantity is the  entropic contribution to stress which also serves   as a measure of thermal expansion   
\begin{equation*}
	\gamma = -\frac{\partial s}{\partial y }=
	 % -\beta^{2}\frac{\partial \mean{\sigma}}{\partial\beta} = 
 \frac{\beta^{2}}{4}\left(y-y_0\right)\left\{\sech\left[\frac{\beta}{2}\left(y-y_0\right)\right]\right\}^{2}.
\end{equation*}
 %=============================================
 \begin{figure}[t]
 	\centering
 	\includegraphics{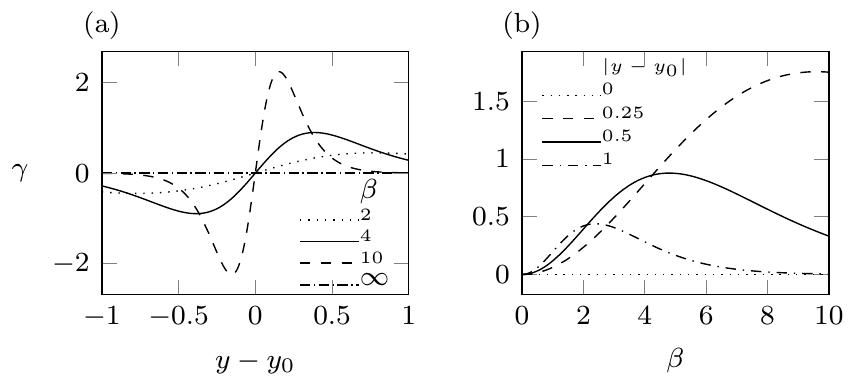}
 	\caption{
 	The dependence of the coefficient $\gamma$   on elongation  (a) and  temperature (b).
 	}
 	\label{fig:pressure_coefficient}
 \end{figure}
 %=============================================
The dependence  of $\gamma$ on elongation and temperature is illustrated in Fig.~\ref{fig:pressure_coefficient}. We observe    that for \( y>y_0 \) (respectively\( y<y_0 \)) the growth of temperature enhances (respectively diminishes) the tension, see   Fig.~\ref{fig:thermal_equilibrium}(b), and  the temperature sensitivity of tension is the highest  at a particular value of the temperature. In  large shortening or stretching regimes and at \( y=y_0 \),  the mechanical response is  temperature insensitive.

 %=============================================
\begin{figure*}[htbp]
	\centering
	\includegraphics{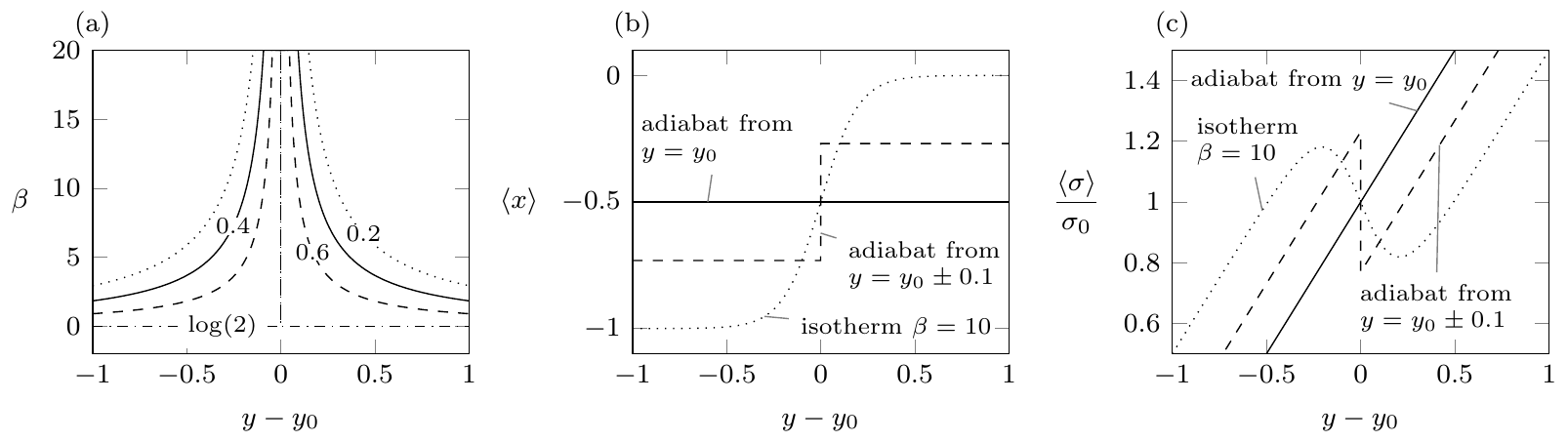}
	\caption{
	Adiabatic response. 
	(a) Evolution of temperature as function of the applied loading along adiabats for \( s=0.2 \) (dotted), \( s=0.4 \) (solid), and \( s=0.6 \) (dashed). 
	(b)  Average conformation following adiabatic length changes from two different thermal equilibrium  initial conditions \( y=y_0 \) (solid) and from \( y=y_{0} \pm 0.1\) (dashed) at \( \beta_{\mbox{\tiny in}}=10 \). The isothermal response is represented by the dotted line. (c) Adiabatic tension-elongation relations. Dotted line, isotherm response for \( \beta=10 \); dashed line, adiabatic response with initial state at \( y-y_0 = \pm 0.1 \) with \( \beta_{\mbox{\tiny in}}=10 \); solid line, adiabatic response with initial state at \( y=y_0 \) with \( \beta_{\mbox{\tiny in}}=10 \).
	}
	\label{fig:adiabats}
\end{figure*}
%=============================================

\paragraph{Adiabatic response.} The knowledge of the thermal  properties of the HS model allows one to address the question of whether the isothermal approximation is justified when applied to  experiments involving folding or unfolding under fast loading.  Below, we consider  an alternative hypothesis that the response is adiabatic, which implies that in this problem  the heat exchange is the rate-limiting process.  To remain within the equilibrium framework,  we replace the task of computing the actual adiabats by computing the isoentropes to which we will be still referring  as adiabats.  We discuss the applicability of the adiabatic assumption for the description of fast force recovery in muscles in Sec.~\ref{sec:skeletal_muscles}.

As the entropy of the system depends solely on \( \beta|y-y_0| \), see Eq.~\eqref{eq:s_def}, the temperature varies along the  adiabats  proportionally to the elongation, see Fig.~\ref{fig:adiabats}(a).  More specifically,  along an adiabat starting at $y=y_{\mathrm{in}}$ with temperature $\beta = \beta_{\mathrm{in}}$, we have
$
	\beta_{\mathrm{ad}} = \beta_{\mathrm{in}}\frac{|y_{\mathrm{in}}-y_0|}{|y-y_0|}.
$
Since, according to  Eq.~\eqref{eq:mean_x}, the average configuration depends only on \( \beta(y-y_0) \), the variation of temperature along adiabats must ensure  that the average configuration \( \mean{x} \) is preserved. This is true for every value of \( y\)  except   \( y=y_0 \) where the adiabat experiences a discontinuity. Along  adiabats the average configuration evolves according to
\begin{equation*}
	\mean{x}_{\mathrm{ad}}(y,\beta) = -\frac{1}{2} + \frac{1}{2}\tanh\left[\frac{\beta_{\mathrm{in}}|y_{\mathrm{in}}-y_{0}|}{2}\mathrm{sign}(y-y_0)\right].
\end{equation*}

The adiabatic response of the microconfiguration of the system  to abrupt ``length steps'' is illustrated in Fig.~\ref{fig:adiabats}(b), where the initial temperature is always \( \beta_{\mbox{\tiny in}}=10 \).
%\deleted{According to  Eq.~\eqref{eq:mean_x}}
Observe that  for the adiabat passing through the point \( y=y_0 \)  the  average configuration is frozen at $\mean{x}=-1/2$ (solid line); the behavior of a microconfiguration along an \emph{isotherm} passing through \( y=y_0 \) drastically differs (dotted line).

Since  equilibrium tension along the adiabats depends  linearly on  $\mean{x}$,  the adiabatic stress response to shortening from \( y=y_0 \)  is  quasilinear elastic, even though the temperature is changing.   More specifically, one can show that outside the point $y=y_0$ the adiabatic stiffness is equal to the purely mechanical stiffness  
  $$\kappa_{\mathrm{ad}}= \frac{\partial^2}{\partial y^2}f(y,\beta_{\mathrm{ ad}}(y,s)) =\kappa_0 \geq \kappa,$$
where the function $\beta_{\mathrm{ad}}(y,s)$ describes  temperature variation with elongation at a given entropy $s$.

Note that at  \( y=y_0 \),  the inverse temperature $\beta$ diverges and  even small adiabatic length change would lead to a dramatic increase of temperature (\( \beta\to 0 \)).
This means, in particular, that reaching  this state  adiabatically brings about infinite cooling. A Similar effect in a paramagnetic spin system is known as ``cooling by adiabatic demagnetization.''
In the HS system  the applied field \( y-y_0 \) can be both positive and  negative and,  in this case, if \( y<y_0 \), a shortening  would lead to a similar ``adiabatic heating,''   which can be, in principle, measured in experiment, see Section~\ref{sec:skeletal_muscles}. 

For the adiabats starting at other points \( y \neq y_0 \)  the average configuration $\mean{x}$ is  frozen at its initial value  until the loading reaches the  point \( y=y_0 \).
At this point the continuity of entropy requires that the configuration changes discontinuously.
Due to adiabatic cooling at \( y-y_0 \)  the temperature  goes to zero and the response becomes discontinuous (quasimechanical, see \cite{Caruel:2015im}).
This is in stark contrast with continuous evolution of the configuration along a typical  isotherm  also shown  in  Fig.~\ref{fig:adiabats}(b) (dotted line).
One can say that, during adiabatic response, the  temperature-induced  smoothing of the force-elongation relation gets overridden by the anomalous cooling around the point \( y=y_0 \). 

The adiabatic tension-elongation relations originating from this behavior of the microconfiguration are piecewise linear with stiffness equal to 1 for  \( y \neq y_0 \),  see  Fig.~\ref{fig:adiabats}(c).
The presence of a discontinuity  at \( y=y_0 \) signifies an extreme  metamaterial-type  behavior. Interestingly, the  adiabat originating exactly from the equilibrium state \( y=y_0 \) can be confused with the purely elastic isothermal force elongation relation.
The associated temperature variation, however, is non-negligible and should be, in principle,  measurable in experiments.

% subsection a_single_hs_element (end)

\subsection{Bundle of HS elements} % (fold)
\label{sub:a_bundle_of_hs_element}

\begin{figure}[t]
	\centering
	\includegraphics{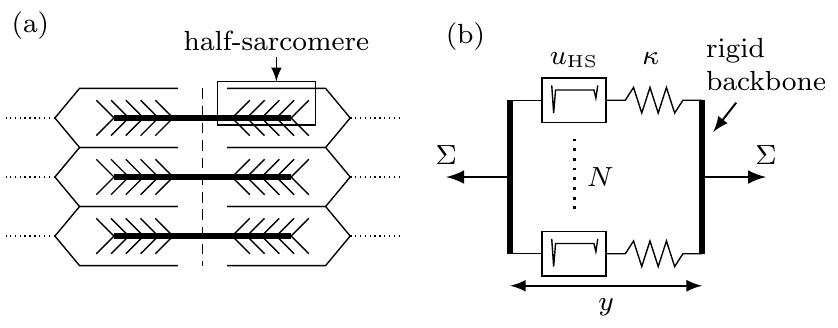}
	\caption{
	(a) Schematic representation of an acto-myosin filaments organization in a superstructure of half-sarcomeres.
	(b) A single half-sarcomere represented as a of cluster containing \( N \) cross-linkers, see Fig.~\ref{fig:Single_cross_bridge}.
	The control parameter is the total elongation \( y \) and the total tension generated by the system is denoted by \( \Sigma \).
	}
	\label{fig:N_cross_bridges}
\end{figure}

  Consider now  a finite number of HS elements attached in parallel between two rigid backbones.
  In the skeletal muscle context, such a bundle represents a minimal acto-myosin complex which we refer to as an elementary  half-sarcomere,  see Fig.~\ref{fig:N_cross_bridges}.
  
  The energy of the system  with $N$ elements can be written as
\begin{equation*}
	e(\boldsymbol{x};y) = \frac{1}{N} \sum_{i=1}^{N} \left[\left(1+x_{i}\right)v_0 + \frac{1}{2}\left(y-x_{i}\right)^{2}\right],
\end{equation*}
where \( \boldsymbol{x} = \{x_{1},\dots,x_{N}\} \).
The  individual  bistable elements do not interact  among themselves while they all interact with the same external field \( y \).  The origin of this mean-field type interaction is a hard device constraint which is not affected by the microconfiguration of the system.  In the language of magnetism, we are dealing here with  a one-dimensional paramagnetic system.  In fact, for such systems, the dimensionality is irrelevant and one  can expect  the results obtained for the zero-dimensional model to remain valid for the case of  $N$ elements.

%=============================================

\paragraph{ Thermal equilibrium.}  In thermal equilibrium, the probability density for a  micro-state \(\boldsymbol{x} \)   reads
\begin{equation}
	\label{eq:rho}
	\rho(\boldsymbol{x};y,\beta) = Z(y,\beta)^{-1}\exp\left[-\beta\, e(\boldsymbol{x};y)\right],
\end{equation}
where the partition function  is  
\begin{equation*}
	Z(y,\beta)=\sum_{\boldsymbol{x}\in\{0,-1\}^{N}}
	\exp\left[-\beta N e(\boldsymbol{x};y) \right].
\end{equation*}
Due to the additivity of the energy  we obtain 
\(
 Z(y,\beta) = 
 \left[
 	Z_{1}(y,\beta)
\right]^{N},
\)
where \( Z_{1} \) is given by (\ref{eq:Z_1}). Therefore   
\(
\rho (\boldsymbol{x};y,\beta) = \prod_{i=1}^{N}\rho_{1}(x_{i};y,\beta)
\)
which shows that the elements are independent.

The total free energy can be written as \(F(y,\beta)= Nf(y,\beta),\) where the expression for the free energy of a single HS element $f$ is given by  (\ref{eq:f}); this formula is analogous to the corresponding result for paramagnetic   Ising  model and other  mean-field-type systems,  e.g., Ref.~\cite{Prados:2012ks}.  Similarly, other extensive  equilibrium variables are also additive  and  it will be convenient to normalize them   by \( N \). 
 
 To shed light on the internal \emph{microconfiguration} of the system we introduce the fraction of HS elements  in the folded  conformation
$$
	p=-\frac{1}{N}\sum_{i=1}^{N} x_{i},
$$
which, in our case, plays the role of an order parameter.  The internal energy (per element)   corresponding to a given \( p \) can be written as
\begin{equation}
	\label{eq:v_bar}
	e(p,y) = p\left(y+1\right)^{2}/2 + (1-p)\left(y^{2}/2+v_{0}\right).
\end{equation}
Due to permutational invariance, we can  write the probability of a given state with \( Np \) elements in the folded state in the form  of the binomial law:
\begin{equation}
	\label{eq:rho_bar}
		\rho(p;y,\beta) = \binom{N}{Np}\left[\rho_{1}(-1;y,\beta)\right]^{Np}\left[\rho_{1}(0;y,\beta)\right]^{N(1-p)},
\end{equation}
where \( \rho_{1} \) is given by  (\ref{eq:rho_1}) and \( \rho_{1}(0;y,\beta)=1-\rho_{1}(-1;y,\beta) \).

Note that the  distribution \eqref{eq:rho_bar} can  be also written as $$\rho(p;y,\beta) = Z(y,\beta)^{-1}\exp[-\beta N\tilde{f}(p;y,\beta)] $$ where 
\begin{equation}
	\label{eq:f_bar}
	\tilde{f}(p;y,\beta) = e(y,p) - \left(1/\beta\right) s(p),
\end{equation}
 is the marginal free energy, \( e \) is the internal energy \eqref{eq:v_bar}, and 
\(
s(p)=\frac{1}{N}\log\binom{N}{Np}
\)
is the ideal entropy, all  corresponding to a fixed value of \( p \) and finite $N$.
%===============================================
\begin{figure}[b]
	\centering
		\includegraphics[]{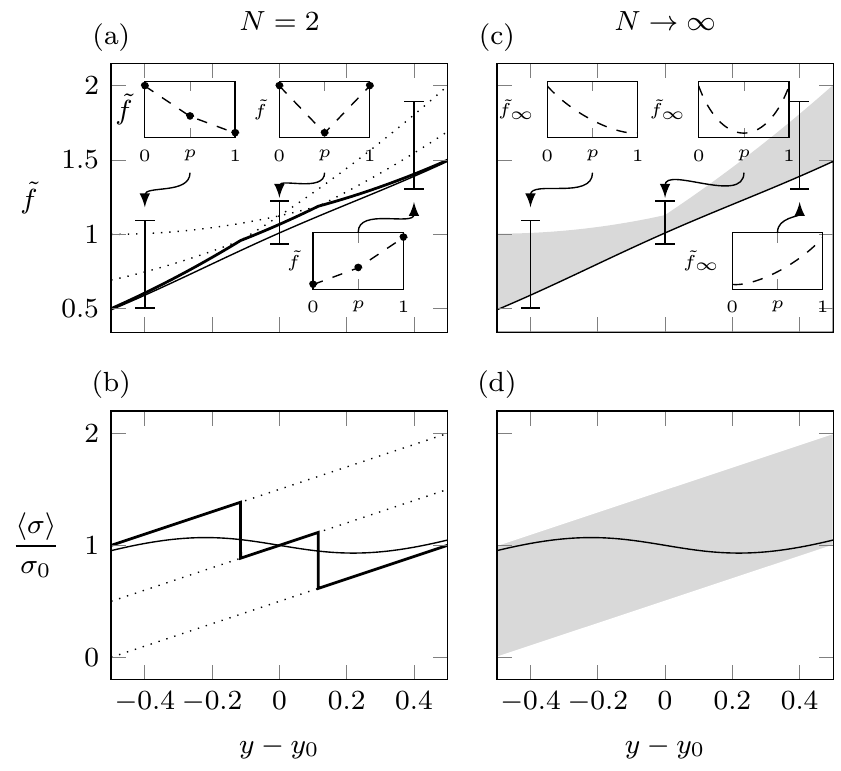}
	\caption{Nonequilibrium free energy landscape and the corresponding tension-elongation relations in a hard device for \( N=2 \) [(a) and (b)] and in the limit \( N\to\infty \) [(c) and (d)].
	Dotted lines, free-energy levels, and tension corresponding to different values of \( p \); thick line, response corresponding to the global minimum of the nonequilibrium free energy; thin lines, response in thermal equilibrium; gray areas, domain of the metastable states in the thermodynamic limit. In (a) and (c), the inserts show the marginal free energy \( \tilde{f} \) as function of \( p \) for \( y-y_0=-0.4, 0, 0.4 \). 
	The plots are obtained with \( \beta=6 \), which explains the presence of negative stiffness.
	}
	\label{fig:marginal_free_energy}
\end{figure}
%===============================================

%=============================================
\begin{figure*}
	\centering
	\includegraphics{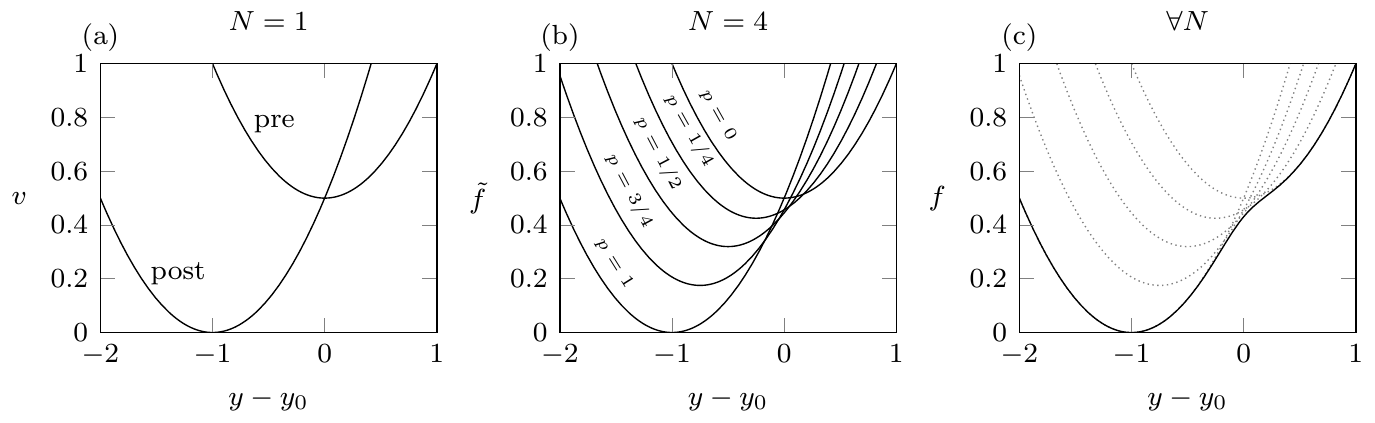}
	\caption{
	Hill-type energy landscapes for \( N=1 \) (a) and \( N=4 \) (b).
	In (c) we show the equilibrium free-energy profile \( f=F/N \) (solid line), which is independent of \( N \) together with the metastable states for \( N=4 \) (dotted lines). Here \( v_0=1/2 \).
	}
	\label{fig:hill_landscape}
\end{figure*}
%=============================================

To illustrate Eq.~\eqref{eq:f_bar},   consider the simplest case  \( N=2 \) when the marginal free energy can take only three values,
\begin{align*}
	\tilde{f}(0;y,\beta) &= e(0;y)= \frac{1}{2}y^{2} + v_{0} \\
	\tilde{f}(1/2;y,\beta) &= \frac{1}{4} (y+1)^{2} + \frac{1}{4}y^{2} + \frac{1}{2}v_{0} - \frac{\log(2)}{2\beta}\\
	\tilde{f}(1;y,\beta) &=e(1;y)= \frac{1}{2}(y+1)^{2},
\end{align*}
which are shown in Fig.~\ref{fig:marginal_free_energy} together with the corresponding tension-elongation relations.  Observe that the global minimum response at finite temperature is   characterized by a series of \emph{jumps} reflecting successive conformational changes in individual elements. 
Between the jumps, the stiffness is positive, which shows that each metastable state has a finite basin of stability even though the overall (global) stiffness is negative.  

The  changes in the marginal free-energy profiles with  increasing $N$ are illustrated in Fig.~\ref{fig:hill_landscape}. At \( N=1 \), we obtain the   representation  of the energy landscape due to T. L. Hill \cite{Eisenberg_1978}.  In this case, the marginal free energy \(\tilde{f} \) and the internal energy \( v \) are identical (no entropic contribution). For finite \( N \) we obtain \( N+1 \) metastable states corresponding to different values of \( p \) with the global minimum represented by a (nonconvex) lower envelope. 

While the lower envelope of the marginal  free energy $\tilde{f}(y,\beta)=\min_{p} \tilde{f}(p,y,\beta)$ is a piece wise smooth function of $y$ with a number of singular points depending on $N$, the equilibrium free energy  $f(y,\beta) = F(y,\beta)/N$ is a smooth function laying strictly below:  $f(y,\beta)\leq \tilde{f}(y,\beta)$. The $N$ independence of  $f(y,\beta)$---~see Eq.~\eqref{eq:f}---shows that for the HS system the equilibrium response is size independent. However, in real experiments for systems with small $N$ conducted at finite deformation rates one can expect to see the steps on the force elongation curves associated with the singularities of   $\tilde{f}(y)$, see Sec.~\ref{sub:bundle_of_hs_elements}. 

The average value of the parameter \( p \) (which is analogous to the variable \( n_{2} \)  in Ref.~\cite{Huxley_1971})  can be found from
\begin{equation}
	\label{eq:average_p}
	\mean{p}(y,\beta)= \sum p\,\rho(p;y,\beta) =
	 -\mean{x}\!(y,\beta) = \rho_{1}(-1;y,\beta)
\end{equation}
This quantity  plays the role of the  average magnetization per spin and  does not depend on \( N \); however, the corresponding variance decreases as $1/N$,
$$ \langle \left[p-\mean{p}(y,\beta)\right]^{2}\rangle = (1/N)
		\langle\left[x-\mean{x}(y,\beta)\right]^{2}\rangle. $$
Our Eq.~\eqref{eq:average_p} also shows that the whole distribution \eqref{eq:rho_bar} can be recovered if  the parameter $N$ is fixed and  \( \mean{p} \) is known as a function of $y$ and $\beta$.  In particular, we can   compute the variance
\begin{equation}
	\label{eq:variance_p}
		 \langle \left[p-\mean{p}(y,\beta)\right]^{2}\rangle =  \mean{p}(1-\mean{p}),
\end{equation}
which gives after substitution
\begin{equation}
	\label{eq:fluctuations_p}
	 \mean{\left[p-\mean{p}(y,\beta)\right]^{2}}^{1/2} =  \frac{1}{2\sqrt{N}}\sech\left[\frac{\beta}{2}(y-y_0)\right].
\end{equation}
By differentiating Eq.~\eqref{eq:average_p} with respect to \( y \) we can also obtain  the equilibrium susceptibility  
\[
		X(y,\beta) = -\dfrac{\partial}{\partial y}\mean{p} = \beta N \langle \left(p-\mean{p}\right)^{2}\rangle =\chi(y,\beta) ,
\]
where $\chi$
is the susceptibility of a  single HS element, see   Eq.~\eqref{eq:chi}. We can similarly  rewrite all other equilibrium characteristics of the system in terms of  \( -\mean{p} \) and \( N\langle\left[p-\mean{p}(y,\beta)\right]^{2}\rangle \).

In the limit \( N\to\infty \)  the expression for the marginal  free energy can be written  explicitly
\begin{equation}\label{eq:hs_f_large_N}
	\tilde{f}_{\infty} (p;y,\beta) =   e (p;y)  - \left(1/\beta\right)s_{\infty}(p),
\end{equation}
where 
$
	s_{\infty}(p) = -\left[ p\log(p) + (1-p)\log(1-p) \right],
$
 is the ideal mixing entropy reflecting the  absence of correlations between the units.
 The function  $\tilde{f}_{\infty} (p)$  is always convex since
\begin{equation*}
	\frac{\partial^{2}}{\partial p^{2}} \tilde{f}_{\infty}(p;y,\beta) = \bigl[\beta\, p (1-p)\bigr]^{-1}>0,
\end{equation*}
which signifies   the  lack of synchronization:  In  a similar ferromagnetic system  the  marginal free energy would be nonconvex.  As \( N  \rightarrow \infty\)  the domain of the phase space occupied by the metastable states becomes compact, see the gray area in Figs.~\ref{fig:marginal_free_energy}(c) and \ref{fig:marginal_free_energy}(d) not shown explicitly in Fig.~\ref{fig:hill_landscape}(c).

At large \( N \), the summation over the set of  discrete values of \( p \) (see Eq.~\eqref{eq:average_p})  can be  approximated by an  integration over  the interval \( [0,1] \).  The   integrals  can be, in turn,  computed by using the Laplace method. Then, for the equilibrium free energy, we can write 
\(
	f(y,\beta) =  \tilde{f}_{\infty}(p_{*}(y,\beta);\beta),
\)
where \( f \) and \( \tilde{f}_{\infty} \) are given by  \eqref{eq:f} and \eqref{eq:hs_f_large_N}, respectively. Here   
$p_{*}(y,\beta)$
is a minimizer of \( \tilde{f}_{\infty} \), which is a solution of the transcendental equation, 
\[
	p_*/(1- p_*) = \exp\bigl[-\beta(y-y_{0})\bigr].
\]
It is easy to check that
\( p_{*}(y,\beta) = \mean{p}(y,\beta) \)  where \( \mean{p}(y,\beta) \)  is given by Eq.~\eqref{eq:average_p} . The resulting free-energy profile is shown in both Fig.~\ref{fig:marginal_free_energy}(c) and Fig.~\ref{fig:hill_landscape}(c).

\paragraph{Mechanical behavior.}   For a given configuration \( \boldsymbol{x} \), the tension in the system can be written as 
\begin{equation*}
	\Sigma(\boldsymbol{x},y)  = N\left[y-(1/N)\sum x_{i}\right]=N\left(y+p\right).
\end{equation*}
The  average tension, conjugate to the control parameter \( y \),  is then
\begin{equation*}
	\mean{\Sigma}(y,\beta)= N\left[y+\mean{p}(y,\beta)\right]= N\mean{\sigma}(y,\beta) ,
\end{equation*}
where \( \mean{\sigma} \) is the average tension of a single element, see Eq.~\eqref{eq:sigma}. The variance of the total tension  can be written as 
\begin{equation*}
		\mean{\left[\Sigma-\mean{\Sigma}(y,\beta)\right]^{2}} 
		= N \mean{\left[\sigma-\mean{\sigma}(y,\beta)\right]^{2}}.
\end{equation*}
The  relative fluctuations,
\begin{align}\label{eq:Sigma_fluctuations}
	\frac{\mean{\left[\Sigma-\mean{\Sigma}(y,\beta)\right]^{2}}^{1/2}}{\mean{\Sigma(y,\beta)}} &=\\
	\frac{1}{2\sqrt{N}}\!
	\left\{\!(y+\frac{1}{2})\cosh\!\left[\frac{\beta}{2}(y-y_0)\right]\right.\!\!\!&\left.-\frac{1}{2}\sinh\left[\frac{\beta}{2}(y-y_0)\right]\right\}^{\!-1}\nonumber
\end{align}
decay as \( 1/N^{1/2} \),  which is a sign that the  measured  force in this model  is  an  extensive quantity.  The formula (\ref{eq:Sigma_fluctuations}) can be used to estimate the number of elements $N$ from the knowledge  of the fluctuations of the force.

If we denote by $K$ the total stiffness of the system, then we can write 
\begin{equation}\label{eq:K}
		K(y,\beta) =N\kappa(y,\beta)
		 = N - \beta \mean{\left[\Sigma-\mean{\Sigma}(y,\beta)\right]^{2}}
\end{equation}
where \( \kappa \) is defined by Eq.~\eqref{eq:kappa}. As in the case of  a single HS element, 
the total stiffness decomposes   into an elastic (or enthalpic) contribution dominating at \( |{y-y_0}|\gg 1 \) and a term containing  entropic contribution which dominates around \( y=y_0 \). 

In  dimensional form  (\ref{eq:K}) becomes
\begin{equation}
\label{eq:Kd}
\begin{split}
	K^{d} &=  N \kappa_0 - \frac{1}{k_{b}T}\mean{\left[\Sigma^{d}-\mean{\Sigma}^{d}\right]^{2}}\\
			&= N \kappa_0\! \left[ 1-  \frac{\kappa_{0}\,a^{2}}{4k_{b}T}\left\{\sech\!\left[\frac{\kappa\, a}{2k_{b}T}\left(y^{d}-y_0^{d}\right)\right]\right\}^{2}\right]\!\!,
\end{split}
\end{equation}
where the superscript \( d \) indicates that  the  normalization has been dropped.  

Note, first, that the   fluctuation-related contribution to stiffness is  an  order of one effect in terms of the number of elements \( N \), which means that the effect of fluctuations does not disappear  in the thermodynamic limit.   Also, since the stiffness in the HS model is an extensive property, the analysis of the temperature and elongation dependence of  \(\kappa(y,\beta)\)  presented in Sec.~\ref{sub:a_single_hs_element} remains valid here as well. 

The fact that the stiffness has an nonthermal, purely mechanical  part, that can be potentially extracted from  structural measurements,  and an equally important  and even dominating fluctuation-related component, that can be measured independently,  has been largely overlooked in the literature on   systems with nonconvex internal degrees of freedom because  in  classical materials, which can be thought to be composed of almost linear  springs, the fluctuational effect on stiffness is usually small.
Here we see that in the presence of internal ``snap-springs'' this effect can be considerable.
For instance, the  difference between the  smaller quasistatic stiffness of myosin II \cite{Veigel_1998,Tyska_2002}  and the larger instantaneous stiffness---believed to be largely unaffected by fluctuations \cite{Brunello:2014fu} ---may be linked to the importance of the second term in our Eq.~\eqref{eq:Kd}.

 On the other hand,  if \( N \) is known   and the   variance of the  total force can be measured, then one  can recover the stiffness of a single element \( \kappa_0 \).  Conversely, knowing \( \kappa_0 \) and measuring fluctuations of the force one can estimate \( N \) from \eqref{eq:Kd}.  We   emphasize  again that the relation \eqref{eq:Kd} is independent of the detailed structure  of the energy landscape \eqref{eq:u} and can therefore be used in the presence  of multiple  power-stroke-type energy wells.

\paragraph{Thermal behavior. } Since the entropy of the finite size bundle of HS elements is extensive $ S(y,\beta)=Ns(y,\beta)$,  
the analysis of the adiabatic response for a single HS element presented in Sec.~\ref{sub:a_single_hs_element} remains valid for the bundle of \( N \) HS elements.

% subsection a_bundle_of_hs_element (end)
% section equilibrium (end)
\section{Kinetics} % (fold)
\label{sec:kinetics}

In this section we study kinetics of the HS system and build  links  between the stochastic dynamics of a single HS element, the evolution of the bundle of   HS elements connected in parallel, and a conventional chemomechanical modeling of such systems in terms of deterministic chemical reactions.  Following the original HS model,  we assume for simplicity that during  loading  the temperature is  kept constant.

\subsection{Single HS element}

In the paper of Huxley and Simmons \cite{Huxley_1971} the relaxation of the system to equilibrium was modeled as  a  deterministic chemical reaction of the  first order.  In their description HS followed  the average population of elements  in  the two  conformational states  without attempting to trace the  dynamics of  individual flips experienced by the spin variables $x$.   
%=============================================
\begin{figure}
	\centering
	\includegraphics{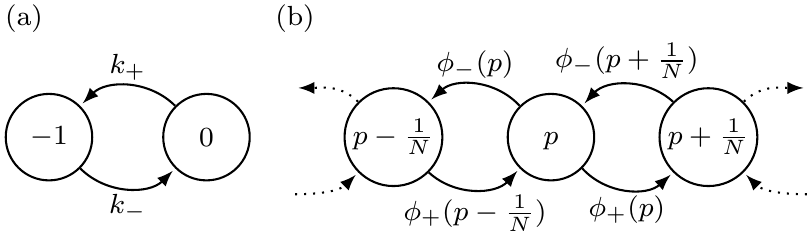}
	\caption{
	One-dimensional Markov chain description for a single HS element (a) and for a system with \( N \) elements (b).
	}
	\label{fig:random_walk}
\end{figure}
%=============================================

To simulate stochastic  dynamics of a single  HS element  we  need to know  the probabilities of the forward and reverse  flips  
\begin{equation}\label{eq:transition_probabilities}
	\begin{split}
		\mathbb{P}\left[\left. x^{t+dt}=-1 \right|x^{t}=0 \right] &= k_{+}(y,\beta)dt\\
		\mathbb{P}\left[\left. x^{t+dt}=0 \right|x^{t}=-1 \right] &= k_{-}(y,\beta)dt.
	\end{split}
\end{equation}
Here \( k_{+} (y,\beta)\) [respectively \( k_{-} (y,\beta) \)] is the transition rate for the jump  from the unfolded state (respectively, folded state) to the folded state (respectively, unfolded state), see Fig.~\ref{fig:random_walk}(a). We assume that the total elongation \( y \) and the inverse temperature \( \beta \) are  the controlling parameters  which   may vary at a time scale much larger that the characteristic time of the individual conformational transitions. As the transition probabilities \eqref{eq:transition_probabilities} depend only on the current  state of the system, the dynamics is described by a discrete Markov chain \cite{vanKampen:2001vs,Erdmann:2013wr}.

%=============================================
\begin{figure}[b]
	\centering
	\includegraphics{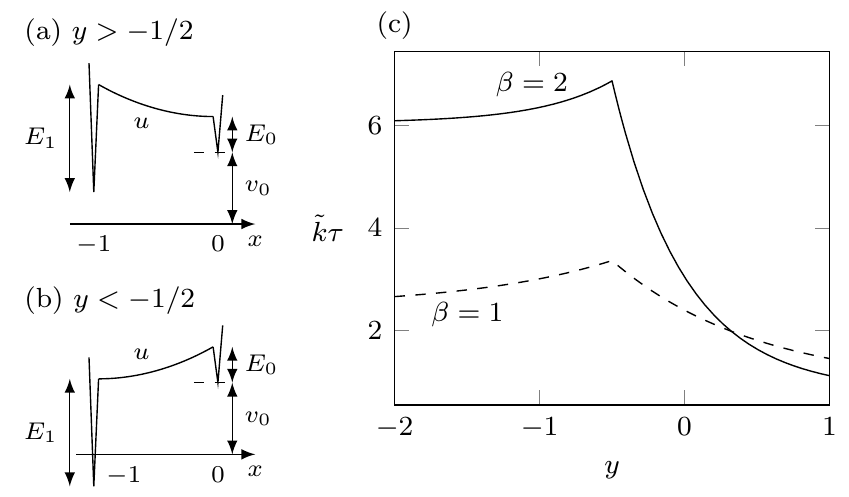}
	\caption{
	Schematic representation of the energy barrier in the HS bistable potential. The energy barriers corresponding to the transition rates in the absence of elastic contribution are denoted \( E_1 \) and \( E_{2} \). We define the characteristic timescale by \( \tau = \exp[\beta E_{1}] \). (a) Energy landscape for \( y>-1/2 \) which is the case considered in Ref.~\cite{Huxley_1971}; (b) energy landscape for \( y<-1/2 \) not considered by HS. (c), Relaxation rate as function of the total elongation \( y \) for \( \beta=1 \) (dashed line) and \( \beta=2 \) (solid line).
	}
	\label{fig:rates}
\end{figure}
%=============================================
 To compute the transition rates \( k_{\pm}(y,\beta) \) we need to know  the structure of the actual energy   landscape separating the  two conformational states.  In their paper \cite{Huxley_1971}, HS simply assumed that  the hypothetical barrier separating the two wells of the potential $v_{\mbox{\tiny HS}}$  is flat and is characterized by the    energy level $E_{1}$, see Fig.~\ref{fig:rates}.  By taking the energy of the elastic spring into account, we can then 
write the transition rates in the form  
\begin{align*}
	k_{+}(y,\beta) & = k \exp\left[-\beta \left[E_{0} + \max\left\{y+1/2,0\right\}\right]\right],\\
	k_{-}(y,\beta) & = k \exp\left[-\beta \left[E_{1} + \max \left\{-y-1/2,0\right\}\right]\right].
\end{align*}
where  \( E_{1} = E_{0} + v_{0} \) and the common pre-factor $k$  defines the characteristic time scale for a single well system.
If \( y>-1/2 \), then only the energy barrier from the unfolded to the folded state depends on \( y \), see Fig.~\ref{fig:rates}(a). In this case, the rates can be written in the form
\begin{subequations}\label{eq:k+_k-}
	\begin{empheq}[]{align}
	\label{eq:k+}
	k_{+} (y,\beta)&= {k_{-}}\exp\left[-\beta\left(y-y_0\right)\right]\\
	\label{eq:k-}
	k_{-} (y,\beta)&= k\exp\left[-\beta\,E_{1}\right] = \text{const}
\end{empheq}
\end{subequations}
and  the timescale of the jump process is  
$ 
\tau = 1/k_{-}=k^{-1}\exp\left[\beta\, E_1\right]. 
$
 If \( y<-1/2 \), see Fig.~\ref{fig:rates}(b), then we obtain instead
\begin{align*}
	k_{+}(y,\beta) &= k_{-} \exp\left[\beta\, v_0\right]= \text{const.}\\
	k_{-}(y,\beta) &= k_{-}\exp\left[\beta \left(y-y_0+v_0\right)\right].
\end{align*}
 
 We see in Fig.~\ref{fig:rates} that, in response to shortening,  the overall  transformation rate \( \tilde{k} = k_{+} + k_{-} \) first increases exponentially as the forward barrier is lowered  (while the reverse barrier remains constant) and then decreases as the   reverse barrier is elevated (while the forward barrier remains constant).
In addition, we see that for large stretching \( \tilde{k}\approx k_- \) and for large shortening \( \tilde{k}\approx k_-\exp[\beta\,v_0] \). 

Note that   HS considered   only  the case \( y>-1/2 \), see Fig.~\ref{fig:rates}(a). To see why the value \( y=-1/2 \) is special in the muscle context, we recall that in this case  the unloaded system  is symmetric, \( \mean{p}=1/2 \).  Then the tension during the  purely elastic phase of the fast  force recovery (when \( \mean{p}\) remains constant)  is $\sigma=y+1/2$, which becomes negative  exactly  at \( y=-1/2 \). In experiments on muscles, the relaxation rates have been  measured only for \( y>-1/2 \) because below  this threshold   muscle fibers are  usually subjected to buckling. Our analysis suggests that near the regimes with \( y=-1/2 \) the step size dependence of the rate of fast force recovery may deviate from exponential.

%=============================================
\begin{figure}[htbp]
	\centering
	\includegraphics{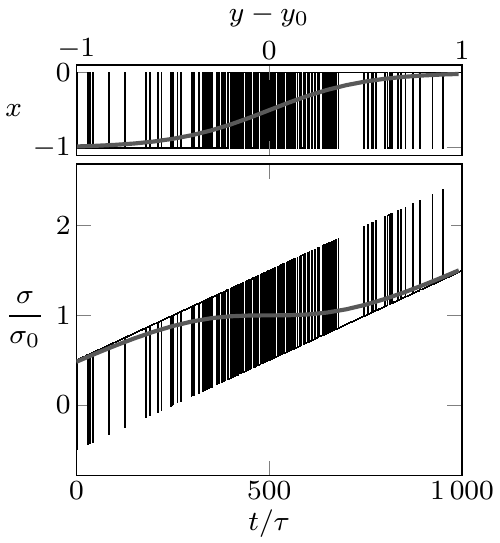}
	\caption{
	 Hopping response to a ramp stretch from \( y=y_0-1 \) to \( y=y_0+1 \)  (single trajectory with jumps hardly distinguishable around \( y=y_0 \) ). The time  step is \( \Delta t=10^{-3}\tau \) and the time is measured in the units of $\tau $. The average  trajectory is shown by the gray line. Here \( \beta=4 \).
	}
	\label{fig:single_xb_dynamics}
\end{figure}
%=============================================

The results of  numerical simulations of stochastic hopping for a single HS element subjected to a quasistatic stretching are shown in  Fig.~\ref{fig:single_xb_dynamics}.
For \( y<y_0 \) the spin variable spends most of the time in the folded conformation (\( x=-1 \)). When the loading device approaches the point \( y=y_0 \), the flips between the wells become more frequent before finally the system stabilizes again in the unfolded configuration (\( x=0 \)). 

 The stochastic  dynamics shown in Fig.~\ref{fig:single_xb_dynamics} can also be seen through the prism  of the deterministic evolution of a single-particle  probability distribution $\rho_{1}(t) $. For a generic test function \( q \) we can write 
 \begin{multline*}
 	d\mean{q(x)} = q(-1)\left[\rho_{1}(-1,t+dt) - \rho_{1}(-1,t)\right]\\ + q(0)\left[\rho_{1}(0,t+dt) - \rho_{1}(0,t)\right],
 \end{multline*}
 then, using \eqref{eq:transition_probabilities}, we obtain
 \begin{multline*}
 	d\mean{q(x)} = q(-1)\left\{k_{+}[1-\rho_{1}(-1,t)] - k_{-}\rho_{1}(-1,t)\right\}dt\\
	 + q(0)\left\{k_{-}[1-\rho_{1}(0,t)]- k_{+}(\rho_{1}(0,t))\right\}dt.
 \end{multline*}
In the limit  \( dt\to0 \), we obtain exactly the HS kinetic equation
\begin{equation}
	\label{eq:HS_kinetic_equation}
	\frac{\partial}{\partial t} \rho_{1}(t) = k_{+}(y)\left[1-\rho_{1}(t)\right] - k_{-}(y)\rho_{1}(t).
\end{equation}
Its general solution can be written as
\begin{multline*}
	\rho_{1}(t) = \rho_{1}(0)\exp\left[-A(t)\right]+\\
	 \int_{0}^{t}\tilde{k}(t')\exp[A(t')-A(t)]\rho_{1}^{\infty}(y(t))dt' ,
\end{multline*}
where \( A(t) = \int_{0}^{t}\tilde{k}(t')dt' \) and \( \rho_{1}^{\infty}=k_{+}/\tilde{k} \) is the stationary distribution  (\ref{eq:rho_1}).
Since  \( \mean{x}=\rho_{1} \), this equation describes the time dependence of the average configuration   shown in Fig.~\ref{fig:single_xb_dynamics} by the gray line.

The comparison of individual stochastic trajectories with the evolution of averages shows that  the information about individual flips, potentially measurable in single molecule experiments,  gets lost in the chemomechanical description. In particular, near the point \( y=y_0 \), fluctuations play a dominant role in the  stochastic  description, as is suggested by our equilibrium theory, while  from the  chemomechanical perspective  this particular state is completely indistinguishable from the other equilibrium states.

 \subsection{Bundle of HS elements} % (fold)
 \label{sub:bundle_of_hs_elements}

 The  isothermal discrete dynamics of a system with \( N \) elements   can be described in terms of the  macroscopic parameter  \( p\in\{0, 1/N,\dots,1\} \). If only one transition  occurs between the time \( t \) and the time \( t+dt \), then the  function \( p(t) \) is  a  one dimensional random walk (see  Fig.~\ref{fig:random_walk}), governed by the jump probabilities
\begin{subequations}\label{eq:random_walk}
\begin{empheq}[left=\empheqlbrace]{align}
	\mathbb{P}\left[p^{t+dt}=p^{t}+1/N\right] &= \phi_{+}(p^{t};y,\beta)dt\\
	\mathbb{P}\left[p^{t+dt}=p^{t}-1/N\right] &= \phi_{-}(p^{t};y,\beta)dt.
\end{empheq}
\end{subequations}
where \( \phi_{+}(p;y,\beta) = N(1-p)\, k_{+}(y,\beta) \) and \( \phi_{-}(p;y,\beta) = Np\, k_{-}(y,\beta) \). 
Following a similar procedure as the one leading to Eq.~\eqref{eq:HS_kinetic_equation}, we obtain the  master equation for the probability distribution  $ \rho(p,t)$  
\begin{multline}
	\label{eq:master_equation}
	\frac{\partial}{\partial t}\rho(p,t;y,\beta) = \phi_{+}\left(1-p+1/N;y\right) \rho\left(p-1/N,t;y,\beta\right)\\
	+ \phi_{-}\left(p+1/N;y\right)
	\rho\left(p+1/N,t;y,\beta\right)\\
	- \left[\phi_{+}(1-p;y)+\phi_{-}(p;y)\right]\rho\left(p,t;y,\beta\right),
\end{multline}
which can be solved numerically since we know the tridiagonal transfer matrix of the process at each time step.
It is clear, however,  that since the transition probabilities \eqref{eq:transition_probabilities}  depend only on the control parameter \( y \), the trajectories  of  individual elements are independent. Hence, at a given   \( y \) each macro-configuration can be viewed as a realization of \( N \)   Bernoulli processes  with the probability of success \( \rho_{1}(t) \) solving Eq.~\eqref{eq:HS_kinetic_equation}. Therefore the probability density \( \rho(p,t)=\mathbb{P}(p^{t}=p) \) is a binomial distribution  with  parameters \( N \) and \( \rho_{1}(t) \):  
\begin{equation}\label{eq:cluster_dynamics}
	\rho(p,t) = \binom{N}{Np}\left[\rho_{1}(t)\right]^{Np}\left[1-\rho_{1}(t)\right]^{N-Np}. 
\end{equation}
One can verify that that \eqref{eq:cluster_dynamics} solves (\ref{eq:master_equation}) and   since  \( \mean{p}(t) = \rho_{1}(t) \), Eq.~\eqref{eq:HS_kinetic_equation} can be  viewed as  the analog of Eq.~(9) in Ref.~\cite{Huxley_1971}. We have then shown that  the dynamics of the entire distribution is enslaved to the dynamics of the order parameter  $\mean{p}(t)$  captured by the original  HS model.  It is also clear that in the long time limit the distribution \eqref{eq:cluster_dynamics} converges to the Boltzmann distribution \eqref{eq:rho}.
 
 %===============================================
 	\begin{figure}[htbp]
 		\centering
 		\includegraphics{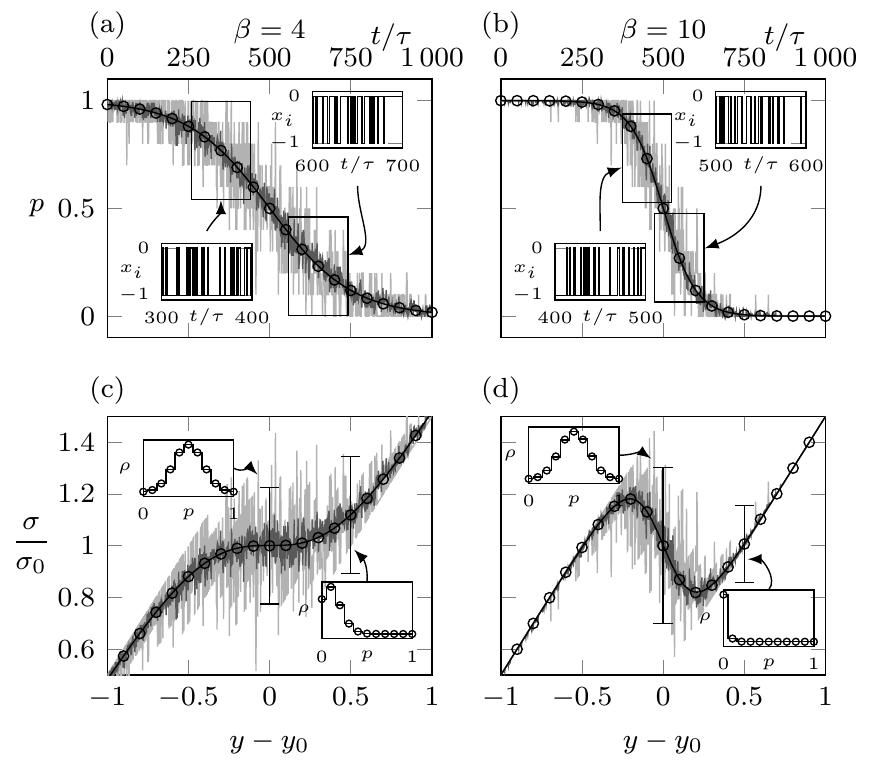} 
 		\caption{
 			Response of the HS model to a ramp loading from   \( y_{0}-1 \) to \( y_{0}+1\) achieved in \( 1000\,\tau \), for \( \beta=4 \) [(a) and (c)] and \( \beta=10 \) [(b) and (d)].
			 Individual  stochastic trajectories are shown for \( N=10 \) (light gray) and \( N=100 \) (dark gray). [(a) and (b)] Evolution of the order parameter \( p \). The inserts show two samples of a single  trajectory around the  point where \( p=1/2 \).
 [(c) and (d)] Tension-elongation relations obtained from \( \sigma=y+p \). The inserts show the marginal distribution \( \rho \) at the two different times indicted by the vertical bars. Solid lines represent the thermal equilibrium averages given by Eq.~\eqref{eq:average_p} and Eq.~\eqref{eq:sigma} and open symbols show the solutions of the HS kinetic equation \eqref{eq:HS_kinetic_equation} and the distribution \eqref{eq:cluster_dynamics}.
 		}
 		\label{fig:kinetics}
 	\end{figure}
 %===============================================

\paragraph{Quasistatic loading.}  To illustrate the fact that   our dynamical  model is fully  compatible with the equilibrium behavior  studied in Sec.~\ref{sub:a_bundle_of_hs_element},  we  now consider the quasistatic driving of a cluster of \( N \) HS elements, see Fig.~\ref{fig:kinetics}.
The behavior of the individual trajectories generated by the stochastic random walk Eq.~\eqref{eq:random_walk} is  shown for \( N=10 \) (light gray) and \( N=100 \) (dark gray).  The system is subjected  to  continuous stretching from \( y=y_0-1 \) to \( y=y_0+1 \)  over the time interval \( [0,10^{3}\,\tau] \) with the temperature remaining constant; this loading protocol  mimics the unzipping tests for biological macromolecules \cite{Liphardt:2001fp,Gupta:2011cp,Bornschlogl:2006we,Wen:2007fg}. The results were obtained using the same numerical procedures as  in the case of a single element. 

The stochastic evolution of the order parameter \( p \) and of the corresponding tension are illustrated in Fig.~\ref{fig:kinetics}. Together with   single trajectories, we show the evolution of the average (solid black line) obtained from Eq.~\eqref{eq:HS_kinetic_equation} and the corresponding equilibrium response curves (open circles). The inserts in Figs.~\ref{fig:kinetics}(a) and \ref{fig:kinetics}(b) show samples of the trajectories for single elements computed from \eqref{eq:transition_probabilities}.  

Observe that individual trajectories reveal at finite \( N \)   a succession of jumps describing individual folding-unfolding events as is suggested by the analysis of the marginally equilibrated system.  As the number of element increases the fluctuations of \( p \) decrease in accordance with our  Eq.~\eqref{eq:fluctuations_p}, and a single realization trajectory (dark gray) gets close to the average trajectory (black). 

In the inserts in Figs.~\ref{fig:kinetics}(c) and \ref{fig:kinetics}(d) we show the probability density \( \rho \) obtained from \eqref{eq:cluster_dynamics} at different times  (solid line) together with the equilibrium density (open circles).
As expected, the distribution does not depend on  temperature at \( y=y_0 \) while  becoming progressively more localized  away from this point.
%=============================================
\begin{figure}[t]
	\centering
	\includegraphics{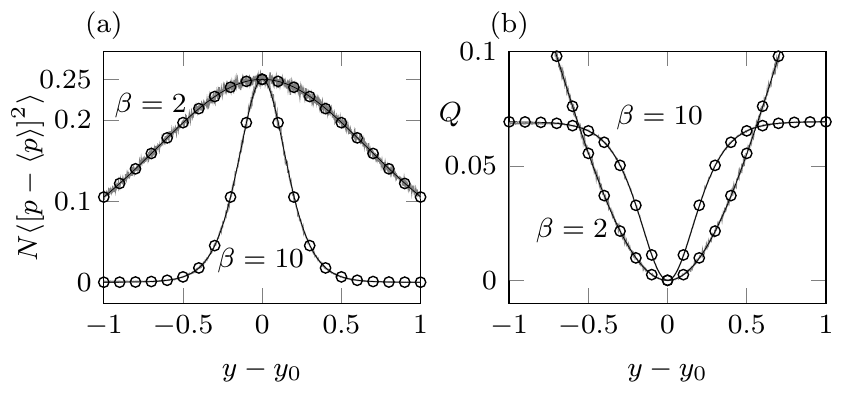}
	\caption{
	Evolution of the normalized variance \( N\langle\left(p-\mean{p}\right)^{2}\rangle \) and the heat release as function of the elongation during a quasistatic stretching between \( t=0 \) and \( t=10^{3}\tau \).
	For each temperature we show the analytic computations from \eqref{eq:fluctuations_p} and \eqref{eq:s_def} (open circles), the results of the stochastic model \eqref{eq:random_walk} corresponding to \( 10^{4} \) independent realizations with \( N=10 \) (gray lines) and the solution based on the solution of the kinetic equations \eqref{eq:HS_kinetic_equation} and \eqref{eq:cluster_dynamics} (thin lines).
	}
	\label{fig:P_statistics}
\end{figure}
%=============================================

To make the stochastic fluctuations more visible, we compare  in Fig.~\ref{fig:P_statistics} the  variance of the order parameter \( p \) obtained from the stochastic model \eqref{eq:random_walk} (gray lines) with the results of the analytic computations  based on   the kinetic equation   (\ref{eq:cluster_dynamics}) (thin lines)  and the  equilibrium model, see Eq.~\eqref{eq:fluctuations_p} (open circles). The system contains $N=10$ elements and each stochastic   trajectory corresponds to \( 10^{4} \) realizations of our random walk.
We see that  stochastic simulations are fully compatible with the predictions of equilibrium theory, in particular, we  see once again that the normalized variance  reaches a maximum at \( y=y_0 \) becoming independent of the temperature.

%=============================================
	\begin{figure}[b]
		\centering
		\includegraphics{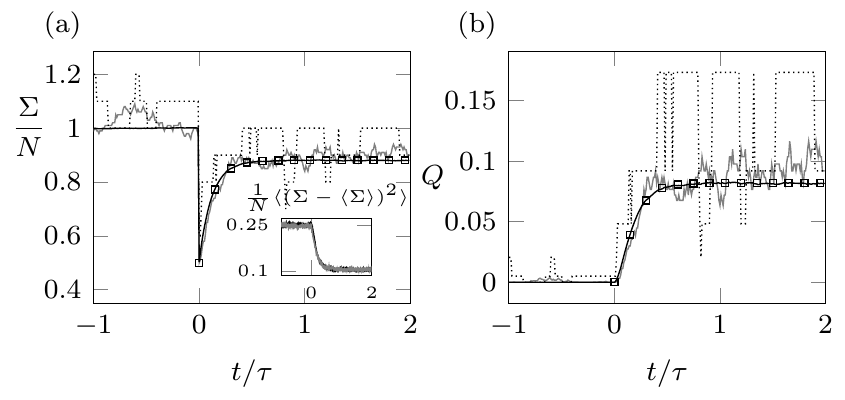}
		\caption{
		Stochastic simulation of a quick force recovery in response to a step of \( y-y_0=-0.5 \). (a) Tension per cross-linker; (b) heat released. Dotted lines: Single stochastic trajectory for a system with \( N=10 \); gray line single trajectory with \( N=100 \); solid line average over 1000 trajectories with \( N=10 \); squares, response obtained using the HS kinetic equation \eqref{eq:HS_kinetic_equation}. Inserts in (a), gray (respectively, solid) line, fluctuations obtained using \( 1000 \) realizations for \( N=100 \) (respectively, \( N=10 \)).
Here \( \beta=4 \) and \( y_0=0.5 \).}
		\label{fig:quick_recovery_dynamics}
	\end{figure}
	%=============================================

\paragraph{Fast loading.} 
In addition to averages, captured already by the chemomechanical  kinetic equation (\ref{eq:HS_kinetic_equation}), the master equation (\ref{eq:master_equation})  allows one to follow the evolution of higher order moments.
To illustrate this point, we now show how the HS system responds to abrupt perturbationss which  is exactly the type of mechanical test conducted  in  Ref.~\cite{Huxley_1971}, see Fig.~\ref{fig:quick_recovery_dynamics}. 
The system is first maintained in equilibrium at \( y=y_0=0.5 \) before an instantaneous length change (to \( y=0. \))  is applied.  This protocol is repeated for  systems with \( N=10 \) (dotted lines) and \( N=100 \) (solid line).
Again, individual realizations may strongly depart, especially at low \( N \), from the average behavior  described  by the HS reaction equation \eqref{eq:HS_kinetic_equation} (symbols).
 
%=============================================
\begin{figure}[t]
	\centering
	\includegraphics{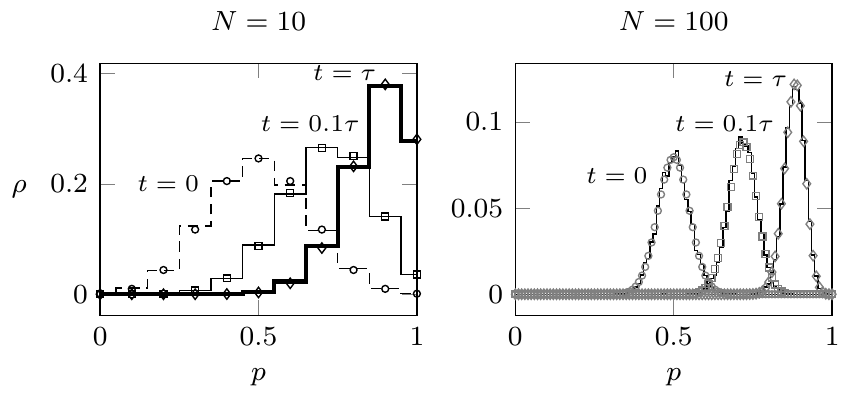}
	\caption{
Time evolution (snapshots) of the probability density $\rho(p,t;y,\beta)$ showing gradual equilibration of the system subjected to an abrupt shortening. Lines: histograms obtained from \( 10^{4} \) trajectories; symbols distributions recovered from the first order kinetic equations \eqref{eq:HS_kinetic_equation} and \eqref{eq:cluster_dynamics}. Parameters are the same as described in the caption to Fig.~\ref{fig:quick_recovery_dynamics}
	}
	\label{fig:proba}
\end{figure}
%=============================================
These fluctuations can also be seen from the dynamics of the density \( \rho \) which is reconstructed from a large number of sample trajectories/experiments in  Fig.~\ref{fig:proba}. We observe that for a system with a small number of elements [see Fig.~\ref{fig:proba}(a), \( N=10 \)) the probability distribution   remains broad even after the recovery while, for a system with large \( N \) [see Fig.~\ref{fig:proba}(b), \( N=10 \)), the distribution is sharply peaked throughout the process. Again, we find a perfect agreement between the distribution obtained from the Monte-Carlo simulations (lines) and the one recovered from the knowledge of the averaged behavior given by  the HS kinetic equation  parametrizing  the binomial distribution \eqref{eq:cluster_dynamics} (symbols).
 % subsection bundle_of_hs_elements (end)
% section kinetics (end)
\section{Skeletal muscles} % (fold)
\label{sec:skeletal_muscles}

  The development of the HS model was originally motivated by the  mechanical experiments involving rapid  shortening of skeletal muscles     with  the   goal  of  distinguishing  passive from active contributions to  tension recovery \cite{Podolsky_1960,Civan_1966,Podolsky:1969wk,Huxley:1970wl, Huxley_1971}.  It was shown  that the first phase  of the response to a quasi-instantaneous shortening imposed on a maximally activated (tetanized) single muscle fiber represents a purely elastic force drop.   During the second phase, the tension recovers to a  level  which  depends nonlinearly on the amplitude of the shortening. This fast force recovery, lasting about 1 ms,  precedes a considerably slower phase at the end of which  the tension fully returns to its original value. The latter,  taking place on a 100-ms time scale,  is usually interpreted as an \emph{active} process driven by ATP hydrolysis \cite{Holmes_2000,Linari_2010a}. 
  
\paragraph{Biochemistry vs mechanics. } In their classical 1971 paper  HS conjectured that the force recovery at the $ms$ time scale must be attributed  to a rapid  folding  in an assembly of attached  cross-bridges  linking actin and myosin filaments. The idea of  bistability in the structure of myosin heads, giving rise to the concept of a  power stroke,  has been later fully  supported by crystallographic studies   \cite{Rayment_1993,Dominguez_1998}. 
 
\begin{figure}
	\centering
	\includegraphics{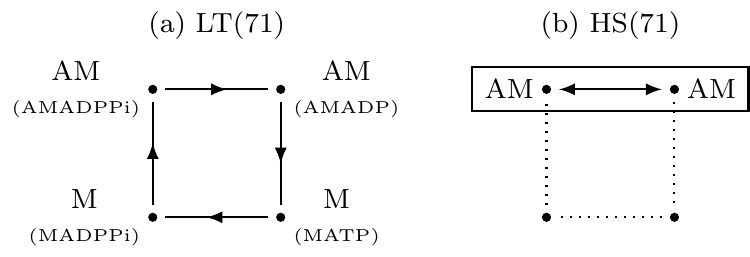}
	\caption{
	Biochemical vs purely mechanistic description of the power stroke in skeletal muscles: (a) The Lymn-Taylor four-state  cycle,  LT(71) and (b)  the Huxley-Simmons two-state cycle,  HS (71).
	}
	\label{fig:LymnTaylorHS}
\end{figure}

While  the   scenario proposed by HS  is  in agreement with the fact that the power stroke is the fastest step in the Lymn-Taylor  (LT) enzymatic cycle  \cite{Lymn_1971,Alberts:2007vj}, there is a subtle formal disagreement with the existing biochemical picture, see Fig.~\ref{fig:LymnTaylorHS}.  Thus, HS  assumed that the mechanism of the fast force recovery is  fully \emph{passive}  and can be reduced  to  a mechanically  induced conformational change.  In contrast, the LT cycle for actomyosin complexes  is based on the  assumption  that   the  power stroke can be  reversed  only \emph{actively} through the completion of the   bio chemical  pathway including  ADP (adenosine diphosphate) release,   myosin unbinding,  binding of uncleaved ATP (adenosine triphosphate), splitting of ATP  into ADP and P\( _{\mbox{\footnotesize i}} \), and then rebinding of myosin to actin \cite{Lymn_1971,Smith_2008}. 

In other words, while HS postulated that thermal  fluctuations experienced by the  attached myosin heads can be   biased by external loading  and   that  the power stroke can be reversed by mechanical means,  most of the biochemical  literature is based on the assumption that   the power-stroke recocking cannot be accomplished  without the presence of  ATP.  In particular,   physiological fluctuations  in muscle response are mostly  addressed   in the  context  of   \emph{active}  behavior  \cite{Hilbert:2013bb,Hilbert:2015bt,Baker:2002gz,Baker:2003kz,Haldeman:2014kx,AnnekaMHooft:2007kp,DelRJackson:2009hg}.

Some authors, however,  follow the HS mechanistic approach in assuming that  the power-stroke-related  leg of the LT cycle  can be  decoupled from the rest of the  biochemical pathway; see, for instance, Refs~\cite{Caremani:2015hr,Woledge:2009cv}. Below we  adopt this perspective, which implies that  at a 1-ms  time scale the mechanism dominating muscle response  is a purely mechanical folding-unfolding.  We then collect specific predictions, generated by our augmented HS model,   and use them as a guidance in designing  new  experiments  aimed, in particular,  at verifying the correctness of the underlying  purely mechanistic model. 

%=============================================
%=============================================

\paragraph{Thermal effects.}  The  knowledge  of  the free energy of the HS system  allows one to assess not only mechanical but also thermal manifestations of the fast force recovery. The latter have been  measured in experiments  employing  calorimetric techniques \cite{Gilbert:1988ir,Linari:1995jr,Woledge:2009cv},  however, a  thermomechanical interpretation of these experiments in the HS framework is still an  open question. 

For instance,  studies of  the heat exchange  following  the application of a fast length drop  showed an increase of temperature at the time scale of the purely elastic response  followed by a  slower cooling during the  force recovery up to a level which is higher than the baseline preceding the step. While, the temperature decay was linked to the equilibrium heat effect of the conformational change, which was assumed to be negative \cite{Gilbert:1988ir}, the HS model predicts a positive heat effect  because a  ``mixed'' state with high entropy  is transformed into a ``pure'' state with low entropy. 

More specifically, the HS-type interpretation of the temperature measurements during the fast force recovery  would be as follows: 

(i) The rapid increase of temperature recorded during the applied length step is a reflection of  an adiabatic temperature increase. To justify this claim,  we mention that several experimental studies of muscles, involving temperature changes due to rapid switching between, solutions showed that the time scale of temperature equilibration within a typical muscle fiber is of the order of \( 10\, \)ms \cite{Pate:1994gy,Linari_2004}, which is 10 times slower than the duration of the fast force recovery process.

(ii) The subsequent  temperature decay is an outcome  of the  cooling due to heat conduction and  the heat release due to the conformational change. The fact that the temperature at the end of the recovery is higher than the baseline temperature is a  signature   of  the exothermic nature of the folding process (of the working-stroke) and the inefficiency of the heat removal mechanism at this time scale.

Several groups have also addressed the influence of temperature on the force generation either by performing mechanical experiments in different solutions \cite{Piazzesi_2003,Decostre_2005} or by applying rapid temperature changes to tetanized muscle fibers \cite{Coupland_2003,Ranatunga_2007,Ranatunga:2010ft}.
In both cases the experiments  show that the isometric tension increases with temperature  while the conformational state of the cross-bridges becomes more homogeneous.  Such a response in the case of  fast  adiabatic changes  can be  explained by the fact that,  in order to maintain the value of the entropy  at higher temperature,  the HS system must evolve towards a more ordered configuration. This effect, however, is not  captured by the HS model where temperature does not affect the value of isometric tension at \( y=y_0 \), see our Fig.~\ref{fig:thermal_equilibrium}.  To describe quantitatively the temperature dependence of the tension-elongation curves,  we may augment the  HS model by assuming phenomenologically that the energy bias   \( v_0 \) is  a function of  temperature \cite{Decostre_2005}. 

While the value of the isometric tension may depend significantly on temperature, experiments show that the slope of the tension-elongation curve  is only weakly  temperature sensitive \citep{Woledge:2009cv}. In general, the HS model predicts considerable dependence of the shape of the tension-elongation relation on temperature, including the possibility of  negative stiffness which is not observed in experiment.  However, in the range of  temperatures  considered  in experiment (no more than 20\textdegree C) this effect is weak.
 For instance, if we assume that the mechanical properties of the cross-bridges are not affected by temperature, as is observed in experiments \cite{Piazzesi_2003}, and take  \( \beta=4 \) at \( T=4 \)\textdegree C,   we obtain   \( \beta\sim 3.8 \) at \( T=24 \)\textdegree C.  In the HS model, the sensitivity of  the  equilibrium tension-elongation curves  to such  variations of \( \beta \)  is negligible.

Among  other interesting thermomechanical effects, invoked by the HS model,   we mention  the  ``infinite'' cooling in the process of reaching the state of  isometric contractions. To observe this effect  the  muscle must be  first equilibrated after shortening (at the $T_2$ state) and then stretched back to the $T_0$  state. 

\paragraph{Fluctuations.}  While in myofibrils half-sarcomeres  fluctuations can be expected to average out,  at the  scale of   an  elementary acto-myosin complex   (our  half-sarcomere, see Fig.~\ref{fig:N_cross_bridges}),   where $N\sim 10^{2} $, fluctuations may interfere with experiments. For instance, as we have seen in our Fig.~\ref{fig:kinetics} and Fig.~\ref{fig:quick_recovery_dynamics},  the abrupt  transitions associated with conformational changes in individual cross-bridges may  produce  measurable steps in the response curves.
 
Another way of assessing the role of fluctuations  is through the measurement of  equilibrium susceptibilities. Thus,  the effective stiffness  of the HS  bundle can be represented  as a sum of an enthalpic term describing zero-temperature elasticity and an entropic term that can be evaluated   from the  measurements of tension fluctuations. More specifically, given that such fluctuations can be  measured, our Eq.~\eqref{eq:Kd} allows one to track the number of the attached elements  at different degrees of stretching and different temperatures.

On the other hand,  in mechanical experiments  conducted on single fibers and involving  x-ray diffraction measurements \cite{Reconditi_2006},  one can, in principle, test the prediction  of the model that the configuration with \( \mean{p}=1/2 \) is the state of maximum disorder.   By studying statistics of the observed fluctuations  in the steady states, one can also search for  deviations from the  static fluctuation-dissipation relation \eqref{eq:chi}.  If found, then they may reveal the presence  of out-of-equilibrium active processes at the time scales of fast force recovery which would then  reconcile the mechanical and the biochemical pictures of this phenomenon \cite{Sheshka:2014jt}.

 \paragraph{Cooperativity.} Statistics of fluctuations  for  groups of myosins has been studied exhausively in experiments involving  \emph{active} contraction.  Considerable coordination between individual elements was detected,  responsible for synchronized oscillations in close to stall conditions  (our point  \( y=y_0 \))  \cite{Podolsky_1960,Huxley:1974we,Armstrong:ta,Edman_1988,Yasuda:1996ta,Edman_2001,Placais_2009}. The cooperative behavior  was explained by the fact  that,  due to the presence of long-range elastic interactions transmited through compliant backbones,  the mechanical state of one motor   influences the kinetics of other motors \cite{Wakabayashi_1994,Huxley_1994}.  The implied   myosin-myosin coupling was taken into consideration in  models addressing active  behavior of  motor groups \cite{Duke_1999,Duke_2000} and  emergent phenomena  characterized by  large-scale entrainment signatures  were identified \cite{Badoual:2002ky,Vilfan_2003,Walcott_2009,Hilbert:2013bb,Hilbert:2015bt}. The claims that activity in such systems is crucial for the emergence of synchronized oscillations were supported by \emph{in vitro} assays \cite{Hilbert:2013bb,Hilbert:2015bt}  showing that  the  finite size scaling of the  fluctuations  is fundamentally differs from the equilibrium one  $N^{-1/2}$. 
  
 It is clear that to capture this effect,  the  original HS model with rigid backbone and controlled displacement  has to be generalized, but the actual role of activity  in synchronization of cross-bridges is  not obvious. Thus, it has been recently argued \cite{Caruel:2013jw}  that  the  dominant factor behind  collective behavior is not activity but  the  long-range  interactions between  cross-bridges.  The simplest way to create such ``cross-talk''   without leaving the   HS framework is to consider the response of a HS  system subjected to a constant   force  (soft device)   rather than  a constant displacement  \cite{Caruel:2015im}. It has been shown that in the systems of this type the nonconvexity of the free energy  can resist thermal fluctuations at sufficiently low temperatures, giving rise to  macroscopic cooperativity. Moreover, in the case of  soft and mixed (soft-hard) loadings, the   pseudocritical point of HS at $\beta=4$ becomes a real critical point  of the Curie type   around which fluctuations diverge in the thermodynamic limit and can show unusual finite-size scaling \cite{Caruel:2013jw}.

% section skeletal_muscles (end)

\section{Nonmuscle systems} % (fold)
\label{sec:non_muscle_applications}

The prototypical nature of the HS model makes it relevant  outside the skeletal muscle context as well. In fact, it  can be viewed as a description of  a large class of biological systems involving  collectively biased multistable elements and exhibiting, as a result,  sigmoidal or ultrasensitive response at finite temperatures  as in  our Fig.~\ref{fig:average_x_and_susceptibility}(a).  The HS model describes,  perhaps,  the most elementary  molecular system capable of  transforming in a Brownian environment a continuous input into a binary, all-or-none output   that is crucial for the fast and efficient, stroke-type behavior.  

We recall that the capacity of multisite systems to flip in a reversible fashion between several metastable conformations is essential for many processes in cellular physiology, including cell signaling, cell movement, chemotaxis, differentiation, and selective expression of genes \cite{Duke:2001uf,Bray:2013gx}. Usually, both the input and the output in such systems, known as allosteric,  are assumed to be of  biochemical origin.  The HS model, dealing with \emph{mechanical} response and relying on \emph{mechanical} driving,  complements biochemical models and presents a different  perspective on allostery.

\paragraph{Hair cell gating.}

Our first example of  hypersensitivity concerns  the transduction channels in hair cells \cite{Bormuth:2014hh}.  Each hair cell contains a bundle of \( N\approx 50 \) stereocilia which are mechanically stimulated by the vibrations in the inner ear. 
The stereocilia possess  transduction channels  closed by  ``gating springs'' which can open (close) in response to  a positive (negative)  shear strain  \( X \), imposed on the cilia from outside.  

The broadly accepted   model of this phenomenon \cite{Howard:1988uu} views the  hair bundle  as a set of $N$ bistable springs arranged in parallel. It  is identical to the HS model  if the   folded (unfolded)  configurations of   cross-bridges are identified  with the closed (opened)  states of the channels.  The applied loading, which tilts the potential and biases in this way the  distribution of closed and open configurations,  is treated in this model as a hard device of HS. 

To stress the equivalence of the results, it is enough to mention  the expression for the total stiffness of a hair bundle obtained  in Ref.~\cite{Howard:1988uu}, which  is a direct analog of our Eq.~\eqref{eq:variance_p}. Moreover,  the mechanical experiments, involving a mechanical solicitation of the hair bundle through an effectively  rigid glass fiber showed that the stiffness of the hair bundle  is  negative around the physiological functioning point of the system \cite{Martin_2000} , which is fully compatible with the predictions of the HS model. 
 
\paragraph{Cell adhesion.} A similar analogy can be drawn between the HS model and the  models of collective unzipping  for  adhesive clusters \cite{Chen:2011go, Yao:2006de, Erdmann_2007, Gao:2011ij, Schwarz:2013em}. At the micro-scale we again deal with $N$  elements representing, for instance, integrins or cadherins, that are  attached  in parallel to a common,  relatively rigid  pad.  The two conformational states, which can be described by a single spin variable,  are the   bound and the  unbound configurations.

The binding-unbinding phenomena in a mechanically biased system of the HS type are usually described by the  Bell model \cite{Bell_1978}, which is  a soft device analog of the HS model with $ \kappa_0=\infty$. In this model the breaking of an adhesive bond represents  an escape  from a metastable state and the corresponding rates are computed by using Kramers' theory \cite{Hanggi_1990,Schwarz:2013em} as in the HS model. In particular, the rebinding rate is often assumed to be constant \cite{Erdmann:2004do,Erdmann:2004hr}, which is also the assumption of HS for the  reverse transition from the post- to the pre-power-stroke state.  More recently, Bell's  model was generalized through the inclusion of ligand tethers, bringing a  finite  value to $ \kappa_0$   and using the master equation for the probability distribution of attached units \cite{Erdmann:2004do, Erdmann_2007}.  

The main difference between the Bell-type  models and  the HS model is that the detached state  cannot bear force while the unfolded conformation can. As a result,  while the cooperative folding-unfolding (ferromagnetic)  behavior in the HS model is possible  in the soft device setting \cite{Caruel:2013jw}, similar cooperative binding-unbinding in the Bell model is  impossible because the rebinding of a fully detached state has zero probablity.  To obtain cooperativity in models of adhesive clusters,  one must use a mixed device,  mimicking the elastic backbone and interpolating between soft and hard driving \cite{Seifert:2000uk,Erdmann_2007, Schwarz:2013em,Caruel:2015im}. 

\paragraph{Synaptic fusion.} While muscle tissues maintain stable architecture over long periods of time, it is feasible that transitory muscle-type structures can be also assembled to perform particular functions. An interesting  example of such assembly is provided by the SNARE proteins responsible for  the  fast  release of  neurotransmitors  from  neurons to  synaptic clefts. The  fusion of synaptic vesicles with the presynaptic plasma membrane \cite{Sudhof:2013fw,Sudhof:2009cw} is achieved  by mechanical zipping of the SNARE complexes which  can in this way transform from opened to  closed conformation \cite{Min:2013kr}. 
 
 To complete the analogy, we mention that  individual SNAREs participating in the collective zipping are attached to an elastic membrane that can be mimicked by an elastic or even rigid  backbone \cite{Li:2007ka}. The presence of  a backbone mediating long-range interactions  allows the SNAREs to cooperate in fast and efficient closing  of the gap between the vesicle and the membrane.  The analogy with muscles is corroborated by the fact  that synaptic fusion takes place at the same time scale as the fast force recovery (1 ms) \cite{Zorman:2014bg}.

\paragraph{Macromolecular hairpins.} Another class of phenomena that can be rationalized  within  the  HS framework   is 
the ubiquitous  flip-flopping of macro-molecular hairpins subjected to mechanical loading \cite{Liphardt:2001fp,Prados:2012ks,Bosaeus:2012kp,Woodside:2008uz}.

We recall that in a typical experiment of this type, a folded (zipped) macromolecule is attached through compliant links to  micron-sized beads trapped in optical  tweezers.  As the distance between the laser beams is increased, the force applied to the molecule rises up to a point where the subdomains start to  unfold.  An individual unfolding  event may  correspond to the collective rupture of $N$ molecular bonds or  an unzipping of a  hairpin. The corresponding  drops in the force accompanied by an abrupt increase in the total stretch  can lead to an overall  negative stiffness response \cite{Gupta:2011cp,Liphardt:2001fp,Wen:2007fg}. 

Realistic examples of unfolding in macromolecules may involve complex  ``fracture'' avalanches  \cite{Srivastava:2013tm} that cannot be modeled by using  the original HS model. However, the HS theoretical framework is general enough to  
accommodate hierarchical meta-structures  whose stability can be also biased by mechanical loading. In fact, the importance of the topology of interconnections among the bonds and the link between the collective nature  of the unfolding and the dominance of the HS-type parallel bonding have been long stressed in the studies  of protein folding \cite{Dietz:2008gj}.   The broad applicability of the HS  mechanical perspective on collective  conformational changes  is also corroborated
by the fact that proteins and nucleic acids  exhibit negative stiffness and behave differently in soft and hard devices \cite{Gerland:2003ie,Bornschlogl:2006we,Thomas:2012ur}. 

The ensemble dependence in these systems suggests that  additional structural information can be  obtained if the unfolding experiments are performed in the mixed device setting. The   type of  loading  may  be  affected through  the variable rigidity of  the  ``handles''  \cite{Cerf:1978vy, Pfitzner:2013ia} or the use of an appropriate  feedback control  that can be  modeled in the HS framework by a variable backbone elasticity.   

\paragraph{Allosteric systems.}

As we have already mentioned, collective conformational changes in  distributed biological systems containing coupled bistable units can be driven not only  mechanically, by applying forces or  displacements, but also  biochemically  by, say, varying  concentrations or  chemical potentials of ligand molecules in the environment \cite{Changeux:1967vf}. Such systems can become ultrasensitive to external stimulations as a result of the interaction between individual units undergoing conformational transformation which gives rise to the phenomenon of allostery also known as conformational spread \cite{Bray:2004iv,Bray:2013gx}.  The   switch-like   input-output relations  are required in a variety of biological applications because they   ensure both robustness in the presence of  external perturbations  and   ability to quickly adjust the configuration in response  to  selected  stimuli \cite{Duke:2001uf,Slepchenko:2004hq}. The mastery of control  of biological machinery through mechanically induced conformational spread is an important step in designing efficient biomimetic nanomachines \cite{Miyashita:2011jb,Yurke:2000ex,Harne:2016gg,Wu:2015hi}. 

To link this behavior to the HS model, we note that the  amplified \emph{dose response},  characteristic of allostery,   is analogous to the sigmoidal \emph{stress response} of the paramagnetic HS system where an  applied displacement plays the role of the  controlled input of a ligand.  Usually, in allosteric protein systems,  the ultrasensitive   behavior is achieved as a result of  nonlocal   interactions   favoring all-or-none types of responses; moreover,  the required long-range coupling is   provided by mechanical forces acting inside membranes and molecular complexes. In the HS model such  coupling is modeled by the parallel arrangement of elements, which  preserves the general idea of nonlocality. Despite its simplicity,  the appropriately generalized HS model   \cite{Caruel:2013jw} captures the main patterns of behavior exhibited by allosteric systems,  including the possibility of a critical point mentioned in Ref.~\cite{Changeux:1967vf}.

% section non_muscle_applications (end)

\section{Conclusions} % (fold)
\label{sec:conclusions}

In this paper we  presented a perspective on the seminal work of Huxley and Simmons by  viewing  their results through the prism of statistical mechanics. This allowed us to  place an emphasis on thermal  effects and equilibrium  fluctuations that cannot be ignored in many biological applications of the HS model.  

The chemomechanical approach of HS  is based on the description of a mechanical system with $N$ elements  in a  thermal bath in terms of a  single deterministic reaction equation.  Instead,  our analysis starts with the  analogy between the HS model in a  hard device  and the paramagnetic Ising model where the average conformation of a cross-bridge viewed as  the counterpart of magnetization.  In view of this analogy,  the  HS model  describes a  size indifferent mean-field statistical mechanical system which explains why the  many-body stochastic dynamics  can be modeled  by  a single chemical reaction.  The analogy with  paramagnetism is, however, not complete,  as is revealed by the phenomena of  negative  susceptibility and   pseudo-criticality that we identify with the HS model.  In particular, we show that while genuine criticality  requires cooperativity,  in the HS model the collective response is  imposed through the rigid backbone  rather than being an  emergent property.

Some of the most interesting findings of this paper concern  the  thermal properties of the HS system. Thus,  our analysis highlights   the previously unnoticed temperature robustness of the pseudocritical state  where  specific heat vanishes and fluctuations become temperature independent. We also quantified the temperature variations during fast  unfolding  and demonstrated   that isothermal and adiabatic responses may differ. These observations point towards the importance of  the nonorthodox  experimental protocols combining mechanical and calorimetric measurements. In particular, the revealed fluctuation  dependence  of equilibrium susceptibilities  suggests  a  noncrystallographic  way for the  evaluation of the number of folding elements.

To account for  fluctuations in kinetics we, following  Refs.~\cite{Vilfan_2003, Erdmann:2013wr},  went beyond the  reaction-based  modeling of the  averages  and studied  the time dependence of  the probability distributions for various parameters during mechanical transients.  The results of the deterministic  model of HS are, of course, recovered from the analysis of the evolution of the first moments of these distributions. 
 
While being the simplest mean-field description of a broad class of biological phenomena, the HS model clearly misrepresents the important elastic coupling between the folding elements, as  has been long realized in muscle physiology  \cite{Wakabayashi_1994,Huxley_1994}.
This drawback can be remedied by taking into account the  mechanical feedback induced by the  backbone  elasticity   \cite{Caruel:2013jw} resulting in  various degrees of synchronization already at zero temperature \cite{Caruel:2015im}. A more  important  limitation of the HS model  is   the neglect of the ATP-fueled activity which can crucially  interfere with  passive folding \cite{Duke_2000,Duke_1999,Badoual:2002ky,Vilfan_2003,Walcott_2009,Guerin_2011,Hilbert:2013bb,Hilbert:2015bt}.
 The account of nonthermal driving in the HS setting  produces new qualitative effects \cite{Sheshka:2016dm,Sheshka:2014jt}  which opens the  possibility to build  a fully mechanistic analog of the enzimatic cycle originating in the work of Lymn and Taylor.
 Quantitative  applications of the HS  model in and outside the muscle context   also call upon the account of  complex  geometry, hierarchical architecture, soft-spin-type multistability, and short-range interactions. All these potential augmentations, however,  will not diminish the role of the original HS model   as    a source of fundamental physical intuition about the behavior of a wide class of biological systems.

 % section conclusions (end)
 
 \section{Acknowledgments} % (fold)
The authors  thank R. Sheshka, P. Recho, T. Leli\`evre and  J.-M. Allain and the anonymous reviewers for helpful comments.

\appendix

 \bibliography{biblio}% Produces the bibliography via BibTeX.

\begin{thebibliography}{114}%
\makeatletter
\providecommand \@ifxundefined [1]{%
 \@ifx{#1\undefined}
}%
\providecommand \@ifnum [1]{%
 \ifnum #1\expandafter \@firstoftwo
 \else \expandafter \@secondoftwo
 \fi
}%
\providecommand \@ifx [1]{%
 \ifx #1\expandafter \@firstoftwo
 \else \expandafter \@secondoftwo
 \fi
}%
\providecommand \natexlab [1]{#1}%
\providecommand \enquote  [1]{``#1''}%
\providecommand \bibnamefont  [1]{#1}%
\providecommand \bibfnamefont [1]{#1}%
\providecommand \citenamefont [1]{#1}%
\providecommand \href@noop [0]{\@secondoftwo}%
\providecommand \href [0]{\begingroup \@sanitize@url \@href}%
\providecommand \@href[1]{\@@startlink{#1}\@@href}%
\providecommand \@@href[1]{\endgroup#1\@@endlink}%
\providecommand \@sanitize@url [0]{\catcode `\\12\catcode `\$12\catcode
  `\&12\catcode `\#12\catcode `\^12\catcode `\_12\catcode `\%12\relax}%
\providecommand \@@startlink[1]{}%
\providecommand \@@endlink[0]{}%
\providecommand \url  [0]{\begingroup\@sanitize@url \@url }%
\providecommand \@url [1]{\endgroup\@href {#1}{\urlprefix }}%
\providecommand \urlprefix  [0]{URL }%
\providecommand \Eprint [0]{\href }%
\providecommand \doibase [0]{http://dx.doi.org/}%
\providecommand \selectlanguage [0]{\@gobble}%
\providecommand \bibinfo  [0]{\@secondoftwo}%
\providecommand \bibfield  [0]{\@secondoftwo}%
\providecommand \translation [1]{[#1]}%
\providecommand \BibitemOpen [0]{}%
\providecommand \bibitemStop [0]{}%
\providecommand \bibitemNoStop [0]{.\EOS\space}%
\providecommand \EOS [0]{\spacefactor3000\relax}%
\providecommand \BibitemShut  [1]{\csname bibitem#1\endcsname}%
\let\auto@bib@innerbib\@empty
%</preamble>
\bibitem [{\citenamefont {Huxley}\ and\ \citenamefont
  {Simmons}(1971)}]{Huxley_1971}%
  \BibitemOpen
  \bibfield  {author} {\bibinfo {author} {\bibfnamefont {A.~F.}\ \bibnamefont
  {Huxley}}\ and\ \bibinfo {author} {\bibfnamefont {R.~M.}\ \bibnamefont
  {Simmons}},\ }\href@noop {} {\bibfield  {journal} {\bibinfo  {journal}
  {Nature}\ }\textbf {\bibinfo {volume} {233}},\ \bibinfo {pages} {533}
  (\bibinfo {year} {1971})}\BibitemShut {NoStop}%
\bibitem [{\citenamefont {Caruel}\ \emph {et~al.}(2013)\citenamefont {Caruel},
  \citenamefont {Allain},\ and\ \citenamefont {Truskinovsky}}]{Caruel:2013jw}%
  \BibitemOpen
  \bibfield  {author} {\bibinfo {author} {\bibfnamefont {M.}~\bibnamefont
  {Caruel}}, \bibinfo {author} {\bibfnamefont {J.~M.}\ \bibnamefont {Allain}},
  \ and\ \bibinfo {author} {\bibfnamefont {L.}~\bibnamefont {Truskinovsky}},\
  }\href@noop {} {\bibfield  {journal} {\bibinfo  {journal} {Phys. Rev. Lett.}\
  }\textbf {\bibinfo {volume} {110}},\ \bibinfo {pages} {248103} (\bibinfo
  {year} {2013})}\BibitemShut {NoStop}%
\bibitem [{\citenamefont {Hill}(1976)}]{Hill:1976gf}%
  \BibitemOpen
  \bibfield  {author} {\bibinfo {author} {\bibfnamefont {T.~L.}\ \bibnamefont
  {Hill}},\ }\href@noop {} {\bibfield  {journal} {\bibinfo  {journal} {Prog.
  Biophys. Molec. Biol.}\ }\textbf {\bibinfo {volume} {29}},\ \bibinfo {pages}
  {105} (\bibinfo {year} {1976})}\BibitemShut {NoStop}%
\bibitem [{\citenamefont {Huxley}\ and\ \citenamefont
  {Tideswell}(1996)}]{Huxley_1996}%
  \BibitemOpen
  \bibfield  {author} {\bibinfo {author} {\bibfnamefont {A.~F.}\ \bibnamefont
  {Huxley}}\ and\ \bibinfo {author} {\bibfnamefont {S.}~\bibnamefont
  {Tideswell}},\ }\href@noop {} {\bibfield  {journal} {\bibinfo  {journal} {J.
  Muscle Re. Cell M.}\ }\textbf {\bibinfo {volume} {17}},\ \bibinfo {pages}
  {507} (\bibinfo {year} {1996})}\BibitemShut {NoStop}%
\bibitem [{\citenamefont {Piazzesi}\ and\ \citenamefont
  {Lombardi}(1995)}]{Piazzesi_1995}%
  \BibitemOpen
  \bibfield  {author} {\bibinfo {author} {\bibfnamefont {G.}~\bibnamefont
  {Piazzesi}}\ and\ \bibinfo {author} {\bibfnamefont {V.}~\bibnamefont
  {Lombardi}},\ }\href@noop {} {\bibfield  {journal} {\bibinfo  {journal}
  {Biophys. J.}\ }\textbf {\bibinfo {volume} {68}},\ \bibinfo {pages} {1966}
  (\bibinfo {year} {1995})}\BibitemShut {NoStop}%
\bibitem [{\citenamefont {Smith}\ \emph {et~al.}(2008)\citenamefont {Smith},
  \citenamefont {Geeves}, \citenamefont {Sleep},\ and\ \citenamefont
  {Mijailovich}}]{Smith_2008}%
  \BibitemOpen
  \bibfield  {author} {\bibinfo {author} {\bibfnamefont {D.}~\bibnamefont
  {Smith}}, \bibinfo {author} {\bibfnamefont {M.}~\bibnamefont {Geeves}},
  \bibinfo {author} {\bibfnamefont {J.}~\bibnamefont {Sleep}}, \ and\ \bibinfo
  {author} {\bibfnamefont {S.}~\bibnamefont {Mijailovich}},\ }\href@noop {}
  {\bibfield  {journal} {\bibinfo  {journal} {Ann. Biomed. Eng.}\ }\textbf
  {\bibinfo {volume} {36}},\ \bibinfo {pages} {1624} (\bibinfo {year}
  {2008})}\BibitemShut {NoStop}%
\bibitem [{\citenamefont {Linari}\ \emph {et~al.}(2010)\citenamefont {Linari},
  \citenamefont {Caremani},\ and\ \citenamefont {Lombardi}}]{Linari_2010a}%
  \BibitemOpen
  \bibfield  {author} {\bibinfo {author} {\bibfnamefont {M.}~\bibnamefont
  {Linari}}, \bibinfo {author} {\bibfnamefont {M.}~\bibnamefont {Caremani}}, \
  and\ \bibinfo {author} {\bibfnamefont {V.}~\bibnamefont {Lombardi}},\
  }\href@noop {} {\bibfield  {journal} {\bibinfo  {journal} {P Roy. Soc. Lond.
  B Bio.}\ }\textbf {\bibinfo {volume} {277}},\ \bibinfo {pages} {19} (\bibinfo
  {year} {2010})}\BibitemShut {NoStop}%
\bibitem [{\citenamefont {Bell}(1978)}]{Bell_1978}%
  \BibitemOpen
  \bibfield  {author} {\bibinfo {author} {\bibfnamefont {G.}~\bibnamefont
  {Bell}},\ }\href@noop {} {\bibfield  {journal} {\bibinfo  {journal}
  {Science}\ }\textbf {\bibinfo {volume} {200}} (\bibinfo {year}
  {1978})}\BibitemShut {NoStop}%
\bibitem [{\citenamefont {Erdmann}\ and\ \citenamefont
  {Schwarz}(2007)}]{Erdmann_2007}%
  \BibitemOpen
  \bibfield  {author} {\bibinfo {author} {\bibfnamefont {T.}~\bibnamefont
  {Erdmann}}\ and\ \bibinfo {author} {\bibfnamefont {U.~S.}\ \bibnamefont
  {Schwarz}},\ }\href@noop {} {\bibfield  {journal} {\bibinfo  {journal} {Eur.
  Phys. J. E}\ }\textbf {\bibinfo {volume} {22}},\ \bibinfo {pages} {123}
  (\bibinfo {year} {2007})}\BibitemShut {NoStop}%
\bibitem [{\citenamefont {Howard}\ and\ \citenamefont
  {Hudspeth}(1988)}]{Howard:1988uu}%
  \BibitemOpen
  \bibfield  {author} {\bibinfo {author} {\bibfnamefont {J.}~\bibnamefont
  {Howard}}\ and\ \bibinfo {author} {\bibfnamefont {A.~J.}\ \bibnamefont
  {Hudspeth}},\ }\href@noop {} {\bibfield  {journal} {\bibinfo  {journal}
  {Neuron}\ }\textbf {\bibinfo {volume} {1}},\ \bibinfo {pages} {189} (\bibinfo
  {year} {1988})}\BibitemShut {NoStop}%
\bibitem [{\citenamefont {Martin}\ \emph {et~al.}(2000)\citenamefont {Martin},
  \citenamefont {Mehta},\ and\ \citenamefont {Hudspeth}}]{Martin_2000}%
  \BibitemOpen
  \bibfield  {author} {\bibinfo {author} {\bibfnamefont {P.}~\bibnamefont
  {Martin}}, \bibinfo {author} {\bibfnamefont {A.}~\bibnamefont {Mehta}}, \
  and\ \bibinfo {author} {\bibfnamefont {A.}~\bibnamefont {Hudspeth}},\
  }\href@noop {} {\bibfield  {journal} {\bibinfo  {journal} {Proc. Natl. Acad.
  Sci. U.S.A.}\ }\textbf {\bibinfo {volume} {97}},\ \bibinfo {pages} {12026}
  (\bibinfo {year} {2000})}\BibitemShut {NoStop}%
\bibitem [{\citenamefont {Liphardt}\ \emph {et~al.}(2001)\citenamefont
  {Liphardt}, \citenamefont {Onoa}, \citenamefont {Smith}, \citenamefont
  {Tinoco},\ and\ \citenamefont {Bustamante}}]{Liphardt:2001fp}%
  \BibitemOpen
  \bibfield  {author} {\bibinfo {author} {\bibfnamefont {J.}~\bibnamefont
  {Liphardt}}, \bibinfo {author} {\bibfnamefont {B.}~\bibnamefont {Onoa}},
  \bibinfo {author} {\bibfnamefont {S.~B.}\ \bibnamefont {Smith}}, \bibinfo
  {author} {\bibfnamefont {I.}~\bibnamefont {Tinoco}}, \ and\ \bibinfo {author}
  {\bibfnamefont {C.}~\bibnamefont {Bustamante}},\ }\href@noop {} {\bibfield
  {journal} {\bibinfo  {journal} {Science}\ }\textbf {\bibinfo {volume}
  {292}},\ \bibinfo {pages} {733} (\bibinfo {year} {2001})}\BibitemShut
  {NoStop}%
\bibitem [{\citenamefont {Bosaeus}\ \emph {et~al.}(2012)\citenamefont
  {Bosaeus}, \citenamefont {El-Sagheer}, \citenamefont {Brown}, \citenamefont
  {Smith}, \citenamefont {{\AA}kerman}, \citenamefont {Bustamante},\ and\
  \citenamefont {Norden}}]{Bosaeus:2012kp}%
  \BibitemOpen
  \bibfield  {author} {\bibinfo {author} {\bibfnamefont {N.}~\bibnamefont
  {Bosaeus}}, \bibinfo {author} {\bibfnamefont {A.~H.}\ \bibnamefont
  {El-Sagheer}}, \bibinfo {author} {\bibfnamefont {T.}~\bibnamefont {Brown}},
  \bibinfo {author} {\bibfnamefont {S.~B.}\ \bibnamefont {Smith}}, \bibinfo
  {author} {\bibfnamefont {B.}~\bibnamefont {{\AA}kerman}}, \bibinfo {author}
  {\bibfnamefont {C.}~\bibnamefont {Bustamante}}, \ and\ \bibinfo {author}
  {\bibfnamefont {B.}~\bibnamefont {Norden}},\ }\href@noop {} {\bibfield
  {journal} {\bibinfo  {journal} {Proc. Natl. Acad. Sci. U.S.A.}\ }\textbf
  {\bibinfo {volume} {109}},\ \bibinfo {pages} {15179} (\bibinfo {year}
  {2012})}\BibitemShut {NoStop}%
\bibitem [{\citenamefont {Woodside}\ \emph {et~al.}(2008)\citenamefont
  {Woodside}, \citenamefont {Garcia-Garcia},\ and\ \citenamefont
  {Block}}]{Woodside:2008uz}%
  \BibitemOpen
  \bibfield  {author} {\bibinfo {author} {\bibfnamefont {M.~T.}\ \bibnamefont
  {Woodside}}, \bibinfo {author} {\bibfnamefont {C.}~\bibnamefont
  {Garcia-Garcia}}, \ and\ \bibinfo {author} {\bibfnamefont {S.~M.}\
  \bibnamefont {Block}},\ }\href@noop {} {\bibfield  {journal} {\bibinfo
  {journal} {Curr. Opin. Chem. Biol.}\ }\textbf {\bibinfo {volume} {12}},\
  \bibinfo {pages} {640} (\bibinfo {year} {2008})}\BibitemShut {NoStop}%
\bibitem [{\citenamefont {Gupta}\ \emph {et~al.}(2011)\citenamefont {Gupta},
  \citenamefont {Vincent}, \citenamefont {Neupane}, \citenamefont {Yu},
  \citenamefont {Wang},\ and\ \citenamefont {Woodside}}]{Gupta:2011cp}%
  \BibitemOpen
  \bibfield  {author} {\bibinfo {author} {\bibfnamefont {A.~N.}\ \bibnamefont
  {Gupta}}, \bibinfo {author} {\bibfnamefont {A.}~\bibnamefont {Vincent}},
  \bibinfo {author} {\bibfnamefont {K.}~\bibnamefont {Neupane}}, \bibinfo
  {author} {\bibfnamefont {H.}~\bibnamefont {Yu}}, \bibinfo {author}
  {\bibfnamefont {F.}~\bibnamefont {Wang}}, \ and\ \bibinfo {author}
  {\bibfnamefont {M.~T.}\ \bibnamefont {Woodside}},\ }\href@noop {} {\bibfield
  {journal} {\bibinfo  {journal} {Nat Phys}\ }\textbf {\bibinfo {volume} {7}},\
  \bibinfo {pages} {631} (\bibinfo {year} {2011})}\BibitemShut {NoStop}%
\bibitem [{\citenamefont {Prados}\ \emph {et~al.}(2012)\citenamefont {Prados},
  \citenamefont {Carpio},\ and\ \citenamefont {Bonilla}}]{Prados:2012ks}%
  \BibitemOpen
  \bibfield  {author} {\bibinfo {author} {\bibfnamefont {A.}~\bibnamefont
  {Prados}}, \bibinfo {author} {\bibfnamefont {A.}~\bibnamefont {Carpio}}, \
  and\ \bibinfo {author} {\bibfnamefont {L.~L.}\ \bibnamefont {Bonilla}},\
  }\href@noop {} {\bibfield  {journal} {\bibinfo  {journal} {Phys. Rev. E}\
  }\textbf {\bibinfo {volume} {86}},\ \bibinfo {pages} {021919} (\bibinfo
  {year} {2012})}\BibitemShut {NoStop}%
\bibitem [{\citenamefont {Mu{\~n}oz}\ \emph {et~al.}(1998)\citenamefont
  {Mu{\~n}oz}, \citenamefont {Henry}, \citenamefont {Hofrichter},\ and\
  \citenamefont {Eaton}}]{Munoz:1998vf}%
  \BibitemOpen
  \bibfield  {author} {\bibinfo {author} {\bibfnamefont {V.}~\bibnamefont
  {Mu{\~n}oz}}, \bibinfo {author} {\bibfnamefont {E.~R.}\ \bibnamefont
  {Henry}}, \bibinfo {author} {\bibfnamefont {J.}~\bibnamefont {Hofrichter}}, \
  and\ \bibinfo {author} {\bibfnamefont {W.~A.}\ \bibnamefont {Eaton}},\
  }\href@noop {} {\bibfield  {journal} {\bibinfo  {journal} {Proc. Natl. Acad.
  Sci. U.S.A.}\ }\textbf {\bibinfo {volume} {95}},\ \bibinfo {pages} {5872}
  (\bibinfo {year} {1998})}\BibitemShut {NoStop}%
\bibitem [{\citenamefont {Bornschl{\"o}gl}\ and\ \citenamefont
  {Rief}(2006)}]{Bornschlogl:2006we}%
  \BibitemOpen
  \bibfield  {author} {\bibinfo {author} {\bibfnamefont {T.}~\bibnamefont
  {Bornschl{\"o}gl}}\ and\ \bibinfo {author} {\bibfnamefont {M.}~\bibnamefont
  {Rief}},\ }\href@noop {} {\bibfield  {journal} {\bibinfo  {journal} {Phys.
  Rev. Lett.}\ }\textbf {\bibinfo {volume} {96}},\ \bibinfo {pages} {118102}
  (\bibinfo {year} {2006})}\BibitemShut {NoStop}%
\bibitem [{\citenamefont {Srivastava}\ and\ \citenamefont
  {Granek}(2013)}]{Srivastava:2013tm}%
  \BibitemOpen
  \bibfield  {author} {\bibinfo {author} {\bibfnamefont {A.}~\bibnamefont
  {Srivastava}}\ and\ \bibinfo {author} {\bibfnamefont {R.}~\bibnamefont
  {Granek}},\ }\href@noop {} {\bibfield  {journal} {\bibinfo  {journal} {Phys.
  Rev. Lett.}\ }\textbf {\bibinfo {volume} {110}},\ \bibinfo {pages} {138101}
  (\bibinfo {year} {2013})}\BibitemShut {NoStop}%
\bibitem [{\citenamefont {Gerland}\ \emph {et~al.}(2003)\citenamefont
  {Gerland}, \citenamefont {Bundschuh},\ and\ \citenamefont
  {Hwa}}]{Gerland:2003ie}%
  \BibitemOpen
  \bibfield  {author} {\bibinfo {author} {\bibfnamefont {U.}~\bibnamefont
  {Gerland}}, \bibinfo {author} {\bibfnamefont {R.}~\bibnamefont {Bundschuh}},
  \ and\ \bibinfo {author} {\bibfnamefont {T.}~\bibnamefont {Hwa}},\
  }\href@noop {} {\bibfield  {journal} {\bibinfo  {journal} {Biophysj}\
  }\textbf {\bibinfo {volume} {84}},\ \bibinfo {pages} {2831} (\bibinfo {year}
  {2003})}\BibitemShut {NoStop}%
\bibitem [{\citenamefont {Thomas}\ and\ \citenamefont
  {Imafuku}(2012)}]{Thomas:2012ur}%
  \BibitemOpen
  \bibfield  {author} {\bibinfo {author} {\bibfnamefont {N.}~\bibnamefont
  {Thomas}}\ and\ \bibinfo {author} {\bibfnamefont {Y.}~\bibnamefont
  {Imafuku}},\ }\href@noop {} {\bibfield  {journal} {\bibinfo  {journal} {J
  Theor Biol}\ }\textbf {\bibinfo {volume} {312}},\ \bibinfo {pages} {96}
  (\bibinfo {year} {2012})}\BibitemShut {NoStop}%
\bibitem [{\citenamefont {S{\"u}dhof}(2013)}]{Sudhof:2013fw}%
  \BibitemOpen
  \bibfield  {author} {\bibinfo {author} {\bibfnamefont {T.~C.}\ \bibnamefont
  {S{\"u}dhof}},\ }\href@noop {} {\bibfield  {journal} {\bibinfo  {journal}
  {Neuron}\ }\textbf {\bibinfo {volume} {80}},\ \bibinfo {pages} {675}
  (\bibinfo {year} {2013})}\BibitemShut {NoStop}%
\bibitem [{\citenamefont {Bonilla}\ \emph {et~al.}(2012)\citenamefont
  {Bonilla}, \citenamefont {Carpio}, \citenamefont {Prados},\ and\
  \citenamefont {Rosales}}]{Bonilla:2012ib}%
  \BibitemOpen
  \bibfield  {author} {\bibinfo {author} {\bibfnamefont {L.~L.}\ \bibnamefont
  {Bonilla}}, \bibinfo {author} {\bibfnamefont {A.}~\bibnamefont {Carpio}},
  \bibinfo {author} {\bibfnamefont {A.}~\bibnamefont {Prados}}, \ and\ \bibinfo
  {author} {\bibfnamefont {R.~R.}\ \bibnamefont {Rosales}},\ }\href@noop {}
  {\bibfield  {journal} {\bibinfo  {journal} {Phys. Rev. E}\ }\textbf {\bibinfo
  {volume} {85}},\ \bibinfo {pages} {031125} (\bibinfo {year}
  {2012})}\BibitemShut {NoStop}%
\bibitem [{\citenamefont {Hill}(1974)}]{Hill:1974ks}%
  \BibitemOpen
  \bibfield  {author} {\bibinfo {author} {\bibfnamefont {T.~L.}\ \bibnamefont
  {Hill}},\ }\href@noop {} {\bibfield  {journal} {\bibinfo  {journal} {Prog.
  Biophys. Molec. Biol.}\ }\textbf {\bibinfo {volume} {28}},\ \bibinfo {pages}
  {267} (\bibinfo {year} {1974})}\BibitemShut {NoStop}%
\bibitem [{\citenamefont {Marcucci}\ and\ \citenamefont
  {Truskinovsky}(2010)}]{Marcucci_2010}%
  \BibitemOpen
  \bibfield  {author} {\bibinfo {author} {\bibfnamefont {L.}~\bibnamefont
  {Marcucci}}\ and\ \bibinfo {author} {\bibfnamefont {L.}~\bibnamefont
  {Truskinovsky}},\ }\href@noop {} {\bibfield  {journal} {\bibinfo  {journal}
  {Phys. Rev. E}\ }\textbf {\bibinfo {volume} {81}} (\bibinfo {year}
  {2010})}\BibitemShut {NoStop}%
\bibitem [{\citenamefont {Caruel}\ \emph {et~al.}(2015)\citenamefont {Caruel},
  \citenamefont {Allain},\ and\ \citenamefont {Truskinovsky}}]{Caruel:2015im}%
  \BibitemOpen
  \bibfield  {author} {\bibinfo {author} {\bibfnamefont {M.}~\bibnamefont
  {Caruel}}, \bibinfo {author} {\bibfnamefont {J.-M.}\ \bibnamefont {Allain}},
  \ and\ \bibinfo {author} {\bibfnamefont {L.}~\bibnamefont {Truskinovsky}},\
  }\href@noop {} {\bibfield  {journal} {\bibinfo  {journal} {J. Mech. Phys.
  Solids}\ }\textbf {\bibinfo {volume} {76}},\ \bibinfo {pages} {237} (\bibinfo
  {year} {2015})}\BibitemShut {NoStop}%
\bibitem [{\citenamefont {Pradhan}\ \emph {et~al.}(2010)\citenamefont
  {Pradhan}, \citenamefont {Hansen},\ and\ \citenamefont
  {Chakrabarti}}]{Pradhan:2010hd}%
  \BibitemOpen
  \bibfield  {author} {\bibinfo {author} {\bibfnamefont {S.}~\bibnamefont
  {Pradhan}}, \bibinfo {author} {\bibfnamefont {A.}~\bibnamefont {Hansen}}, \
  and\ \bibinfo {author} {\bibfnamefont {B.~K.}\ \bibnamefont {Chakrabarti}},\
  }\href@noop {} {\bibfield  {journal} {\bibinfo  {journal} {Rev. Mod. Phys.}\
  }\textbf {\bibinfo {volume} {82}},\ \bibinfo {pages} {499} (\bibinfo {year}
  {2010})}\BibitemShut {NoStop}%
\bibitem [{\citenamefont {Balian}(2006)}]{Balian:2006wd}%
  \BibitemOpen
  \bibfield  {author} {\bibinfo {author} {\bibfnamefont {R.}~\bibnamefont
  {Balian}},\ }\href@noop {} {\emph {\bibinfo {title} {{From Microphysics to
  Macrophysics}}}},\ Methods and Applications of Statistical Physics\ (\bibinfo
   {publisher} {Springer Science {\&} Business Media},\ \bibinfo {year}
  {2006})\BibitemShut {NoStop}%
\bibitem [{\citenamefont {Vilfan}\ and\ \citenamefont
  {Duke}(2003)}]{Vilfan_2003}%
  \BibitemOpen
  \bibfield  {author} {\bibinfo {author} {\bibfnamefont {A.}~\bibnamefont
  {Vilfan}}\ and\ \bibinfo {author} {\bibfnamefont {T.}~\bibnamefont {Duke}},\
  }\href@noop {} {\bibfield  {journal} {\bibinfo  {journal} {Biophys. J.}\
  }\textbf {\bibinfo {volume} {85}},\ \bibinfo {pages} {818} (\bibinfo {year}
  {2003})}\BibitemShut {NoStop}%
\bibitem [{\citenamefont {Erdmann}\ \emph {et~al.}(2013)\citenamefont
  {Erdmann}, \citenamefont {Albert},\ and\ \citenamefont
  {Schwarz}}]{Erdmann:2013wr}%
  \BibitemOpen
  \bibfield  {author} {\bibinfo {author} {\bibfnamefont {T.}~\bibnamefont
  {Erdmann}}, \bibinfo {author} {\bibfnamefont {P.~J.}\ \bibnamefont {Albert}},
  \ and\ \bibinfo {author} {\bibfnamefont {U.~S.}\ \bibnamefont {Schwarz}},\
  }\href@noop {} {\bibfield  {journal} {\bibinfo  {journal} {J. Chem. Phys.}\
  }\textbf {\bibinfo {volume} {139}},\ \bibinfo {pages} {175104} (\bibinfo
  {year} {2013})}\BibitemShut {NoStop}%
\bibitem [{\citenamefont {Rowlinson}(1959)}]{Rowlinson:511572}%
  \BibitemOpen
  \bibfield  {author} {\bibinfo {author} {\bibfnamefont {J.~S.}\ \bibnamefont
  {Rowlinson}},\ }\href@noop {} {\emph {\bibinfo {title} {{Liquids and liquid
  mixtures}}}},\ Modern aspects series of chemistry\ (\bibinfo  {publisher}
  {Butterworths},\ \bibinfo {address} {London},\ \bibinfo {year}
  {1959})\BibitemShut {NoStop}%
\bibitem [{\citenamefont {Squire}\ \emph {et~al.}(1969)\citenamefont {Squire},
  \citenamefont {Holt},\ and\ \citenamefont {Hoover}}]{Squire:1969dc}%
  \BibitemOpen
  \bibfield  {author} {\bibinfo {author} {\bibfnamefont {D.~R.}\ \bibnamefont
  {Squire}}, \bibinfo {author} {\bibfnamefont {A.~C.}\ \bibnamefont {Holt}}, \
  and\ \bibinfo {author} {\bibfnamefont {W.~G.}\ \bibnamefont {Hoover}},\
  }\href@noop {} {\bibfield  {journal} {\bibinfo  {journal} {Physica}\ }\textbf
  {\bibinfo {volume} {42}},\ \bibinfo {pages} {388} (\bibinfo {year}
  {1969})}\BibitemShut {NoStop}%
\bibitem [{\citenamefont {Lutsko}(1989)}]{Lutsko:1989bj}%
  \BibitemOpen
  \bibfield  {author} {\bibinfo {author} {\bibfnamefont {J.~F.}\ \bibnamefont
  {Lutsko}},\ }\href@noop {} {\bibfield  {journal} {\bibinfo  {journal} {J.
  Appl. Physiol.}\ }\textbf {\bibinfo {volume} {65}},\ \bibinfo {pages} {2991}
  (\bibinfo {year} {1989})}\BibitemShut {NoStop}%
\bibitem [{\citenamefont {Wittmer}\ \emph {et~al.}(2013)\citenamefont
  {Wittmer}, \citenamefont {Xu}, \citenamefont {Poli{\'{n}}ska}, \citenamefont
  {Weysser},\ and\ \citenamefont {Baschnagel}}]{Wittmer:2013gz}%
  \BibitemOpen
  \bibfield  {author} {\bibinfo {author} {\bibfnamefont {J.~P.}\ \bibnamefont
  {Wittmer}}, \bibinfo {author} {\bibfnamefont {H.}~\bibnamefont {Xu}},
  \bibinfo {author} {\bibfnamefont {P.}~\bibnamefont {Poli{\'{n}}ska}},
  \bibinfo {author} {\bibfnamefont {F.}~\bibnamefont {Weysser}}, \ and\
  \bibinfo {author} {\bibfnamefont {J.}~\bibnamefont {Baschnagel}},\
  }\href@noop {} {\bibfield  {journal} {\bibinfo  {journal} {J. Chem. Phys.}\
  }\textbf {\bibinfo {volume} {138}},\ \bibinfo {pages} {191101} (\bibinfo
  {year} {2013})}\BibitemShut {NoStop}%
\bibitem [{\citenamefont {Wu}\ \emph {et~al.}(2016)\citenamefont {Wu},
  \citenamefont {Harne},\ and\ \citenamefont {Wang}}]{Wu:2015hi}%
  \BibitemOpen
  \bibfield  {author} {\bibinfo {author} {\bibfnamefont {Z.}~\bibnamefont
  {Wu}}, \bibinfo {author} {\bibfnamefont {R.}~\bibnamefont {Harne}}, \ and\
  \bibinfo {author} {\bibfnamefont {K.}~\bibnamefont {Wang}},\ }\href@noop {}
  {\bibfield  {journal} {\bibinfo  {journal} {J. Intell. Mat. Syst. Struct.}\
  }\textbf {\bibinfo {volume} {27}},\ \bibinfo {pages} {1189} (\bibinfo {year}
  {2016})}\BibitemShut {NoStop}%
\bibitem [{\citenamefont {Eisenberg}\ and\ \citenamefont
  {Hill}(1978)}]{Eisenberg_1978}%
  \BibitemOpen
  \bibfield  {author} {\bibinfo {author} {\bibfnamefont {E.}~\bibnamefont
  {Eisenberg}}\ and\ \bibinfo {author} {\bibfnamefont {T.}~\bibnamefont
  {Hill}},\ }\href@noop {} {\bibfield  {journal} {\bibinfo  {journal} {Prog.
  Biophys. Molec. Biol.}\ }\textbf {\bibinfo {volume} {33}},\ \bibinfo {pages}
  {55} (\bibinfo {year} {1978})}\BibitemShut {NoStop}%
\bibitem [{\citenamefont {Ford}\ \emph {et~al.}(1977)\citenamefont {Ford},
  \citenamefont {Huxley},\ and\ \citenamefont {Simmons}}]{Ford_1977}%
  \BibitemOpen
  \bibfield  {author} {\bibinfo {author} {\bibfnamefont {L.~E.}\ \bibnamefont
  {Ford}}, \bibinfo {author} {\bibfnamefont {A.~F.}\ \bibnamefont {Huxley}}, \
  and\ \bibinfo {author} {\bibfnamefont {R.~M.}\ \bibnamefont {Simmons}},\
  }\href@noop {} {\bibfield  {journal} {\bibinfo  {journal} {J. Physiol.}\
  }\textbf {\bibinfo {volume} {269}},\ \bibinfo {pages} {441} (\bibinfo {year}
  {1977})}\BibitemShut {NoStop}%
\bibitem [{\citenamefont {Ford}\ \emph {et~al.}(1981)\citenamefont {Ford},
  \citenamefont {Huxley},\ and\ \citenamefont {Simmons}}]{Ford:1981dl}%
  \BibitemOpen
  \bibfield  {author} {\bibinfo {author} {\bibfnamefont {L.~E.}\ \bibnamefont
  {Ford}}, \bibinfo {author} {\bibfnamefont {A.~F.}\ \bibnamefont {Huxley}}, \
  and\ \bibinfo {author} {\bibfnamefont {R.~M.}\ \bibnamefont {Simmons}},\
  }\href@noop {} {\bibfield  {journal} {\bibinfo  {journal} {The Journal of
  physiology}\ }\textbf {\bibinfo {volume} {311}},\ \bibinfo {pages} {219}
  (\bibinfo {year} {1981})}\BibitemShut {NoStop}%
\bibitem [{\citenamefont {Kojima}\ \emph {et~al.}(1994)\citenamefont {Kojima},
  \citenamefont {Ishuima},\ and\ \citenamefont {Yanagida}}]{Kojima_1994}%
  \BibitemOpen
  \bibfield  {author} {\bibinfo {author} {\bibfnamefont {H.}~\bibnamefont
  {Kojima}}, \bibinfo {author} {\bibfnamefont {A.}~\bibnamefont {Ishuima}}, \
  and\ \bibinfo {author} {\bibfnamefont {T.}~\bibnamefont {Yanagida}},\
  }\href@noop {} {\bibfield  {journal} {\bibinfo  {journal} {Proc. Natl. Acad.
  Sci. U.S.A.}\ }\textbf {\bibinfo {volume} {91}},\ \bibinfo {pages} {12962}
  (\bibinfo {year} {1994})}\BibitemShut {NoStop}%
\bibitem [{\citenamefont {Wakabayashi}\ \emph {et~al.}(1994)\citenamefont
  {Wakabayashi}, \citenamefont {Sugimoto}, \citenamefont {Tanake},
  \citenamefont {Ueno}, \citenamefont {Takezawa},\ and\ \citenamefont
  {Amemiya}}]{Wakabayashi_1994}%
  \BibitemOpen
  \bibfield  {author} {\bibinfo {author} {\bibfnamefont {K.}~\bibnamefont
  {Wakabayashi}}, \bibinfo {author} {\bibfnamefont {Y.}~\bibnamefont
  {Sugimoto}}, \bibinfo {author} {\bibfnamefont {H.}~\bibnamefont {Tanake}},
  \bibinfo {author} {\bibfnamefont {Y.}~\bibnamefont {Ueno}}, \bibinfo {author}
  {\bibfnamefont {Y.}~\bibnamefont {Takezawa}}, \ and\ \bibinfo {author}
  {\bibfnamefont {Y.}~\bibnamefont {Amemiya}},\ }\href@noop {} {\bibfield
  {journal} {\bibinfo  {journal} {Biophys. J.}\ }\textbf {\bibinfo {volume}
  {67}},\ \bibinfo {pages} {2422} (\bibinfo {year} {1994})}\BibitemShut
  {NoStop}%
\bibitem [{\citenamefont {Huxley}\ \emph {et~al.}(1994)\citenamefont {Huxley},
  \citenamefont {Stewart}, \citenamefont {Sosa},\ and\ \citenamefont
  {Irving}}]{Huxley_1994}%
  \BibitemOpen
  \bibfield  {author} {\bibinfo {author} {\bibfnamefont {H.}~\bibnamefont
  {Huxley}}, \bibinfo {author} {\bibfnamefont {A.}~\bibnamefont {Stewart}},
  \bibinfo {author} {\bibfnamefont {H.}~\bibnamefont {Sosa}}, \ and\ \bibinfo
  {author} {\bibfnamefont {T.}~\bibnamefont {Irving}},\ }\href@noop {}
  {\bibfield  {journal} {\bibinfo  {journal} {Biophys. J.}\ }\textbf {\bibinfo
  {volume} {67}},\ \bibinfo {pages} {2411} (\bibinfo {year}
  {1994})}\BibitemShut {NoStop}%
\bibitem [{\citenamefont {Piazzesi}\ \emph {et~al.}(2007)\citenamefont
  {Piazzesi}, \citenamefont {Reconditi}, \citenamefont {Linari}, \citenamefont
  {Lucii}, \citenamefont {Bianco}, \citenamefont {Brunello}, \citenamefont
  {Decostre}, \citenamefont {Stewart}, \citenamefont {Gore}, \citenamefont
  {Irving}, \citenamefont {Irving},\ and\ \citenamefont
  {Lombardi}}]{Piazzesi_2007}%
  \BibitemOpen
  \bibfield  {author} {\bibinfo {author} {\bibfnamefont {G.}~\bibnamefont
  {Piazzesi}}, \bibinfo {author} {\bibfnamefont {M.}~\bibnamefont {Reconditi}},
  \bibinfo {author} {\bibfnamefont {M.}~\bibnamefont {Linari}}, \bibinfo
  {author} {\bibfnamefont {L.}~\bibnamefont {Lucii}}, \bibinfo {author}
  {\bibfnamefont {P.}~\bibnamefont {Bianco}}, \bibinfo {author} {\bibfnamefont
  {E.}~\bibnamefont {Brunello}}, \bibinfo {author} {\bibfnamefont
  {V.}~\bibnamefont {Decostre}}, \bibinfo {author} {\bibfnamefont
  {A.}~\bibnamefont {Stewart}}, \bibinfo {author} {\bibfnamefont {D.~B.}\
  \bibnamefont {Gore}}, \bibinfo {author} {\bibfnamefont {T.~C.}\ \bibnamefont
  {Irving}}, \bibinfo {author} {\bibfnamefont {M.}~\bibnamefont {Irving}}, \
  and\ \bibinfo {author} {\bibfnamefont {V.}~\bibnamefont {Lombardi}},\
  }\href@noop {} {\bibfield  {journal} {\bibinfo  {journal} {Cell}\ }\textbf
  {\bibinfo {volume} {131}},\ \bibinfo {pages} {784} (\bibinfo {year}
  {2007})}\BibitemShut {NoStop}%
\bibitem [{\citenamefont {Irving}\ \emph {et~al.}(2000)\citenamefont {Irving},
  \citenamefont {Piazzesi}, \citenamefont {Lucii}, \citenamefont {Sun},
  \citenamefont {Harford}, \citenamefont {Dobbie}, \citenamefont {Ferenczi},
  \citenamefont {Reconditi},\ and\ \citenamefont {Lombardi}}]{Irving_2000}%
  \BibitemOpen
  \bibfield  {author} {\bibinfo {author} {\bibfnamefont {M.}~\bibnamefont
  {Irving}}, \bibinfo {author} {\bibfnamefont {G.}~\bibnamefont {Piazzesi}},
  \bibinfo {author} {\bibfnamefont {L.}~\bibnamefont {Lucii}}, \bibinfo
  {author} {\bibfnamefont {Y.}~\bibnamefont {Sun}}, \bibinfo {author}
  {\bibfnamefont {J.}~\bibnamefont {Harford}}, \bibinfo {author} {\bibfnamefont
  {I.}~\bibnamefont {Dobbie}}, \bibinfo {author} {\bibfnamefont
  {M.}~\bibnamefont {Ferenczi}}, \bibinfo {author} {\bibfnamefont
  {M.}~\bibnamefont {Reconditi}}, \ and\ \bibinfo {author} {\bibfnamefont
  {V.}~\bibnamefont {Lombardi}},\ }\href@noop {} {\bibfield  {journal}
  {\bibinfo  {journal} {Nat. Struct. Biol.}\ }\textbf {\bibinfo {volume} {7}},\
  \bibinfo {pages} {482} (\bibinfo {year} {2000})}\BibitemShut {NoStop}%
\bibitem [{\citenamefont {Hill}(1938)}]{Hill_1938}%
  \BibitemOpen
  \bibfield  {author} {\bibinfo {author} {\bibfnamefont {A.~V.}\ \bibnamefont
  {Hill}},\ }\href@noop {} {\bibfield  {journal} {\bibinfo  {journal} {P Roy.
  Soc. Lond. B Bio.}\ }\textbf {\bibinfo {volume} {126}},\ \bibinfo {pages}
  {136} (\bibinfo {year} {1938})}\BibitemShut {NoStop}%
\bibitem [{\citenamefont {Piazzesi}\ \emph {et~al.}(2003)\citenamefont
  {Piazzesi}, \citenamefont {Reconditi}, \citenamefont {Koubassova},
  \citenamefont {Decostre}, \citenamefont {Linari}, \citenamefont {Lucii},\
  and\ \citenamefont {Lombardi}}]{Piazzesi_2003}%
  \BibitemOpen
  \bibfield  {author} {\bibinfo {author} {\bibfnamefont {G.}~\bibnamefont
  {Piazzesi}}, \bibinfo {author} {\bibfnamefont {A.}~\bibnamefont {Reconditi}},
  \bibinfo {author} {\bibfnamefont {N.}~\bibnamefont {Koubassova}}, \bibinfo
  {author} {\bibfnamefont {V.}~\bibnamefont {Decostre}}, \bibinfo {author}
  {\bibfnamefont {M.}~\bibnamefont {Linari}}, \bibinfo {author} {\bibfnamefont
  {L.}~\bibnamefont {Lucii}}, \ and\ \bibinfo {author} {\bibfnamefont
  {V.}~\bibnamefont {Lombardi}},\ }\href@noop {} {\bibfield  {journal}
  {\bibinfo  {journal} {J. Physiol.}\ }\textbf {\bibinfo {volume} {549}},\
  \bibinfo {pages} {93} (\bibinfo {year} {2003})}\BibitemShut {NoStop}%
\bibitem [{\citenamefont {Decostre}\ \emph {et~al.}(2005)\citenamefont
  {Decostre}, \citenamefont {Bianco}, \citenamefont {Lombardi},\ and\
  \citenamefont {Piazzesi}}]{Decostre_2005}%
  \BibitemOpen
  \bibfield  {author} {\bibinfo {author} {\bibfnamefont {V.}~\bibnamefont
  {Decostre}}, \bibinfo {author} {\bibfnamefont {P.}~\bibnamefont {Bianco}},
  \bibinfo {author} {\bibfnamefont {V.}~\bibnamefont {Lombardi}}, \ and\
  \bibinfo {author} {\bibfnamefont {G.}~\bibnamefont {Piazzesi}},\ }\href@noop
  {} {\bibfield  {journal} {\bibinfo  {journal} {Proc. Natl. Acad. Sci. USA}\
  }\textbf {\bibinfo {volume} {102}},\ \bibinfo {pages} {13927} (\bibinfo
  {year} {2005})}\BibitemShut {NoStop}%
\bibitem [{\citenamefont {Ranatunga}\ \emph {et~al.}(2007)\citenamefont
  {Ranatunga}, \citenamefont {Coupland}, \citenamefont {Pinniger},
  \citenamefont {Roots},\ and\ \citenamefont {Offer}}]{Ranatunga_2007}%
  \BibitemOpen
  \bibfield  {author} {\bibinfo {author} {\bibfnamefont {K.~W.}\ \bibnamefont
  {Ranatunga}}, \bibinfo {author} {\bibfnamefont {M.~E.}\ \bibnamefont
  {Coupland}}, \bibinfo {author} {\bibfnamefont {G.~J.}\ \bibnamefont
  {Pinniger}}, \bibinfo {author} {\bibfnamefont {H.}~\bibnamefont {Roots}}, \
  and\ \bibinfo {author} {\bibfnamefont {G.~W.}\ \bibnamefont {Offer}},\
  }\href@noop {} {\bibfield  {journal} {\bibinfo  {journal} {J. Physiol.}\
  }\textbf {\bibinfo {volume} {585}},\ \bibinfo {pages} {263} (\bibinfo {year}
  {2007})}\BibitemShut {NoStop}%
\bibitem [{\citenamefont {Coupland}\ and\ \citenamefont
  {Ranatunga}(2003)}]{Coupland_2003}%
  \BibitemOpen
  \bibfield  {author} {\bibinfo {author} {\bibfnamefont {M.}~\bibnamefont
  {Coupland}}\ and\ \bibinfo {author} {\bibfnamefont {K.~W.}\ \bibnamefont
  {Ranatunga}},\ }\href@noop {} {\bibfield  {journal} {\bibinfo  {journal} {J.
  Physiol.}\ }\textbf {\bibinfo {volume} {548}},\ \bibinfo {pages} {439}
  (\bibinfo {year} {2003})}\BibitemShut {NoStop}%
\bibitem [{\citenamefont {Woledge}\ \emph {et~al.}(2009)\citenamefont
  {Woledge}, \citenamefont {Barclay},\ and\ \citenamefont
  {Curtin}}]{Woledge:2009cv}%
  \BibitemOpen
  \bibfield  {author} {\bibinfo {author} {\bibfnamefont {R.~C.}\ \bibnamefont
  {Woledge}}, \bibinfo {author} {\bibfnamefont {C.~J.}\ \bibnamefont
  {Barclay}}, \ and\ \bibinfo {author} {\bibfnamefont {N.~A.}\ \bibnamefont
  {Curtin}},\ }\href@noop {} {\bibfield  {journal} {\bibinfo  {journal} {P Roy.
  Soc. Lond. B Bio.}\ }\textbf {\bibinfo {volume} {276}},\ \bibinfo {pages}
  {2685} (\bibinfo {year} {2009})}\BibitemShut {NoStop}%
\bibitem [{\citenamefont {Linari}\ and\ \citenamefont
  {Woledge}(1995)}]{Linari:1995jr}%
  \BibitemOpen
  \bibfield  {author} {\bibinfo {author} {\bibfnamefont {M.}~\bibnamefont
  {Linari}}\ and\ \bibinfo {author} {\bibfnamefont {R.~C.}\ \bibnamefont
  {Woledge}},\ }\href@noop {} {\bibfield  {journal} {\bibinfo  {journal} {J.
  Physiol.}\ }\textbf {\bibinfo {volume} {487}},\ \bibinfo {pages} {699}
  (\bibinfo {year} {1995})}\BibitemShut {NoStop}%
\bibitem [{\citenamefont {Veigel}\ \emph {et~al.}(1998)\citenamefont {Veigel},
  \citenamefont {Bartoo}, \citenamefont {White}, \citenamefont {Sparrow},\ and\
  \citenamefont {Molloy}}]{Veigel_1998}%
  \BibitemOpen
  \bibfield  {author} {\bibinfo {author} {\bibfnamefont {C.}~\bibnamefont
  {Veigel}}, \bibinfo {author} {\bibfnamefont {M.}~\bibnamefont {Bartoo}},
  \bibinfo {author} {\bibfnamefont {D.}~\bibnamefont {White}}, \bibinfo
  {author} {\bibfnamefont {J.}~\bibnamefont {Sparrow}}, \ and\ \bibinfo
  {author} {\bibfnamefont {J.}~\bibnamefont {Molloy}},\ }\href@noop {}
  {\bibfield  {journal} {\bibinfo  {journal} {Biophys. J.}\ }\textbf {\bibinfo
  {volume} {75}},\ \bibinfo {pages} {1424} (\bibinfo {year}
  {1998})}\BibitemShut {NoStop}%
\bibitem [{\citenamefont {Tyska}\ and\ \citenamefont
  {Warshaw}(2002)}]{Tyska_2002}%
  \BibitemOpen
  \bibfield  {author} {\bibinfo {author} {\bibfnamefont {M.}~\bibnamefont
  {Tyska}}\ and\ \bibinfo {author} {\bibfnamefont {D.}~\bibnamefont
  {Warshaw}},\ }\href@noop {} {\bibfield  {journal} {\bibinfo  {journal} {Cell
  Motil. Cytoskel.}\ }\textbf {\bibinfo {volume} {51}},\ \bibinfo {pages} {1}
  (\bibinfo {year} {2002})}\BibitemShut {NoStop}%
\bibitem [{\citenamefont {Brunello}\ \emph {et~al.}(2014)\citenamefont
  {Brunello}, \citenamefont {Caremani}, \citenamefont {Melli}, \citenamefont
  {Linari}, \citenamefont {Fernandez-Martinez}, \citenamefont {Narayanan},
  \citenamefont {Irving}, \citenamefont {Piazzesi}, \citenamefont {Lombardi},\
  and\ \citenamefont {Reconditi}}]{Brunello:2014fu}%
  \BibitemOpen
  \bibfield  {author} {\bibinfo {author} {\bibfnamefont {E.}~\bibnamefont
  {Brunello}}, \bibinfo {author} {\bibfnamefont {M.}~\bibnamefont {Caremani}},
  \bibinfo {author} {\bibfnamefont {L.}~\bibnamefont {Melli}}, \bibinfo
  {author} {\bibfnamefont {M.}~\bibnamefont {Linari}}, \bibinfo {author}
  {\bibfnamefont {M.}~\bibnamefont {Fernandez-Martinez}}, \bibinfo {author}
  {\bibfnamefont {T.}~\bibnamefont {Narayanan}}, \bibinfo {author}
  {\bibfnamefont {M.}~\bibnamefont {Irving}}, \bibinfo {author} {\bibfnamefont
  {G.}~\bibnamefont {Piazzesi}}, \bibinfo {author} {\bibfnamefont
  {V.}~\bibnamefont {Lombardi}}, \ and\ \bibinfo {author} {\bibfnamefont
  {M.}~\bibnamefont {Reconditi}},\ }\href@noop {} {\bibfield  {journal}
  {\bibinfo  {journal} {The Journal of physiology}\ }\textbf {\bibinfo {volume}
  {592}},\ \bibinfo {pages} {3881} (\bibinfo {year} {2014})}\BibitemShut
  {NoStop}%
\bibitem [{\citenamefont {van Kampen}(2001)}]{vanKampen:2001vs}%
  \BibitemOpen
  \bibfield  {author} {\bibinfo {author} {\bibfnamefont {N.}~\bibnamefont {van
  Kampen}},\ }\href@noop {} {\emph {\bibinfo {title} {Stochastic Processes in
  Physics and Chemistry}}}\ (\bibinfo  {publisher} {North-Holland, Amsterdam},\
  \bibinfo {year} {2001})\BibitemShut {NoStop}%
\bibitem [{\citenamefont {Wen}\ \emph {et~al.}(2007)\citenamefont {Wen},
  \citenamefont {Manosas}, \citenamefont {Li}, \citenamefont {Smith},
  \citenamefont {Bustamante}, \citenamefont {Ritort},\ and\ \citenamefont
  {Tinoco}}]{Wen:2007fg}%
  \BibitemOpen
  \bibfield  {author} {\bibinfo {author} {\bibfnamefont {J.-D.}\ \bibnamefont
  {Wen}}, \bibinfo {author} {\bibfnamefont {M.}~\bibnamefont {Manosas}},
  \bibinfo {author} {\bibfnamefont {P.~T.~X.}\ \bibnamefont {Li}}, \bibinfo
  {author} {\bibfnamefont {S.~B.}\ \bibnamefont {Smith}}, \bibinfo {author}
  {\bibfnamefont {C.}~\bibnamefont {Bustamante}}, \bibinfo {author}
  {\bibfnamefont {F.}~\bibnamefont {Ritort}}, \ and\ \bibinfo {author}
  {\bibfnamefont {I.}~\bibnamefont {Tinoco}},\ }\href@noop {} {\bibfield
  {journal} {\bibinfo  {journal} {Biophys. J.}\ }\textbf {\bibinfo {volume}
  {92}},\ \bibinfo {pages} {2996} (\bibinfo {year} {2007})}\BibitemShut
  {NoStop}%
\bibitem [{\citenamefont {Podolsky}(1960)}]{Podolsky_1960}%
  \BibitemOpen
  \bibfield  {author} {\bibinfo {author} {\bibfnamefont {R.}~\bibnamefont
  {Podolsky}},\ }\href@noop {} {\bibfield  {journal} {\bibinfo  {journal}
  {Nature}\ }\textbf {\bibinfo {volume} {188}},\ \bibinfo {pages} {666}
  (\bibinfo {year} {1960})}\BibitemShut {NoStop}%
\bibitem [{\citenamefont {Civan}\ and\ \citenamefont
  {Podolsky}(1966)}]{Civan_1966}%
  \BibitemOpen
  \bibfield  {author} {\bibinfo {author} {\bibfnamefont {M.}~\bibnamefont
  {Civan}}\ and\ \bibinfo {author} {\bibfnamefont {R.}~\bibnamefont
  {Podolsky}},\ }\href@noop {} {\bibfield  {journal} {\bibinfo  {journal} {J.
  Physiol.}\ }\textbf {\bibinfo {volume} {184}},\ \bibinfo {pages} {511}
  (\bibinfo {year} {1966})}\BibitemShut {NoStop}%
\bibitem [{\citenamefont {Podolsky}\ \emph {et~al.}(1969)\citenamefont
  {Podolsky}, \citenamefont {Nolan},\ and\ \citenamefont
  {Zaveler}}]{Podolsky:1969wk}%
  \BibitemOpen
  \bibfield  {author} {\bibinfo {author} {\bibfnamefont {R.}~\bibnamefont
  {Podolsky}}, \bibinfo {author} {\bibfnamefont {A.~C.}\ \bibnamefont {Nolan}},
  \ and\ \bibinfo {author} {\bibfnamefont {S.~A.}\ \bibnamefont {Zaveler}},\
  }\href@noop {} {\bibfield  {journal} {\bibinfo  {journal} {Proc. Natl. Acad.
  Sci. U.S.A.}\ }\textbf {\bibinfo {volume} {64}},\ \bibinfo {pages} {504}
  (\bibinfo {year} {1969})}\BibitemShut {NoStop}%
\bibitem [{\citenamefont {Huxley}\ and\ \citenamefont
  {Simmons}(1970)}]{Huxley:1970wl}%
  \BibitemOpen
  \bibfield  {author} {\bibinfo {author} {\bibfnamefont {A.~F.}\ \bibnamefont
  {Huxley}}\ and\ \bibinfo {author} {\bibfnamefont {R.~M.}\ \bibnamefont
  {Simmons}},\ }\href@noop {} {\bibfield  {journal} {\bibinfo  {journal} {J.
  Physiol.}\ }\textbf {\bibinfo {volume} {208}},\ \bibinfo {pages} {52P}
  (\bibinfo {year} {1970})}\BibitemShut {NoStop}%
\bibitem [{\citenamefont {Holmes}\ and\ \citenamefont
  {Geeves}(2000)}]{Holmes_2000}%
  \BibitemOpen
  \bibfield  {author} {\bibinfo {author} {\bibfnamefont {K.}~\bibnamefont
  {Holmes}}\ and\ \bibinfo {author} {\bibfnamefont {M.}~\bibnamefont
  {Geeves}},\ }\href@noop {} {\bibfield  {journal} {\bibinfo  {journal}
  {Philos. T. Roy. Soc. B}\ }\textbf {\bibinfo {volume} {355}},\ \bibinfo
  {pages} {419} (\bibinfo {year} {2000})}\BibitemShut {NoStop}%
\bibitem [{\citenamefont {Rayment}\ \emph {et~al.}(1993)\citenamefont
  {Rayment}, \citenamefont {Holden}, \citenamefont {Whittaker}, \citenamefont
  {Yohn}, \citenamefont {Lorenz}, \citenamefont {Holmes},\ and\ \citenamefont
  {Milligan}}]{Rayment_1993}%
  \BibitemOpen
  \bibfield  {author} {\bibinfo {author} {\bibfnamefont {I.}~\bibnamefont
  {Rayment}}, \bibinfo {author} {\bibfnamefont {H.}~\bibnamefont {Holden}},
  \bibinfo {author} {\bibfnamefont {M.}~\bibnamefont {Whittaker}}, \bibinfo
  {author} {\bibfnamefont {C.}~\bibnamefont {Yohn}}, \bibinfo {author}
  {\bibfnamefont {M.}~\bibnamefont {Lorenz}}, \bibinfo {author} {\bibfnamefont
  {K.}~\bibnamefont {Holmes}}, \ and\ \bibinfo {author} {\bibfnamefont
  {R.}~\bibnamefont {Milligan}},\ }\href@noop {} {\bibfield  {journal}
  {\bibinfo  {journal} {Science}\ }\textbf {\bibinfo {volume} {261}},\ \bibinfo
  {pages} {58} (\bibinfo {year} {1993})}\BibitemShut {NoStop}%
\bibitem [{\citenamefont {Dominguez}\ \emph {et~al.}(1998)\citenamefont
  {Dominguez}, \citenamefont {Freyzon}, \citenamefont {Trybus},\ and\
  \citenamefont {Cohen}}]{Dominguez_1998}%
  \BibitemOpen
  \bibfield  {author} {\bibinfo {author} {\bibfnamefont {R.}~\bibnamefont
  {Dominguez}}, \bibinfo {author} {\bibfnamefont {Y.}~\bibnamefont {Freyzon}},
  \bibinfo {author} {\bibfnamefont {K.~M.}\ \bibnamefont {Trybus}}, \ and\
  \bibinfo {author} {\bibfnamefont {C.}~\bibnamefont {Cohen}},\ }\href@noop {}
  {\bibfield  {journal} {\bibinfo  {journal} {Cell}\ }\textbf {\bibinfo
  {volume} {94}},\ \bibinfo {pages} {559} (\bibinfo {year} {1998})}\BibitemShut
  {NoStop}%
\bibitem [{\citenamefont {Lymn}\ and\ \citenamefont
  {Taylor}(1971)}]{Lymn_1971}%
  \BibitemOpen
  \bibfield  {author} {\bibinfo {author} {\bibfnamefont {R.}~\bibnamefont
  {Lymn}}\ and\ \bibinfo {author} {\bibfnamefont {E.}~\bibnamefont {Taylor}},\
  }\href@noop {} {\bibfield  {journal} {\bibinfo  {journal} {Biochemistry}\
  }\textbf {\bibinfo {volume} {10}},\ \bibinfo {pages} {4617} (\bibinfo {year}
  {1971})}\BibitemShut {NoStop}%
\bibitem [{\citenamefont {Alberts}\ \emph {et~al.}(2007)\citenamefont
  {Alberts}, \citenamefont {Johnson}, \citenamefont {Lewis}, \citenamefont
  {Raff}, \citenamefont {Roberts},\ and\ \citenamefont
  {Walter}}]{Alberts:2007vj}%
  \BibitemOpen
  \bibfield  {author} {\bibinfo {author} {\bibfnamefont {B.}~\bibnamefont
  {Alberts}}, \bibinfo {author} {\bibfnamefont {A.}~\bibnamefont {Johnson}},
  \bibinfo {author} {\bibfnamefont {J.}~\bibnamefont {Lewis}}, \bibinfo
  {author} {\bibfnamefont {M.}~\bibnamefont {Raff}}, \bibinfo {author}
  {\bibfnamefont {K.}~\bibnamefont {Roberts}}, \ and\ \bibinfo {author}
  {\bibfnamefont {P.}~\bibnamefont {Walter}},\ }\href@noop {} {\emph {\bibinfo
  {title} {{Molecular Biology of the Cell}}}},\ \bibinfo {edition} {5th}\ ed.\
  (\bibinfo  {publisher} {Garland Science, New York},\ \bibinfo {year}
  {2007})\BibitemShut {NoStop}%
\bibitem [{\citenamefont {Hilbert}\ \emph {et~al.}(2013)\citenamefont
  {Hilbert}, \citenamefont {Cumarasamy}, \citenamefont {Zitouni}, \citenamefont
  {Mackey},\ and\ \citenamefont {Lauzon}}]{Hilbert:2013bb}%
  \BibitemOpen
  \bibfield  {author} {\bibinfo {author} {\bibfnamefont {L.}~\bibnamefont
  {Hilbert}}, \bibinfo {author} {\bibfnamefont {S.}~\bibnamefont {Cumarasamy}},
  \bibinfo {author} {\bibfnamefont {N.~B.}\ \bibnamefont {Zitouni}}, \bibinfo
  {author} {\bibfnamefont {M.~C.}\ \bibnamefont {Mackey}}, \ and\ \bibinfo
  {author} {\bibfnamefont {A.-M.}\ \bibnamefont {Lauzon}},\ }\href@noop {}
  {\bibfield  {journal} {\bibinfo  {journal} {Biophysj}\ }\textbf {\bibinfo
  {volume} {105}},\ \bibinfo {pages} {1466} (\bibinfo {year}
  {2013})}\BibitemShut {NoStop}%
\bibitem [{\citenamefont {Hilbert}\ \emph {et~al.}(2015)\citenamefont
  {Hilbert}, \citenamefont {Balassy}, \citenamefont {Zitouni}, \citenamefont
  {Mackey},\ and\ \citenamefont {Lauzon}}]{Hilbert:2015bt}%
  \BibitemOpen
  \bibfield  {author} {\bibinfo {author} {\bibfnamefont {L.}~\bibnamefont
  {Hilbert}}, \bibinfo {author} {\bibfnamefont {Z.}~\bibnamefont {Balassy}},
  \bibinfo {author} {\bibfnamefont {N.~B.}\ \bibnamefont {Zitouni}}, \bibinfo
  {author} {\bibfnamefont {M.~C.}\ \bibnamefont {Mackey}}, \ and\ \bibinfo
  {author} {\bibfnamefont {A.-M.}\ \bibnamefont {Lauzon}},\ }\href@noop {}
  {\bibfield  {journal} {\bibinfo  {journal} {Biophys. J.}\ }\textbf {\bibinfo
  {volume} {108}},\ \bibinfo {pages} {622} (\bibinfo {year}
  {2015})}\BibitemShut {NoStop}%
\bibitem [{\citenamefont {Baker}\ \emph {et~al.}(2002)\citenamefont {Baker},
  \citenamefont {Brosseau}, \citenamefont {Joel},\ and\ \citenamefont
  {Warshaw}}]{Baker:2002gz}%
  \BibitemOpen
  \bibfield  {author} {\bibinfo {author} {\bibfnamefont {J.~E.}\ \bibnamefont
  {Baker}}, \bibinfo {author} {\bibfnamefont {C.}~\bibnamefont {Brosseau}},
  \bibinfo {author} {\bibfnamefont {P.~B.}\ \bibnamefont {Joel}}, \ and\
  \bibinfo {author} {\bibfnamefont {D.~M.}\ \bibnamefont {Warshaw}},\
  }\href@noop {} {\bibfield  {journal} {\bibinfo  {journal} {Biophysj}\
  }\textbf {\bibinfo {volume} {82}},\ \bibinfo {pages} {2134} (\bibinfo {year}
  {2002})}\BibitemShut {NoStop}%
\bibitem [{\citenamefont {Baker}\ \emph {et~al.}(2003)\citenamefont {Baker},
  \citenamefont {Brosseau}, \citenamefont {Fagnant},\ and\ \citenamefont
  {Warshaw}}]{Baker:2003kz}%
  \BibitemOpen
  \bibfield  {author} {\bibinfo {author} {\bibfnamefont {J.~E.}\ \bibnamefont
  {Baker}}, \bibinfo {author} {\bibfnamefont {C.}~\bibnamefont {Brosseau}},
  \bibinfo {author} {\bibfnamefont {P.}~\bibnamefont {Fagnant}}, \ and\
  \bibinfo {author} {\bibfnamefont {D.~M.}\ \bibnamefont {Warshaw}},\
  }\href@noop {} {\bibfield  {journal} {\bibinfo  {journal} {J. Biol. Chem.}\
  }\textbf {\bibinfo {volume} {278}},\ \bibinfo {pages} {28533} (\bibinfo
  {year} {2003})}\BibitemShut {NoStop}%
\bibitem [{\citenamefont {Haldeman}\ \emph {et~al.}(2014)\citenamefont
  {Haldeman}, \citenamefont {Brizendine}, \citenamefont {Facemyer},
  \citenamefont {Baker},\ and\ \citenamefont {Cremo}}]{Haldeman:2014kx}%
  \BibitemOpen
  \bibfield  {author} {\bibinfo {author} {\bibfnamefont {B.~D.}\ \bibnamefont
  {Haldeman}}, \bibinfo {author} {\bibfnamefont {R.~K.}\ \bibnamefont
  {Brizendine}}, \bibinfo {author} {\bibfnamefont {K.~C.}\ \bibnamefont
  {Facemyer}}, \bibinfo {author} {\bibfnamefont {J.~E.}\ \bibnamefont {Baker}},
  \ and\ \bibinfo {author} {\bibfnamefont {C.~R.}\ \bibnamefont {Cremo}},\
  }\href@noop {} {\bibfield  {journal} {\bibinfo  {journal} {J. Biol. Chem.}\
  }\textbf {\bibinfo {volume} {289}},\ \bibinfo {pages} {21055} (\bibinfo
  {year} {2014})}\BibitemShut {NoStop}%
\bibitem [{\citenamefont {{Anneka M Hooft}}\ \emph {et~al.}(2007)\citenamefont
  {{Anneka M Hooft}}, \citenamefont {{Erik J Maki}}, \citenamefont {Cox},\ and\
  \citenamefont {Baker}}]{AnnekaMHooft:2007kp}%
  \BibitemOpen
  \bibfield  {author} {\bibinfo {author} {\bibnamefont {{Anneka M Hooft}}},
  \bibinfo {author} {\bibnamefont {{Erik J Maki}}}, \bibinfo {author}
  {\bibfnamefont {K.~K.}\ \bibnamefont {Cox}}, \ and\ \bibinfo {author}
  {\bibfnamefont {J.~E.}\ \bibnamefont {Baker}},\ }\href@noop {} {\bibfield
  {journal} {\bibinfo  {journal} {Biochemistry}\ }\textbf {\bibinfo {volume}
  {46}},\ \bibinfo {pages} {3513} (\bibinfo {year} {2007})}\BibitemShut
  {NoStop}%
\bibitem [{\citenamefont {Del R~Jackson}\ and\ \citenamefont
  {Baker}(2009)}]{DelRJackson:2009hg}%
  \BibitemOpen
  \bibfield  {author} {\bibinfo {author} {\bibfnamefont {J.}~\bibnamefont {Del
  R~Jackson}}\ and\ \bibinfo {author} {\bibfnamefont {J.~E.}\ \bibnamefont
  {Baker}},\ }\href@noop {} {\bibfield  {journal} {\bibinfo  {journal}
  {Physical Chemistry Chemical Physics}\ }\textbf {\bibinfo {volume} {11}},\
  \bibinfo {pages} {4808} (\bibinfo {year} {2009})}\BibitemShut {NoStop}%
\bibitem [{\citenamefont {Caremani}\ \emph {et~al.}(2015)\citenamefont
  {Caremani}, \citenamefont {Melli}, \citenamefont {Dolfi}, \citenamefont
  {Lombardi},\ and\ \citenamefont {Linari}}]{Caremani:2015hr}%
  \BibitemOpen
  \bibfield  {author} {\bibinfo {author} {\bibfnamefont {M.}~\bibnamefont
  {Caremani}}, \bibinfo {author} {\bibfnamefont {L.}~\bibnamefont {Melli}},
  \bibinfo {author} {\bibfnamefont {M.}~\bibnamefont {Dolfi}}, \bibinfo
  {author} {\bibfnamefont {V.}~\bibnamefont {Lombardi}}, \ and\ \bibinfo
  {author} {\bibfnamefont {M.}~\bibnamefont {Linari}},\ }\href@noop {}
  {\bibfield  {journal} {\bibinfo  {journal} {J. Physiol.}\ }\textbf {\bibinfo
  {volume} {593}},\ \bibinfo {pages} {3313} (\bibinfo {year}
  {2015})}\BibitemShut {NoStop}%
\bibitem [{\citenamefont {Gilbert}\ and\ \citenamefont
  {Ford}(1988)}]{Gilbert:1988ir}%
  \BibitemOpen
  \bibfield  {author} {\bibinfo {author} {\bibfnamefont {S.~H.}\ \bibnamefont
  {Gilbert}}\ and\ \bibinfo {author} {\bibfnamefont {L.~E.}\ \bibnamefont
  {Ford}},\ }\href@noop {} {\bibfield  {journal} {\bibinfo  {journal}
  {Biophysj}\ }\textbf {\bibinfo {volume} {54}},\ \bibinfo {pages} {611}
  (\bibinfo {year} {1988})}\BibitemShut {NoStop}%
\bibitem [{\citenamefont {Pate}\ \emph {et~al.}(1994)\citenamefont {Pate},
  \citenamefont {Wilson}, \citenamefont {Bhimani},\ and\ \citenamefont
  {Cooke}}]{Pate:1994gy}%
  \BibitemOpen
  \bibfield  {author} {\bibinfo {author} {\bibfnamefont {E.}~\bibnamefont
  {Pate}}, \bibinfo {author} {\bibfnamefont {G.~J.}\ \bibnamefont {Wilson}},
  \bibinfo {author} {\bibfnamefont {M.}~\bibnamefont {Bhimani}}, \ and\
  \bibinfo {author} {\bibfnamefont {R.}~\bibnamefont {Cooke}},\ }\href@noop {}
  {\bibfield  {journal} {\bibinfo  {journal} {Biophysj}\ }\textbf {\bibinfo
  {volume} {66}},\ \bibinfo {pages} {1554} (\bibinfo {year}
  {1994})}\BibitemShut {NoStop}%
\bibitem [{\citenamefont {Linari}\ \emph {et~al.}(2004)\citenamefont {Linari},
  \citenamefont {Bottinelli}, \citenamefont {Pellegrino}, \citenamefont
  {Reconditi}, \citenamefont {Reggiani},\ and\ \citenamefont
  {Lombardi}}]{Linari_2004}%
  \BibitemOpen
  \bibfield  {author} {\bibinfo {author} {\bibfnamefont {M.}~\bibnamefont
  {Linari}}, \bibinfo {author} {\bibfnamefont {R.}~\bibnamefont {Bottinelli}},
  \bibinfo {author} {\bibfnamefont {M.}~\bibnamefont {Pellegrino}}, \bibinfo
  {author} {\bibfnamefont {M.}~\bibnamefont {Reconditi}}, \bibinfo {author}
  {\bibfnamefont {C.}~\bibnamefont {Reggiani}}, \ and\ \bibinfo {author}
  {\bibfnamefont {V.}~\bibnamefont {Lombardi}},\ }\href@noop {} {\bibfield
  {journal} {\bibinfo  {journal} {J. Physiol. - London}\ }\textbf {\bibinfo
  {volume} {555}},\ \bibinfo {pages} {851} (\bibinfo {year}
  {2004})}\BibitemShut {NoStop}%
\bibitem [{\citenamefont {Ranatunga}(2010)}]{Ranatunga:2010ft}%
  \BibitemOpen
  \bibfield  {author} {\bibinfo {author} {\bibfnamefont {K.~W.}\ \bibnamefont
  {Ranatunga}},\ }\href@noop {} {\bibfield  {journal} {\bibinfo  {journal} {The
  Journal of physiology}\ }\textbf {\bibinfo {volume} {588}},\ \bibinfo {pages}
  {3657} (\bibinfo {year} {2010})}\BibitemShut {NoStop}%
\bibitem [{\citenamefont {Reconditi}(2006)}]{Reconditi_2006}%
  \BibitemOpen
  \bibfield  {author} {\bibinfo {author} {\bibfnamefont {M.}~\bibnamefont
  {Reconditi}},\ }\href@noop {} {\bibfield  {journal} {\bibinfo  {journal}
  {Rep. Prog. Phys.}\ }\textbf {\bibinfo {volume} {69}},\ \bibinfo {pages}
  {2709} (\bibinfo {year} {2006})}\BibitemShut {NoStop}%
\bibitem [{\citenamefont {Sheshka}\ and\ \citenamefont
  {Truskinovsky}(2014)}]{Sheshka:2014jt}%
  \BibitemOpen
  \bibfield  {author} {\bibinfo {author} {\bibfnamefont {R.}~\bibnamefont
  {Sheshka}}\ and\ \bibinfo {author} {\bibfnamefont {L.}~\bibnamefont
  {Truskinovsky}},\ }\href@noop {} {\bibfield  {journal} {\bibinfo  {journal}
  {Phys. Rev. E}\ }\textbf {\bibinfo {volume} {89}},\ \bibinfo {pages} {012708}
  (\bibinfo {year} {2014})}\BibitemShut {NoStop}%
\bibitem [{\citenamefont {Huxley}(1974)}]{Huxley:1974we}%
  \BibitemOpen
  \bibfield  {author} {\bibinfo {author} {\bibfnamefont {A.~F.}\ \bibnamefont
  {Huxley}},\ }\href@noop {} {\bibfield  {journal} {\bibinfo  {journal} {J.
  Physiol.}\ }\textbf {\bibinfo {volume} {243}},\ \bibinfo {pages} {1}
  (\bibinfo {year} {1974})}\BibitemShut {NoStop}%
\bibitem [{\citenamefont {Armstrong}\ \emph {et~al.}(1966)\citenamefont
  {Armstrong}, \citenamefont {Huxley},\ and\ \citenamefont
  {Julian}}]{Armstrong:ta}%
  \BibitemOpen
  \bibfield  {author} {\bibinfo {author} {\bibfnamefont {C.~F.}\ \bibnamefont
  {Armstrong}}, \bibinfo {author} {\bibfnamefont {A.~F.}\ \bibnamefont
  {Huxley}}, \ and\ \bibinfo {author} {\bibfnamefont {F.~J.}\ \bibnamefont
  {Julian}},\ }\href@noop {} {\bibfield  {journal} {\bibinfo  {journal} {J.
  Physiol.}\ }\textbf {\bibinfo {volume} {186}},\ \bibinfo {pages} {22P}
  (\bibinfo {year} {1966})}\BibitemShut {NoStop}%
\bibitem [{\citenamefont {Edman}(1988)}]{Edman_1988}%
  \BibitemOpen
  \bibfield  {author} {\bibinfo {author} {\bibfnamefont {K.~A.~P.}\
  \bibnamefont {Edman}},\ }\href@noop {} {\bibfield  {journal} {\bibinfo
  {journal} {J. Physiol. - London}\ }\textbf {\bibinfo {volume} {404}},\
  \bibinfo {pages} {301} (\bibinfo {year} {1988})}\BibitemShut {NoStop}%
\bibitem [{\citenamefont {Yasuda}\ \emph {et~al.}(1996)\citenamefont {Yasuda},
  \citenamefont {Shindo},\ and\ \citenamefont {Ishiwata}}]{Yasuda:1996ta}%
  \BibitemOpen
  \bibfield  {author} {\bibinfo {author} {\bibfnamefont {K.}~\bibnamefont
  {Yasuda}}, \bibinfo {author} {\bibfnamefont {Y.}~\bibnamefont {Shindo}}, \
  and\ \bibinfo {author} {\bibfnamefont {S.}~\bibnamefont {Ishiwata}},\
  }\href@noop {} {\bibfield  {journal} {\bibinfo  {journal} {Biophys. J.}\
  }\textbf {\bibinfo {volume} {70}},\ \bibinfo {pages} {1823} (\bibinfo {year}
  {1996})}\BibitemShut {NoStop}%
\bibitem [{\citenamefont {Edman}\ and\ \citenamefont
  {Curtin}(2001)}]{Edman_2001}%
  \BibitemOpen
  \bibfield  {author} {\bibinfo {author} {\bibfnamefont {K.~A.~P.}\
  \bibnamefont {Edman}}\ and\ \bibinfo {author} {\bibfnamefont
  {N.}~\bibnamefont {Curtin}},\ }\href@noop {} {\bibfield  {journal} {\bibinfo
  {journal} {J. Physiol. - London}\ }\textbf {\bibinfo {volume} {534}},\
  \bibinfo {pages} {553} (\bibinfo {year} {2001})}\BibitemShut {NoStop}%
\bibitem [{\citenamefont {Pla{\c c}ais}\ \emph {et~al.}(2009)\citenamefont
  {Pla{\c c}ais}, \citenamefont {Balland}, \citenamefont {Guerin},
  \citenamefont {Joanny},\ and\ \citenamefont {Martin}}]{Placais_2009}%
  \BibitemOpen
  \bibfield  {author} {\bibinfo {author} {\bibfnamefont {P.~Y.}\ \bibnamefont
  {Pla{\c c}ais}}, \bibinfo {author} {\bibfnamefont {M.}~\bibnamefont
  {Balland}}, \bibinfo {author} {\bibfnamefont {T.}~\bibnamefont {Guerin}},
  \bibinfo {author} {\bibfnamefont {J.~F.}\ \bibnamefont {Joanny}}, \ and\
  \bibinfo {author} {\bibfnamefont {P.}~\bibnamefont {Martin}},\ }\href@noop {}
  {\bibfield  {journal} {\bibinfo  {journal} {Phys. Rev. Lett.}\ }\textbf
  {\bibinfo {volume} {103}} (\bibinfo {year} {2009})}\BibitemShut {NoStop}%
\bibitem [{\citenamefont {Duke}(1999)}]{Duke_1999}%
  \BibitemOpen
  \bibfield  {author} {\bibinfo {author} {\bibfnamefont {T.~A.~J.}\
  \bibnamefont {Duke}},\ }\href@noop {} {\bibfield  {journal} {\bibinfo
  {journal} {Proc. Natl. Acad. Sci. USA}\ }\textbf {\bibinfo {volume} {96}},\
  \bibinfo {pages} {2770} (\bibinfo {year} {1999})}\BibitemShut {NoStop}%
\bibitem [{\citenamefont {Duke}(2000)}]{Duke_2000}%
  \BibitemOpen
  \bibfield  {author} {\bibinfo {author} {\bibfnamefont {T.}~\bibnamefont
  {Duke}},\ }\href@noop {} {\bibfield  {journal} {\bibinfo  {journal} {Philos.
  T. Roy. Soc. B}\ }\textbf {\bibinfo {volume} {355}},\ \bibinfo {pages} {529}
  (\bibinfo {year} {2000})}\BibitemShut {NoStop}%
\bibitem [{\citenamefont {Badoual}\ \emph {et~al.}(2002)\citenamefont
  {Badoual}, \citenamefont {J{\"u}licher},\ and\ \citenamefont
  {Prost}}]{Badoual:2002ky}%
  \BibitemOpen
  \bibfield  {author} {\bibinfo {author} {\bibfnamefont {M.}~\bibnamefont
  {Badoual}}, \bibinfo {author} {\bibfnamefont {F.}~\bibnamefont
  {J{\"u}licher}}, \ and\ \bibinfo {author} {\bibfnamefont {J.}~\bibnamefont
  {Prost}},\ }\href@noop {} {\bibfield  {journal} {\bibinfo  {journal} {Proc.
  Natl. Acad. Sci. U.S.A.}\ }\textbf {\bibinfo {volume} {99}},\ \bibinfo
  {pages} {6696} (\bibinfo {year} {2002})}\BibitemShut {NoStop}%
\bibitem [{\citenamefont {Walcott}\ and\ \citenamefont
  {Sun}(2009)}]{Walcott_2009}%
  \BibitemOpen
  \bibfield  {author} {\bibinfo {author} {\bibfnamefont {S.}~\bibnamefont
  {Walcott}}\ and\ \bibinfo {author} {\bibfnamefont {S.~X.}\ \bibnamefont
  {Sun}},\ }\href@noop {} {\bibfield  {journal} {\bibinfo  {journal} {Physical
  Chemistry Chemical Physics}\ }\textbf {\bibinfo {volume} {11}},\ \bibinfo
  {pages} {4871} (\bibinfo {year} {2009})}\BibitemShut {NoStop}%
\bibitem [{\citenamefont {Duke}\ \emph {et~al.}(2001)\citenamefont {Duke},
  \citenamefont {Le~Novere},\ and\ \citenamefont {Bray}}]{Duke:2001uf}%
  \BibitemOpen
  \bibfield  {author} {\bibinfo {author} {\bibfnamefont {T.~A.}\ \bibnamefont
  {Duke}}, \bibinfo {author} {\bibfnamefont {N.}~\bibnamefont {Le~Novere}}, \
  and\ \bibinfo {author} {\bibfnamefont {D.}~\bibnamefont {Bray}},\ }\href@noop
  {} {\bibfield  {journal} {\bibinfo  {journal} {J. Mol. Biol.}\ }\textbf
  {\bibinfo {volume} {308}},\ \bibinfo {pages} {541} (\bibinfo {year}
  {2001})}\BibitemShut {NoStop}%
\bibitem [{\citenamefont {Bray}(2013)}]{Bray:2013gx}%
  \BibitemOpen
  \bibfield  {author} {\bibinfo {author} {\bibfnamefont {D.}~\bibnamefont
  {Bray}},\ }\href@noop {} {\bibfield  {journal} {\bibinfo  {journal} {Journal
  of Molecular Biology}\ }\textbf {\bibinfo {volume} {425}},\ \bibinfo {pages}
  {1410} (\bibinfo {year} {2013})}\BibitemShut {NoStop}%
\bibitem [{\citenamefont {Bormuth}\ \emph {et~al.}(2014)\citenamefont
  {Bormuth}, \citenamefont {Barral}, \citenamefont {Joanny}, \citenamefont
  {J{\"u}licher},\ and\ \citenamefont {Martin}}]{Bormuth:2014hh}%
  \BibitemOpen
  \bibfield  {author} {\bibinfo {author} {\bibfnamefont {V.}~\bibnamefont
  {Bormuth}}, \bibinfo {author} {\bibfnamefont {J.}~\bibnamefont {Barral}},
  \bibinfo {author} {\bibfnamefont {J.~F.}\ \bibnamefont {Joanny}}, \bibinfo
  {author} {\bibfnamefont {F.}~\bibnamefont {J{\"u}licher}}, \ and\ \bibinfo
  {author} {\bibfnamefont {P.}~\bibnamefont {Martin}},\ }\href@noop {}
  {\bibfield  {journal} {\bibinfo  {journal} {Proc. Natl. Acad. Sci. U.S.A.}\
  }\textbf {\bibinfo {volume} {111}},\ \bibinfo {pages} {7185} (\bibinfo {year}
  {2014})}\BibitemShut {NoStop}%
\bibitem [{\citenamefont {Chen}\ and\ \citenamefont {Gao}(2011)}]{Chen:2011go}%
  \BibitemOpen
  \bibfield  {author} {\bibinfo {author} {\bibfnamefont {B.}~\bibnamefont
  {Chen}}\ and\ \bibinfo {author} {\bibfnamefont {H.}~\bibnamefont {Gao}},\
  }\href@noop {} {\bibfield  {journal} {\bibinfo  {journal} {Biophys. J.}\
  }\textbf {\bibinfo {volume} {101}},\ \bibinfo {pages} {396} (\bibinfo {year}
  {2011})}\BibitemShut {NoStop}%
\bibitem [{\citenamefont {Yao}\ and\ \citenamefont {Gao}(2006)}]{Yao:2006de}%
  \BibitemOpen
  \bibfield  {author} {\bibinfo {author} {\bibfnamefont {H.}~\bibnamefont
  {Yao}}\ and\ \bibinfo {author} {\bibfnamefont {H.}~\bibnamefont {Gao}},\
  }\href@noop {} {\bibfield  {journal} {\bibinfo  {journal} {J. Mech. Phys.
  Solids}\ }\textbf {\bibinfo {volume} {54}},\ \bibinfo {pages} {1120}
  (\bibinfo {year} {2006})}\BibitemShut {NoStop}%
\bibitem [{\citenamefont {Gao}\ \emph {et~al.}(2011)\citenamefont {Gao},
  \citenamefont {Qian},\ and\ \citenamefont {Chen}}]{Gao:2011ij}%
  \BibitemOpen
  \bibfield  {author} {\bibinfo {author} {\bibfnamefont {H.}~\bibnamefont
  {Gao}}, \bibinfo {author} {\bibfnamefont {J.}~\bibnamefont {Qian}}, \ and\
  \bibinfo {author} {\bibfnamefont {B.}~\bibnamefont {Chen}},\ }\href@noop {}
  {\bibfield  {journal} {\bibinfo  {journal} {J. R. Soc. Interface}\ }\textbf
  {\bibinfo {volume} {8}},\ \bibinfo {pages} {1217} (\bibinfo {year}
  {2011})}\BibitemShut {NoStop}%
\bibitem [{\citenamefont {Schwarz}\ and\ \citenamefont
  {Safran}(2013)}]{Schwarz:2013em}%
  \BibitemOpen
  \bibfield  {author} {\bibinfo {author} {\bibfnamefont {U.~S.}\ \bibnamefont
  {Schwarz}}\ and\ \bibinfo {author} {\bibfnamefont {S.~A.}\ \bibnamefont
  {Safran}},\ }\href@noop {} {\bibfield  {journal} {\bibinfo  {journal}
  {Reviews of Modern Physics}\ }\textbf {\bibinfo {volume} {85}},\ \bibinfo
  {pages} {1327} (\bibinfo {year} {2013})}\BibitemShut {NoStop}%
\bibitem [{\citenamefont {Hanggi}\ \emph {et~al.}(1990)\citenamefont {Hanggi},
  \citenamefont {Talkner},\ and\ \citenamefont {Borkovec}}]{Hanggi_1990}%
  \BibitemOpen
  \bibfield  {author} {\bibinfo {author} {\bibfnamefont {P.}~\bibnamefont
  {Hanggi}}, \bibinfo {author} {\bibfnamefont {P.}~\bibnamefont {Talkner}}, \
  and\ \bibinfo {author} {\bibfnamefont {M.}~\bibnamefont {Borkovec}},\
  }\href@noop {} {\bibfield  {journal} {\bibinfo  {journal} {Reviews of Modern
  Physics}\ }\textbf {\bibinfo {volume} {62}},\ \bibinfo {pages} {251}
  (\bibinfo {year} {1990})}\BibitemShut {NoStop}%
\bibitem [{\citenamefont {Erdmann}\ and\ \citenamefont
  {Schwarz}(2004{\natexlab{a}})}]{Erdmann:2004do}%
  \BibitemOpen
  \bibfield  {author} {\bibinfo {author} {\bibfnamefont {T.}~\bibnamefont
  {Erdmann}}\ and\ \bibinfo {author} {\bibfnamefont {U.~S.}\ \bibnamefont
  {Schwarz}},\ }\href@noop {} {\bibfield  {journal} {\bibinfo  {journal} {Phys.
  Rev. Lett.}\ }\textbf {\bibinfo {volume} {92}},\ \bibinfo {pages} {108102}
  (\bibinfo {year} {2004}{\natexlab{a}})}\BibitemShut {NoStop}%
\bibitem [{\citenamefont {Erdmann}\ and\ \citenamefont
  {Schwarz}(2004{\natexlab{b}})}]{Erdmann:2004hr}%
  \BibitemOpen
  \bibfield  {author} {\bibinfo {author} {\bibfnamefont {T.}~\bibnamefont
  {Erdmann}}\ and\ \bibinfo {author} {\bibfnamefont {U.~S.}\ \bibnamefont
  {Schwarz}},\ }\href@noop {} {\bibfield  {journal} {\bibinfo  {journal} {J.
  Chem. Phys.}\ }\textbf {\bibinfo {volume} {121}},\ \bibinfo {pages} {8997}
  (\bibinfo {year} {2004}{\natexlab{b}})}\BibitemShut {NoStop}%
\bibitem [{\citenamefont {Seifert}(2000)}]{Seifert:2000uk}%
  \BibitemOpen
  \bibfield  {author} {\bibinfo {author} {\bibfnamefont {U.}~\bibnamefont
  {Seifert}},\ }\href@noop {} {\bibfield  {journal} {\bibinfo  {journal} {Phys.
  Rev. Lett.}\ }\textbf {\bibinfo {volume} {84}},\ \bibinfo {pages} {2750}
  (\bibinfo {year} {2000})}\BibitemShut {NoStop}%
\bibitem [{\citenamefont {S{\"u}dhof}\ and\ \citenamefont
  {Rothman}(2009)}]{Sudhof:2009cw}%
  \BibitemOpen
  \bibfield  {author} {\bibinfo {author} {\bibfnamefont {T.~C.}\ \bibnamefont
  {S{\"u}dhof}}\ and\ \bibinfo {author} {\bibfnamefont {J.~E.}\ \bibnamefont
  {Rothman}},\ }\href@noop {} {\bibfield  {journal} {\bibinfo  {journal}
  {Science}\ }\textbf {\bibinfo {volume} {323}},\ \bibinfo {pages} {474}
  (\bibinfo {year} {2009})}\BibitemShut {NoStop}%
\bibitem [{\citenamefont {Min}\ \emph {et~al.}(2013)\citenamefont {Min},
  \citenamefont {Kim}, \citenamefont {Hyeon}, \citenamefont {Cho},
  \citenamefont {Shin},\ and\ \citenamefont {Yoon}}]{Min:2013kr}%
  \BibitemOpen
  \bibfield  {author} {\bibinfo {author} {\bibfnamefont {D.}~\bibnamefont
  {Min}}, \bibinfo {author} {\bibfnamefont {K.}~\bibnamefont {Kim}}, \bibinfo
  {author} {\bibfnamefont {C.}~\bibnamefont {Hyeon}}, \bibinfo {author}
  {\bibfnamefont {Y.~H.}\ \bibnamefont {Cho}}, \bibinfo {author} {\bibfnamefont
  {Y.-K.}\ \bibnamefont {Shin}}, \ and\ \bibinfo {author} {\bibfnamefont
  {T.-Y.}\ \bibnamefont {Yoon}},\ }\href@noop {} {\bibfield  {journal}
  {\bibinfo  {journal} {Nat. Commun.}\ }\textbf {\bibinfo {volume} {4}},\
  \bibinfo {pages} {1705} (\bibinfo {year} {2013})}\BibitemShut {NoStop}%
\bibitem [{\citenamefont {Li}\ \emph {et~al.}(2007)\citenamefont {Li},
  \citenamefont {Pincet}, \citenamefont {Perez}, \citenamefont {Eng},
  \citenamefont {Melia}, \citenamefont {Rothman},\ and\ \citenamefont
  {Tareste}}]{Li:2007ka}%
  \BibitemOpen
  \bibfield  {author} {\bibinfo {author} {\bibfnamefont {F.}~\bibnamefont
  {Li}}, \bibinfo {author} {\bibfnamefont {F.}~\bibnamefont {Pincet}}, \bibinfo
  {author} {\bibfnamefont {E.}~\bibnamefont {Perez}}, \bibinfo {author}
  {\bibfnamefont {W.~S.}\ \bibnamefont {Eng}}, \bibinfo {author} {\bibfnamefont
  {T.~J.}\ \bibnamefont {Melia}}, \bibinfo {author} {\bibfnamefont {J.~E.}\
  \bibnamefont {Rothman}}, \ and\ \bibinfo {author} {\bibfnamefont
  {D.}~\bibnamefont {Tareste}},\ }\href@noop {} {\bibfield  {journal} {\bibinfo
   {journal} {Nature Structural {\&} Molecular Biology}\ }\textbf {\bibinfo
  {volume} {14}},\ \bibinfo {pages} {890} (\bibinfo {year} {2007})}\BibitemShut
  {NoStop}%
\bibitem [{\citenamefont {Zorman}\ \emph {et~al.}(2014)\citenamefont {Zorman},
  \citenamefont {Rebane}, \citenamefont {Ma}, \citenamefont {Yang},
  \citenamefont {Molski}, \citenamefont {Coleman}, \citenamefont {Pincet},
  \citenamefont {Rothman},\ and\ \citenamefont {Zhang}}]{Zorman:2014bg}%
  \BibitemOpen
  \bibfield  {author} {\bibinfo {author} {\bibfnamefont {S.}~\bibnamefont
  {Zorman}}, \bibinfo {author} {\bibfnamefont {A.~A.}\ \bibnamefont {Rebane}},
  \bibinfo {author} {\bibfnamefont {L.}~\bibnamefont {Ma}}, \bibinfo {author}
  {\bibfnamefont {G.}~\bibnamefont {Yang}}, \bibinfo {author} {\bibfnamefont
  {M.~A.}\ \bibnamefont {Molski}}, \bibinfo {author} {\bibfnamefont
  {J.}~\bibnamefont {Coleman}}, \bibinfo {author} {\bibfnamefont
  {F.}~\bibnamefont {Pincet}}, \bibinfo {author} {\bibfnamefont {J.~E.}\
  \bibnamefont {Rothman}}, \ and\ \bibinfo {author} {\bibfnamefont
  {Y.}~\bibnamefont {Zhang}},\ }\href@noop {} {\bibfield  {journal} {\bibinfo
  {journal} {Elife}\ }\textbf {\bibinfo {volume} {3}},\ \bibinfo {pages}
  {e03348} (\bibinfo {year} {2014})}\BibitemShut {NoStop}%
\bibitem [{\citenamefont {Dietz}\ and\ \citenamefont
  {Rief}(2008)}]{Dietz:2008gj}%
  \BibitemOpen
  \bibfield  {author} {\bibinfo {author} {\bibfnamefont {H.}~\bibnamefont
  {Dietz}}\ and\ \bibinfo {author} {\bibfnamefont {M.}~\bibnamefont {Rief}},\
  }\href@noop {} {\bibfield  {journal} {\bibinfo  {journal} {Phys. Rev. Lett.}\
  }\textbf {\bibinfo {volume} {100}},\ \bibinfo {pages} {098101} (\bibinfo
  {year} {2008})}\BibitemShut {NoStop}%
\bibitem [{\citenamefont {Cerf}(1978)}]{Cerf:1978vy}%
  \BibitemOpen
  \bibfield  {author} {\bibinfo {author} {\bibfnamefont {R.}~\bibnamefont
  {Cerf}},\ }\href@noop {} {\bibfield  {journal} {\bibinfo  {journal} {Proc.
  Natl. Acad. Sci. U.S.A.}\ }\textbf {\bibinfo {volume} {75}},\ \bibinfo
  {pages} {2755} (\bibinfo {year} {1978})}\BibitemShut {NoStop}%
\bibitem [{\citenamefont {Pfitzner}\ \emph {et~al.}(2013)\citenamefont
  {Pfitzner}, \citenamefont {Wachauf}, \citenamefont {Kilchherr},\ and\
  \citenamefont {Pelz}}]{Pfitzner:2013ia}%
  \BibitemOpen
  \bibfield  {author} {\bibinfo {author} {\bibfnamefont {E.}~\bibnamefont
  {Pfitzner}}, \bibinfo {author} {\bibfnamefont {C.}~\bibnamefont {Wachauf}},
  \bibinfo {author} {\bibfnamefont {F.}~\bibnamefont {Kilchherr}}, \ and\
  \bibinfo {author} {\bibfnamefont {B.}~\bibnamefont {Pelz}},\ }\href@noop {}
  {\bibfield  {journal} {\bibinfo  {journal} {Angew. Chem. Int. Ed.}\ }\textbf
  {\bibinfo {volume} {52}},\ \bibinfo {pages} {335} (\bibinfo {year}
  {2013})}\BibitemShut {NoStop}%
\bibitem [{\citenamefont {Changeux}\ \emph {et~al.}(1967)\citenamefont
  {Changeux}, \citenamefont {Thi{\'e}ry}, \citenamefont {Tung},\ and\
  \citenamefont {Kittel}}]{Changeux:1967vf}%
  \BibitemOpen
  \bibfield  {author} {\bibinfo {author} {\bibfnamefont {J.~P.}\ \bibnamefont
  {Changeux}}, \bibinfo {author} {\bibfnamefont {J.}~\bibnamefont
  {Thi{\'e}ry}}, \bibinfo {author} {\bibfnamefont {Y.}~\bibnamefont {Tung}}, \
  and\ \bibinfo {author} {\bibfnamefont {C.}~\bibnamefont {Kittel}},\
  }\href@noop {} {\bibfield  {journal} {\bibinfo  {journal} {Proc. Natl. Acad.
  Sci. USA}\ }\textbf {\bibinfo {volume} {57}},\ \bibinfo {pages} {335}
  (\bibinfo {year} {1967})}\BibitemShut {NoStop}%
\bibitem [{\citenamefont {Bray}\ and\ \citenamefont
  {Duke}(2004)}]{Bray:2004iv}%
  \BibitemOpen
  \bibfield  {author} {\bibinfo {author} {\bibfnamefont {D.}~\bibnamefont
  {Bray}}\ and\ \bibinfo {author} {\bibfnamefont {T.}~\bibnamefont {Duke}},\
  }\href@noop {} {\bibfield  {journal} {\bibinfo  {journal} {Annu Rev Biophys
  Biomol Struct}\ }\textbf {\bibinfo {volume} {33}},\ \bibinfo {pages} {53}
  (\bibinfo {year} {2004})}\BibitemShut {NoStop}%
\bibitem [{\citenamefont {Slepchenko}\ and\ \citenamefont
  {Terasaki}(2004)}]{Slepchenko:2004hq}%
  \BibitemOpen
  \bibfield  {author} {\bibinfo {author} {\bibfnamefont {B.~M.}\ \bibnamefont
  {Slepchenko}}\ and\ \bibinfo {author} {\bibfnamefont {M.}~\bibnamefont
  {Terasaki}},\ }\href@noop {} {\bibfield  {journal} {\bibinfo  {journal}
  {Current Opinion in Genetics {\&} Development}\ }\textbf {\bibinfo {volume}
  {14}},\ \bibinfo {pages} {428} (\bibinfo {year} {2004})}\BibitemShut
  {NoStop}%
\bibitem [{\citenamefont {Miyashita}\ \emph {et~al.}(2011)\citenamefont
  {Miyashita}, \citenamefont {Onuchic},\ and\ \citenamefont
  {Wolynes}}]{Miyashita:2011jb}%
  \BibitemOpen
  \bibfield  {author} {\bibinfo {author} {\bibfnamefont {O.}~\bibnamefont
  {Miyashita}}, \bibinfo {author} {\bibfnamefont {J.~N.}\ \bibnamefont
  {Onuchic}}, \ and\ \bibinfo {author} {\bibfnamefont {P.~G.}\ \bibnamefont
  {Wolynes}},\ }\href@noop {} {\bibfield  {journal} {\bibinfo  {journal}
  {PNAS}\ }\textbf {\bibinfo {volume} {100}},\ \bibinfo {pages} {12570}
  (\bibinfo {year} {2011})}\BibitemShut {NoStop}%
\bibitem [{\citenamefont {Yurke}\ \emph {et~al.}(2000)\citenamefont {Yurke},
  \citenamefont {Turberfield}, \citenamefont {Mills}, \citenamefont {Simmel},\
  and\ \citenamefont {Neumann}}]{Yurke:2000ex}%
  \BibitemOpen
  \bibfield  {author} {\bibinfo {author} {\bibfnamefont {B.}~\bibnamefont
  {Yurke}}, \bibinfo {author} {\bibfnamefont {A.~J.}\ \bibnamefont
  {Turberfield}}, \bibinfo {author} {\bibfnamefont {A.~P.}\ \bibnamefont
  {Mills}}, \bibinfo {author} {\bibfnamefont {F.~C.}\ \bibnamefont {Simmel}}, \
  and\ \bibinfo {author} {\bibfnamefont {J.~L.}\ \bibnamefont {Neumann}},\
  }\href@noop {} {\bibfield  {journal} {\bibinfo  {journal} {Nature}\ }\textbf
  {\bibinfo {volume} {406}},\ \bibinfo {pages} {605} (\bibinfo {year}
  {2000})}\BibitemShut {NoStop}%
\bibitem [{\citenamefont {Harne}\ \emph {et~al.}(2016)\citenamefont {Harne},
  \citenamefont {Wu},\ and\ \citenamefont {Wang}}]{Harne:2016gg}%
  \BibitemOpen
  \bibfield  {author} {\bibinfo {author} {\bibfnamefont {R.~L.}\ \bibnamefont
  {Harne}}, \bibinfo {author} {\bibfnamefont {Z.}~\bibnamefont {Wu}}, \ and\
  \bibinfo {author} {\bibfnamefont {K.~W.}\ \bibnamefont {Wang}},\ }\href@noop
  {} {\bibfield  {journal} {\bibinfo  {journal} {J. Mech. Des}\ }\textbf
  {\bibinfo {volume} {138}},\ \bibinfo {pages} {021402} (\bibinfo {year}
  {2016})}\BibitemShut {NoStop}%
\bibitem [{\citenamefont {Guerin}\ \emph {et~al.}(2011)\citenamefont {Guerin},
  \citenamefont {Prost},\ and\ \citenamefont {Joanny}}]{Guerin_2011}%
  \BibitemOpen
  \bibfield  {author} {\bibinfo {author} {\bibfnamefont {T.}~\bibnamefont
  {Guerin}}, \bibinfo {author} {\bibfnamefont {J.}~\bibnamefont {Prost}}, \
  and\ \bibinfo {author} {\bibfnamefont {J.~F.}\ \bibnamefont {Joanny}},\
  }\href@noop {} {\bibfield  {journal} {\bibinfo  {journal} {Eur. Phys. J.}\
  }\textbf {\bibinfo {volume} {34}},\ \bibinfo {pages} {60} (\bibinfo {year}
  {2011})}\BibitemShut {NoStop}%
\bibitem [{\citenamefont {{Sheshka, R}}\ \emph {et~al.}(2016)\citenamefont
  {{Sheshka, R}}, \citenamefont {{Recho, P}},\ and\ \citenamefont
  {{Truskinovsky, Lev}}}]{Sheshka:2016dm}%
  \BibitemOpen
  \bibfield  {author} {\bibinfo {author} {\bibnamefont {{Sheshka, R}}},
  \bibinfo {author} {\bibnamefont {{Recho, P}}}, \ and\ \bibinfo {author}
  {\bibnamefont {{Truskinovsky, Lev}}},\ }\href@noop {} {\bibfield  {journal}
  {\bibinfo  {journal} {Phys. Rev. E}\ }\textbf {\bibinfo {volume} {93}},\
  \bibinfo {pages} {052604} (\bibinfo {year} {2016})}\BibitemShut {NoStop}%
\end{thebibliography}%
 \bibliographystyle{apsrev4-1}

\end{document}